\documentclass{aa}  
\usepackage{graphicx}
\usepackage{txfonts}
\usepackage[colorinlistoftodos]{todonotes}
\usepackage{hyperref}  
\def\ts     {\thinspace} 
\usepackage{amsmath}

\def\kms  {\ifmmode{{\rm \ts km\ts s}^{-1}}\else{\ts km\ts s$^{-1}$\ts}\fi}
\def\msol {\ifmmode{{\rm M}_{\odot}}\else{M$_{\odot}$\ts}\fi}
\def\lsun {\ifmmode{{\rm L}_{\odot}}\else{L$_{\odot}$\ts}\fi}
\def\cii  {\ifmmode{{\rm [C}{\rm \small II}]}\else{[C\ts {\scriptsize II}]\ts}\fi}
\def\m    {\ifmmode{\mu {\rm m}}\else{$\mu$m}\fi}
\def\hi   {\ifmmode{{\rm H}{\rm \small I}}\else{H\ts {\scriptsize I}\ts}\fi}
\def\hii  {\ifmmode{{\rm H}{\rm \small II}}\else{H\ts {\scriptsize II}\ts}\fi}
\def\nii  {\ifmmode{{\rm [N}{\rm \small II}]}\else{[N\ts {\scriptsize II}]\ts}\fi}
\def\oiii {\ifmmode{{\rm [O}{\rm \small III}]}\else{[O\ts {\scriptsize III}]\ts}\fi}
\def\hh   {\ifmmode{{\rm H}_2}\else{H$_2$\ts}\fi}
\def\nhh  {\ifmmode{N({\rm H}_2)}\else{$N$(H$_2$)\ts}\fi}
\def\microns {\ifmmode{\mu{\rm m}}\else{$\mu$m\ts}\fi}

\begin{document} 
\title{The ALMA Frontier Fields survey}

\subtitle{I: 1.1\,mm continuum detections in Abell\,2744, MACSJ0416.1-2403 and MACSJ1149.5+2223}

\author{J. Gonz{\'{a}}lez-L{\'{o}}pez\inst{1}
\and
{F. E. Bauer}\inst{1,2,3}
\and
{C. Romero-Ca\~nizales}\inst{2,1,7}
\and
{R. Kneissl}\inst{4,5}
\and
{E. Villard}\inst{4,5}
\and
{R. Carvajal}\inst{1}
\and
{S. Kim}\inst{1}
\and
{N. Laporte}\inst{1,17}
\and
{T. Anguita}\inst{6,2}
\and
{M. Aravena}\inst{7}
\and
{R. J. Bouwens}\inst{8}
\and
{L. Bradley}\inst{9}
\and
{M. Carrasco}\inst{10}
\and
{R. Demarco}\inst{11}
\and
{H. Ford}\inst{12}
\and
{E. Ibar}\inst{13}
\and
{L. Infante}\inst{1}
\and
{H. Messias}\inst{14}
\and
{A. M. Mu{\~{n}}oz Arancibia}\inst{13,1}
\and
{N. Nagar}\inst{11}
\and
{N. Padilla}\inst{1}
\and
{E. Treister}\inst{1,11}
\and
{P. Troncoso}\inst{1}
\and
{A. Zitrin}\inst{15,16}
}
\institute{Instituto de Astrof\'{\i}sica and Centro de Astroingenier{\'{\i}}a, Facultad de F\'{i}sica, Pontificia Universidad Cat\'{o}lica de Chile, Casilla 306, Santiago 22, Chile\\
              \email{jgonzal@astro.puc.cl}
         \and
{Millennium Institute of Astrophysics, Chile} 
         \and
{Space Science Institute, 4750 Walnut Street, Suite 205, Boulder, Colorado 80301} 
         \and
{Joint ALMA Observatory, Alonso de C\'{o}rdova 3107, Vitacura, Santiago, Chile}
         \and
{European Southern Observatory, Alonso de C\'{o}rdova 3107, Vitacura, Casilla 19001, Santiago, Chile}
		  \and
{Departamento de Ciencias F\'{i}sicas, Universidad Andres Bello, 252 Avenida Rep\'{u}blica, Santiago, Chile}
         \and
{N\'{u}cleo de Astronom\'{i}a, Facultad de Ingenier\'{i}a, Universidad Diego Portales, Av. Ej\'{e}rcito 441, Santiago, Chile.}
		 \and
{Leiden Observatory, Leiden University, NL-2300 RA Leiden, Netherlands}
		 \and
{Space Telescope Science Institute, 3700 San Martin Dr., Baltimore, MD 21218 USA}
		 \and
{Universit\"{a}t Heidelberg, Zentrum f\"{u}r Astronomie, Institut f\"{u}r Theoretische Astrophysik, Philosophenweg 12,
69120 Heidelberg, Germany}
		 \and
{Department of Astronomy, Universidad de Concepcion, Casilla 160-C, Concepci\'{o}n, Chile}
		 \and
{Department of Physics and Astronomy, Johns Hopkins University, Baltimore, MD 21218, USA}
		 \and
{Instituto de F\'isica y Astronom\'ia, Universidad de Valpara\'iso, Avda. Gran Breta\~na 1111, Valpara\'iso, Chile}
		 \and
{Instituto de Astrof\'isica e Ci\^{e}ncias do Espa\c{c}o, Universidade de Lisboa, OAL, Tapada da Ajuda, PT 1349-018 Lisboa, Portugal}
		 \and
{Cahill Center for Astronomy and Astrophysics, California Institute of Technology, MC 249-17, Pasadena, CA 91125, USA}
		 \and
{Hubble Fellow}
		\and
{Department of Physics and Astronomy, University College London, Gower Street, London WC1E 6BT, UK}
}


 
  \abstract
   {Dusty star-forming galaxies are among the most prodigious systems at high redshift  ($z>1$), characterized by high star-formation rates and huge dust reservoirs. The bright end of this population has been well characterized in recent years, but considerable uncertainties remain for fainter dusty star-forming galaxies, which are responsible for the bulk of star formation at high redshift and thus  play a key role in galaxy growth and evolution.}
   {
   In this first paper of our series, we describe our methods for finding high redshift faint dusty galaxies using millimeter observations with ALMA.}
   {We obtained ALMA 1.1\,mm mosaic images for three strong-lensing galaxy clusters from the Frontier Fields survey, which constitute some of the best studied gravitational lenses to date. The $\approx$2$'$$\times$2$'$ mosaics overlap with the deep {\it HST} WFC3/IR footprints and encompass the high magnification regions of each cluster for maximum intrinsic source sensitivity. The combination of extremely high ALMA sensitivity and the magnification power of these clusters allows us to systematically probe the sub-mJy population of dusty star-forming galaxies over a large surveyed area.}
   {We present a description of the reduction and analysis of the ALMA continuum observations for the galaxy clusters Abell 2744 ($z=0.308$), MACSJ0416.1-2403 ($z=0.396$) and MACSJ1149.5+2223 ($z=0.543$), for which we reach observed rms sensitivities of 55, 59 and 71 ${\rm \mu Jy}\,{\rm beam}^{-1}$ respectively. We detect 12 dusty star-forming galaxies at $S/N\geq5.0$ across the three clusters, all of them presenting coincidence with near-infrared detected counterparts in the {\it HST} images. None of the sources fall close to the lensing caustics, thus they are not strongly lensed. The observed 1.1\,mm flux densities for the total sample of galaxies range from 0.41 to 2.82 mJy, with observed effective radii spanning $\lesssim$0\farcs05 to $0\farcs37\pm0\farcs21$. The lensing-corrected sizes of the detected sources appear to be in the same range as those measured in brighter samples, albeit with possibly larger dispersion.
   } 
   {}


   \maketitle

\section{Introduction}

\hspace{1cm}Past studies of dusty, star-forming galaxies (DSFGs) at infrared (IR) through radio wavelengths have firmly established their role in the growth and evolution of massive galaxies across cosmic time \citetext{see reviews by \citealp{Blain2002} and \citealp{Casey2014}}. At the bright end, DSFGs are observed to have IR luminosities in excess of $10^{13}$ $L_{\odot}$, yet  appear extraordinarily compact, with typical resolved half-light radii of $R_{\rm e}$$\approx$1--1.5\,kpc \citetext{e.g.,  \citealp{Bussmann2013,Ikarashi2014,Simpson2015,Miettinen2015};  hereafter B13, I14, S15 and M15, respectively}. The bulk of their IR emission is thought to be powered by star formation \citep{Alexander2005}, with  unobscured star-formation rates (SFRs) of up to several thousands of $M_{\odot}$\,yr$^{-1}$ and  SFR densities of $\gtrsim$100 $M_{\odot}$\,yr$^{-1}$\,kpc$^{-2}$. As such, DSFGs represent the most intense starbursts in the Universe. While little is known about fainter DSFGs (sub-mJy population), extrapolation of the DSFG population is estimated to account for roughly half of the IR background light in aggregate \citep[e.g.,][]{Magnelli2011, Viero2013}. Obtaining a working knowledge of the underlying physics that drives the distribution of faint lensed DSFGs thus appears to be critical for understanding cosmic galaxy assembly.

Many DSFGs are optically faint ($I_{AB}$$\gtrsim$24\,mag, $K_{AB}$$\sim$21-22), due to a combination of high redshift and strong dust-extinction \citep[e.g.,][]{Barger2000, Smail2004, Chapman2005}. This has made unbiased estimates of their population statistics and physical properties (i.e., dust, gas and stellar contents and morphologies) challenging and expensive with current technology. It  also suggests that finding and characterizing the fainter DSFG population may prove even harder. Nonetheless, the fainter population is particularly interesting because their SFRs are in the same range as those found for ultraviolet (UV) and optically selected samples such as star-forming $BzK$ galaxies, BX/BM galaxies, and Lyman break galaxies (LBGs), which comprise the normal galaxy ‘main sequence’ \citep[e.g.,][]{Noeske2007}. Meaningful comparisons between the more abundant faint DSFGs and these unobscured main sequence populations could help to elucidate the factors that determine the dust content in galaxies with comparable properties. 
 
One way to make progress toward the study of faint DSFGs, in the face of instrumental confusion limits at far-IR (FIR) and submillimeter (submm) wavelengths ($\sim$2--7\,mJy beam$^{-1}$ between 0.25--1.3\,mm) as well as traditional difficulties associated with expensive multi-wavelength follow-up, is to leverage with the power of gravitational lensing. Such studies have helped provide detailed characterizations of many dozens of intrinsically faint DSFGs selected either in wide area submm surveys as galaxy-galaxy lenses \citep[e.g.,][]{Blain1996, Negrello2007, Negrello2010, Wardlow2013} or behind massive lensing galaxy clusters \citep[e.g.,][]{Smail1997, Cowie2002, Swinbank2010}. Although immensely insightful, such studies can suffer from additional uncertainties due to the quality of the mass models, microlensing, and potential biases stemming from the cross-section of the DSFG population which is more easily lensed (i.e., compact starbursts). Robustly determined mass models and observations over long timescales can limit the impact of these problems.

Building on past works of lensed DSFGs, here we employ the novel sensitivity of the Atacama Large Millimeter/submillimeter Array (ALMA) to detect DSFGs roughly 1 dex fainter than the aforementioned confusion limits, behind three strong-lensing galaxy clusters: Abell 2744 ($z=0.308$), MACSJ0416.1-2403 ($z=0.396$) and MACSJ1149.5+2223 ($z=0.543$) (hereafter A2744, MACSJ0416, MACSJ1149, respectively).\footnote{Approved ALMA Cycle 3 observations of the remaining three FFs clusters, namely MACSJ0717.5+3745, Abell 370, and Abell 1063S, are in progress.} Importantly, these clusters are part of the Frontier Fields (FFs) Survey,\footnote{\url{http://www.stsci.edu/hst/campaigns/frontier-fields/}} a legacy project which combines the power of gravitational lensing by massive clusters (with magnifications of $\mu$$>$5--10 over up to several arcmin$^{2}$ regions and 100's of multiple images) with extremely deep multi-band {\it HST} and {\it Spitzer} imaging of six lensing clusters and adjacent parallel fields \citep{Coe2015}.
While the primary goal of these observations is to potentially detect and characterize $z\geq1$ galaxies 10--50 times intrinsically fainter than any seen before \citep[e.g.,][]{Atek2014, Zitrin2014, Zheng2014, Laporte2015, Mcleod2015, Infante2015}, the {\it HST}+{\it Spitzer} observations and their associated ancillary data enable several ALMA related science cases.

The FFs campaign is comprised of allocations of 840 {\it HST} orbits and 1000 {\it Spitzer} hrs, respectively, taking advantage of the two Great Observatories unsurpassed spatial resolution and/or depth. The fields have amassed a wealth of multi-wavelength ancillary data. Notably, each cluster already has extensive space+ground-based archival data (e.g., 16-band {\it HST}, {\it XMM-Newton}, {\it Chandra}, {\it Spitzer}, {\it Herschel}, and VLA imaging) and more than 800 spectroscopically confirmed sources (cluster+background galaxies), including VLT/VIMOS confirmations for most of the major gravitational arcs/images down to $\sim$26 ABmag (PI Rosati). To this, the FFs campaign adds deep ACS ($F435W$$=$$F606W$$=$$F814W$$\approx$28.4--29.0\,ABmag), WFC3 ($F105W$$=$$F125W$$=$$F140W$$=$$F160$$\approx$29.1--29.4\,ABmag), and IRAC1/IRAC2 ($\approx$25.0\,ABmag) imaging \citep[e.g.,][]{Coe2015, Lotz2016}. These are complemented by new and growing data from {\it HST}, {\it Chandra}, JVLA, VLT HAWK-I/VIMOS/MUSE, Keck DEIMOS/LRIS/MOSFIRE, and Magellan FourSTAR/IMACS/MMIRS. The current data already allow the assembly of 10,000's of accurate $z_{\rm ph}$'s, 400/200/10's of Lyman Break dropouts at $z$$>$6/7/8, and spectacularly resolved images of $>$600 $z$$=$1--6 gravitational arcs (2--10") and multiple-lenses in a central region of each cluster \citep[e.g.,][]{Richard2014}. The FFs clusters represent the best studied gravitational lenses, with the best available estimates of magnifications and uncertainties (ever-improving lens models are available from many teams).\footnote{\url{http://www.stsci.edu/hst/campaigns/frontier-fields/Lensing-Models}} Thus they are key regions on the sky to observe normal galaxies at high-$z$, and warrant strong mm constraints.
The search for DSFGs on the FFs using {\it Herschel} data has already found nine sources with magnification $\mu\geq4$ at $z<1.5$ and other nine sources at $z>2$ \citep{Rawle2016}. 

Each of the three FFs clusters is being complemented with a $\approx$2\farcm1$\times$2\farcm2 1.1\,mm ALMA mosaic, coincident with the deep {\it HST} WFC3/IR imaging region, achieving an rms depth of 55--71 $\mu$Jy\,${\rm beam}^{-1}$, depending on the field. The 1.1\,mm band benefits from the negative K-correction due to the typical galaxy spectral energy distribution (SED), and thus effectively provides a redshift unbiased census between $z$$\sim$1--10 \citep[e.g.,][]{Blain2002}. The data allow us to pinpoint locations of extreme star formation and probe factors of 2--10$\times$ fainter into the $L_{\rm IR}$ and SFR distributions compared to similar blank-field surveys due to typical lensing boosts.  The data also allow constraints on the SFRs and line emission for 1000's of optical and IR objects undetected in our 1.1\,mm survey (individually and in aggregate). 

This paper is the first in a series, and primarily describes the reduction and analysis of the ALMA 1.1\,mm data for A2744, MACSJ0416, and MACSJ1149 (hereafter, the ALMA-FFs). We focus this paper on the observed sources properties in the lensed image plane. To fully characterize the detected sources in the source plane, a combination of redshifts estimates and mass models for the galaxy clusters is needed.
Separate companion papers will detail the identification, redshift estimation and initial characterization of multi-wavelength counterparts (N. Laporte et al. 2016, in prep.), physical properties based on SED-fitting and morphological studies in the source plane (J. Gonz\'alez-L\'opez et al. 2016, in prep.), number counts (A. Mu{\~{n}}oz Arancibia et al. 2016, in prep.), and stacking analyses of multiply imaged sources and dropout candidates (R. Carvajal et al. 2016, in prep.). This paper is organized as follows: in $\S$2 we outline the observations and describe the reduction and imaging procedures; in $\S$3 we discuss our source detection methods and sensitivity estimates; in $\S$4 we assess the fluxes and angular extents of the detected sources; and in $\S$5 we summarize our results. Throughout the paper we assume a concordance cosmology and quote errors at 1$\sigma$ unless stated otherwise.

\section{ALMA Data}\label{sec:data}

\subsection{Observations}\label{sec:obs}

\begin{table*}
\caption[]{ALMA observation log. 
{\it Col. 1}: Cluster name.
{\it Col. 2}: Cluster redshift.
{\it Cols. 3--4}: Central J2000 position of mosaic in hh:mm:ss.ss+dd:mm:ss.ss.
{\it Col. 5}: First and last dates of observations.
{\it Col. 6}: Minimum projected baseline in meters.
{\it Col. 7}: Maximum projected baseline in meters.
{\it Column 8}: Mean projected baseline and standard deviation ($\sigma$) in meters.
\label{tab:obs_information}}
\centering
\begin{tabular}{lcccccccc}
\hline                    
Cluster Name & z & R.A. [J2000] & DEC [J2000] & Date of observations & Min$_{\rm B}$& Max$_{\rm B}$& Mean$_{\rm B}\,|\,{\rm \sigma}_{\rm B}$\\ 
&&&&&[m]&[m]&[m]\\ 
\hline                                 
Abell 2744 			&	0.308	&	00:14:21.2	&	   $-$30:23:50.1	&	29-Jun-2014/31-Dec-2014	&	15.1 &	783.5 &	$170.0\,|\,115.6$\\
MACSJ0416.1$-$2403 	&	0.396	&	04:16:08.9	&	   $-$24:04:28.7 &	04-Jan-2015/02-May-2015	&	15.1 &	348.5 &	$114.6\,|\,62.8$\\
MACSJ1149.5$+$2223	&	0.543	&	11:49:36.3	&	$+$22:23:58.1 &	14-Jan-2015/22-Apr-2015	&	15.1 &	348.5 &	$110.5\,|\,61.0$\\
\hline
\end{tabular}
\end{table*}

Three galaxy clusters belonging to the FFs sample were observed with ALMA (Table \ref{tab:obs_information}) using the array of 12\,m antennas as part of the project ADS/JAO.ALMA\#2013.1.00999.S\footnote{\url{https://almascience.nrao.edu/aq}}\footnote{\url{http://www.astro.puc.cl/~jgonzal/ALMA_FF.html}}. Each field was covered with 126 pointings to create mosaics spanning an area of $\approx4.6$ square arcminutes within the half power region formed by the individual beams at the edge (i.e. their HPBW).
For uniformity, the same spectral configuration setup was chosen for the three fields. The Local Oscillator frequency was set to 263.14 GHz ($\approx$1.1\,mm), with two spectral windows placed in each sideband. The correlator was set to frequency division mode (FDM) with a bandwidth of 1875\,MHz and a channel spacing of 0.488\,MHz. This setup yielded a total frequency coverage of 7.45\,GHz. 

We requested channel averaging inside the correlator with an averaging factor of $N=16$, resulting in a final spectral resolution of 7.813 MHz ($\sim9\kms$). This width decreases the data rate of the observations while still allowing a spectral resolution high enough to detect any possible narrow emission lines.

The observations for each of the three fields consisted of six independent executions, with each execution aiming to observe the whole mosaic. The last two executions for A2744 and all executions for MACSJ0416 and MACSJ1149 were performed in the second most compact array configuration, C36-2. The first four executions on A2744 were carried out in a more extended configuration. Table \ref{tab:obs_information} presents basic observing information for each  cluster, such as the pointing center, observation span, and projected baseline properties. The fact that A2744 was partially observed in a more extended configuration resulted in a longer mean projected baselines and a wider range of UV coverage compared to the other two clusters.

The final sensitivity depends on various factors such as the precipitable water vapor (PWV) and elevations of the field in the sky. As we will see in $\S$\ref{sec:PB}, these have important repercussions on the uniformity of the mosaics.

\subsection{Reduction}\label{sec:reduction}

The reduction of the datasets was done using the common astronomy software applications package \citep[{\sc casa};][]{Mcmullin2007}. Our reduction procedure closely followed the reduction scripts delivered by joint ALMA observatory (JAO) together with the datasets. 

Out of the three clusters, MACSJ0416 is the only one where the observations were reduced using the ALMA pipeline included in {\sc casa} version 4.2.2. The manual ("script") reduction of A2744 and MACSJ1149, as well as all imaging and analysis of the clusters, were performed using {\sc casa} version 4.3.1. The clusters A2744 and MACSJ1149 fall under the "non-standard" category (multiple array configurations and unfinished executions), therefore were not reduced using the pipeline.

We adopted most of the manual data flagging implemented by the JAO staff for our final reductions. These flags correspond to bad antennas, bad channels and time dependent bad behaviors. These manual flags were used in the reduction scripts as well as in the ALMA pipeline reduction. 
However, after our initial reductions, we found that the manual reductions scripts for A2744 and MACSJ1149 were overly conservative in flagging the edge channels of the spectral windows, corresponding to a $\approx$12\% loss in the total frequency coverage. After visual inspection of the calibrator spectra for MACSJ0416, we found that the ALMA pipeline, on the other hand, correctly left these edge channels unflagged for this cluster. After reinstating  the flagged edge channels for A2744 and MACSJ1149, the rms values of the new images improved by $\approx$5\%
compared to the initial images delivered by ALMA. A final inspection of the calibrator spectra confirmed that such flagging was not needed. 
Incidentally, the ALMA Cycle 2 Technical handbook states that for the Frequency Division Mode (FDM) spectral windows, the usable bandwidth is equal to the total bandwidth (1875\,MHz), consistent with our data. 

For A2744, the amplitude calibrators were J2258$-$279 and Uranus, the bandpass calibrators were J2258$-$2758 and J2258$-$279, and the phase calibrators were J0011$-$2612 and J2359$-$3133. For MACSJ0416, the amplitude calibrators were J0334$-$401, Ganymede and Uranus, the bandpass calibrators were J0334$-$4008, J0519$-$4546, J0334$-$401, J0423$-$0120 and J0416$-$2056, and the phase calibrator was J0416-2056. Finally, for MACSJ1149, amplitude calibrators were  3C273, Titan and Callisto, the bandpass calibrator was J1256$-$0547 and the phase calibrator was J1159$+$2914.
We estimate that the flux calibration for each field is accurate to $\sim$8\% based on a comparison of the amplitudes for the different flux calibrators.

\subsection{Imaging}\label{sec:imaging}

\begin{table*}
\caption[]{Mosaic properties for the different galaxy clusters. 
{\it Col. 1}: Cluster name.
{\it Col. 2}: Image weighting method used.
{\it Col. 3}: Major axis of synthesized beam, $b_{\rm max}$, in arcseconds.
{\it Col. 4}: Minor axis of synthesized beam, $b_{\rm min}$, in arcseconds.
{\it Col. 5}: Position angle of synthesized beam, $b_{\rm PA}$, in degrees.
{\it Col. 6}: Highest sensitivity (Lowest rms) achieved in each synthesized image, in $\mu$Jy beam$^{-1}$.
\label{tab:imaging_results}}
\centering
\begin{tabular}{cccccc}
\hline     
Cluster Name& {Map} & {b$_{\rm max}$ [$\arcsec$]} & {b$_{\rm min}$ [$\arcsec$]}& {b$_{\rm PA}$ [$^{\circ}$]} & {rms [$\mu$Jy beam$^{-1}$]}\\
\hline
\noalign{\smallskip}
Abell 2744 			&	Natural	&	0.63	&	0.49	&	86.14 & 55\\
MACSJ0416 			&	Natural	&	1.52	&	0.85	&	-85.13 & 59\\
MACSJ1149 			&	Natural	&	1.22	&	1.08	&	-43.46 & 71\\
\hline
\end{tabular}
\end{table*}

Mosaicked images were created using natural weighting, which is generally the most sensitive since it optimally weights all of the baselines \citep{Briggs1999}. Image details are provided in Table~\ref{tab:imaging_results}; the unique synthesized beams and rms sensitivities depended on the specific observations of each cluster. The A2744 dataset has the lowest rms (55 $\mu$Jy beam$^{-1}$) amongst the observed clusters, as well as the smallest beam due to the more extended array configuration in which it was partially observed; MACSJ0416 and MACSJ1149 have natural weighted beams which are nearly twice as large as that of A2744. 

The data were imaged using the multi-frequency synthesis algorithm in \texttt{CLEAN}, adopting the \texttt{clark} psfmode, the \texttt{mosaic} imagermode, and a 0\farcs1 pixel size, which is small enough to sample the synthesized beams listed in Table~\ref{tab:imaging_results}. Dirty images were made using niter=0, providing the rms values presented in the last column in Table~\ref{tab:imaging_results}.
The cleaned images were generated using \texttt{niter}=1000 and a threshold of four times the rms measured in the dirty images, with no masking applied during cleaning. The cleaned images have the same rms values as the dirty ones.

\begin{figure*}[!htbp]
\centering
\includegraphics[width=\hsize]{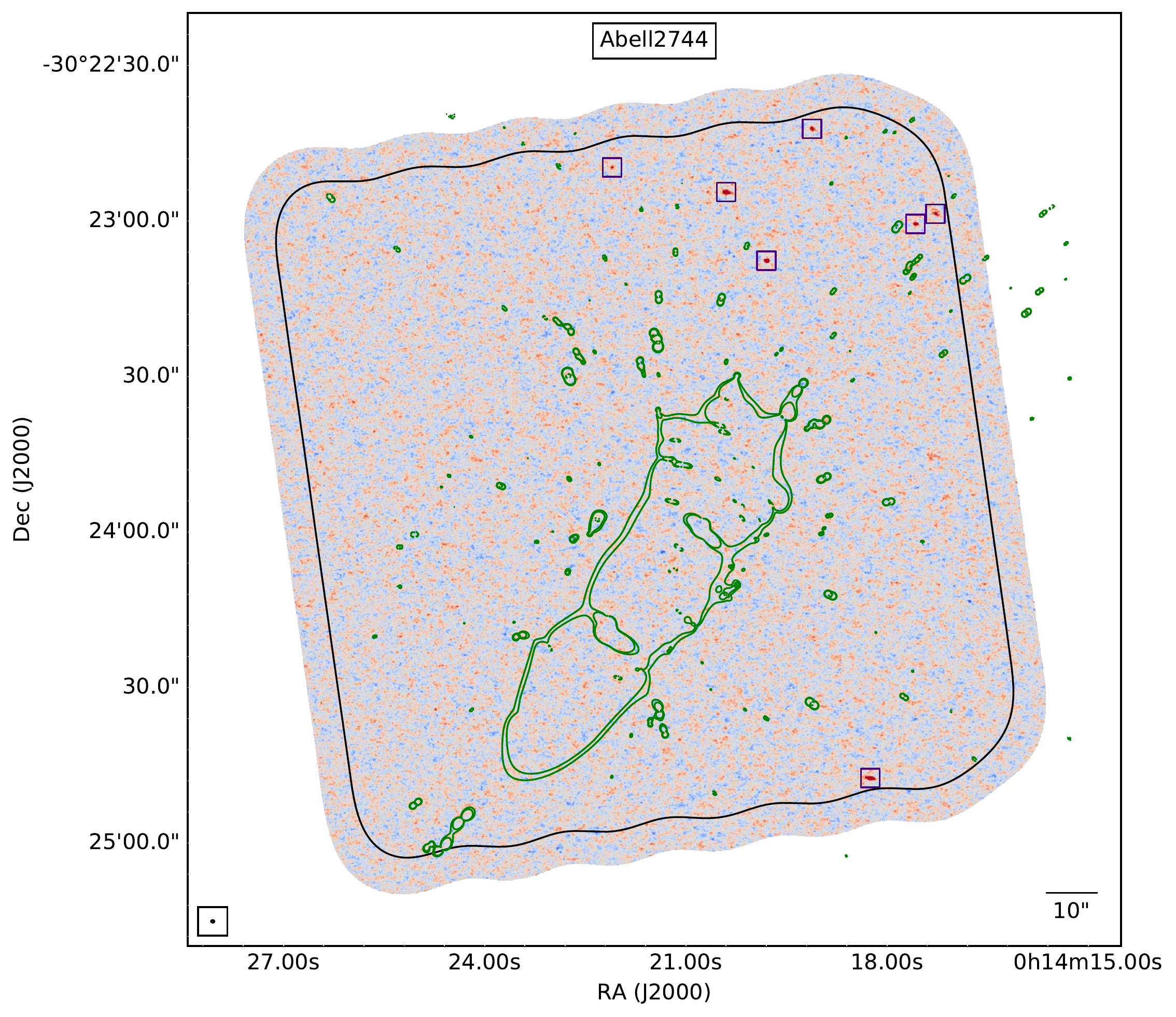}
\caption{1.1\,mm continuum map for the galaxy cluster A2744 made with natural weighting. The image has an rms of 55\,$\mu$Jy beam$^{-1}$ and has not been corrected by the primary beam sensitivity for visualization purposes. The color scale corresponds to $-5\sigma$ to $5\sigma$ from blue to red. The black curve corresponds to the point where the primary beam correction is equal to 0.5, meaning where the sensitivity is $2\sigma$. Squares show the positions of the sources with $S/N\geq5$. In the small box bottom left corner we show the synthesized beam of 0\farcs63$\times$0\farcs49 and position angle of 86\fdg16.
The critical curves (i.g. where magnification is infinite) for a source at $z=2$ given by the model Zitrin-NFWv3 \citep{Zitrin2009,Zitrin2013} are shown in green.
\label{fig:map_A2744}}
\end{figure*}

\begin{figure*}[!htbp]
\centering
\includegraphics[width=\hsize]{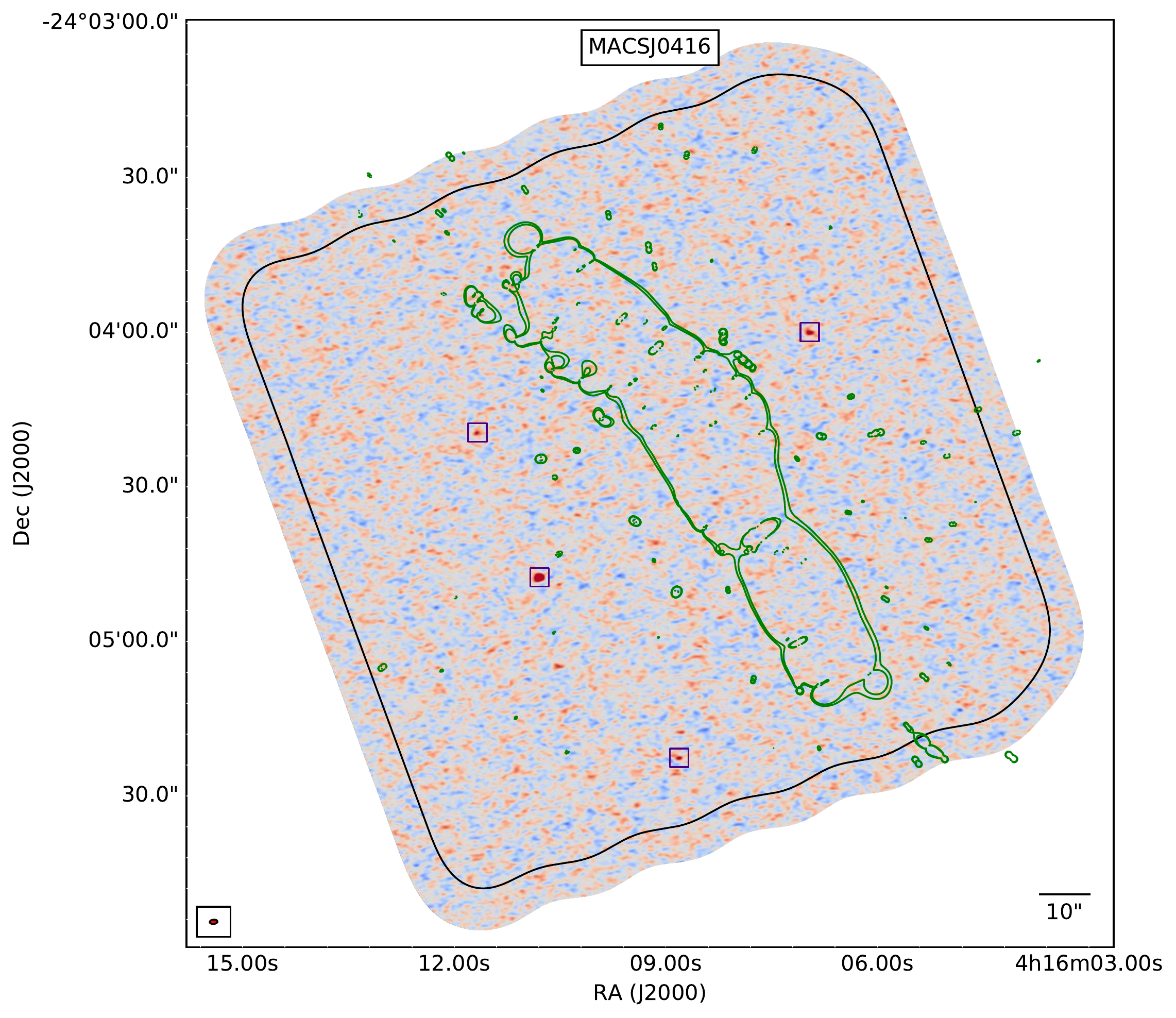}
\caption{1.1\,mm continuum map for the MACSJ0416 made with natural weighting. The image has an rms of 59\,$\mu$Jy beam$^{-1}$ and has not been corrected by the primary beam sensitivity for visualization purposes. The color scale and black curve are the same as in Figure\,\ref{fig:map_A2744}. Squares show the positions of the sources with $S/N\geq5$. In the small box bottom left corner we show the synthesized beam of 1\farcs52$\times$0\farcs85 and position angle of $-$85\fdg13. The critical curves (i.g. where magnification is infinite) for a source at $z=2$ given by the model Zitrin-NFWv3 \citep{Zitrin2009,Zitrin2013} are shown in green.
\label{fig:map_MACS0416}}
\end{figure*}

\begin{figure*}[!htbp]
\centering
\includegraphics[width=\hsize]{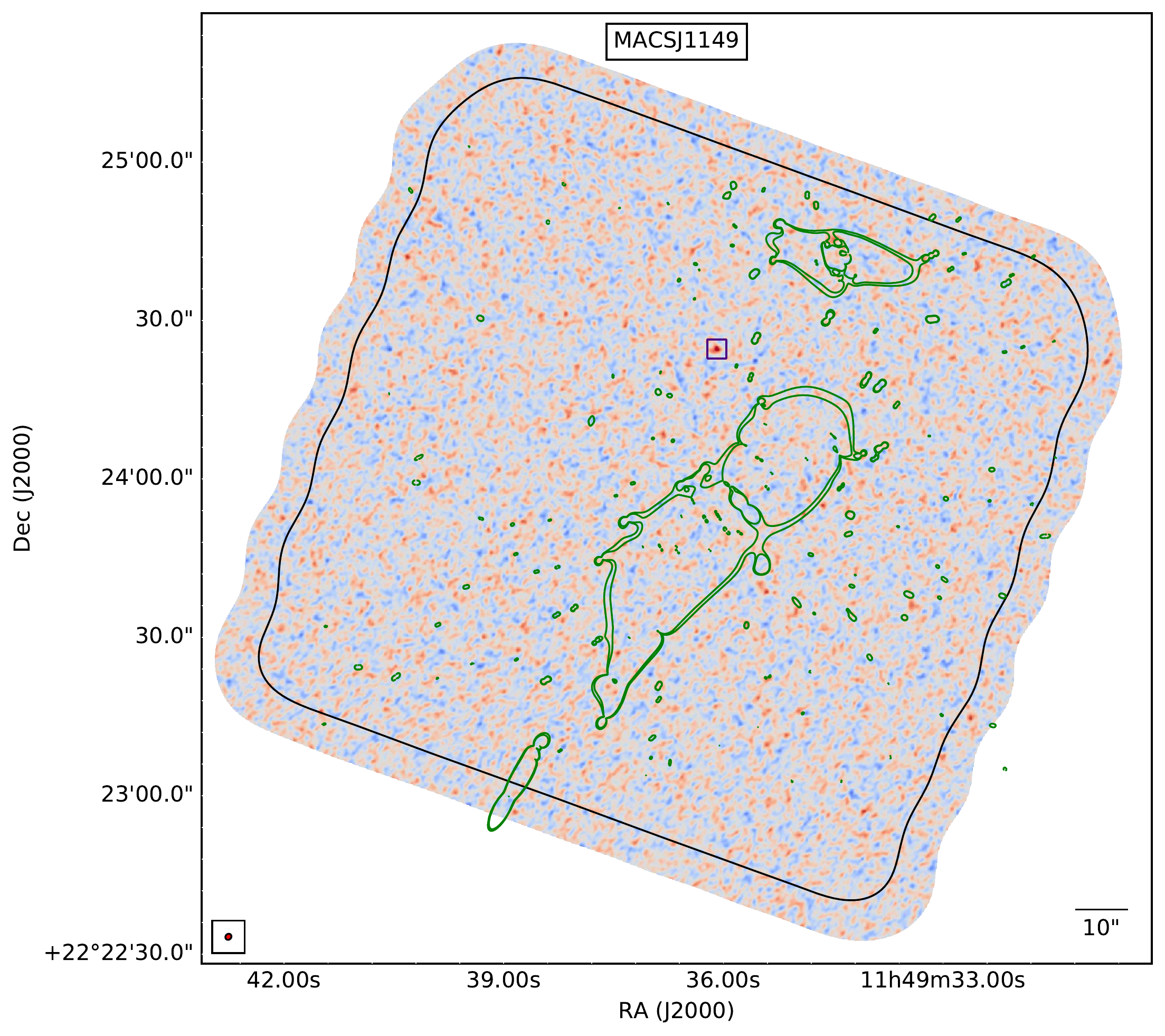}
\caption{1.1\,mm continuum map for MACSJ1149 made with natural weighting. The image has an rms of 71\,$\mu$Jy beam$^{-1}$ and has not been corrected by the primary beam sensitivity for visualization purposes. The color scale and black curve are the same as in Figure\,\ref{fig:map_A2744}. Squares show the positions of the sources with $S/N\geq5$. In the small box bottom left corner we show the synthesized beam of 1\farcs2$\times$1\farcs08 and position angle of $-$43\fdg46. The critical curves (i.g. where magnification is infinite) for a source at $z=2$ given by the model GLAFICv3 \citep{Kawamata2016} are shown in green.
\label{fig:map_MACS1149}}
\end{figure*}

Figures\,\ref{fig:map_A2744}, \ref{fig:map_MACS0416} and \ref{fig:map_MACS1149} present the natural-weighted clean images for the three observed FFs clusters, respectively. These images have not been corrected for the primary beam (PB) sensitivity response, which is discussed below.

\subsection{Mosaic sensitivity.}\label{sec:PB}

\begin{figure*}[!htbp]
\centering
\includegraphics[width=\hsize]{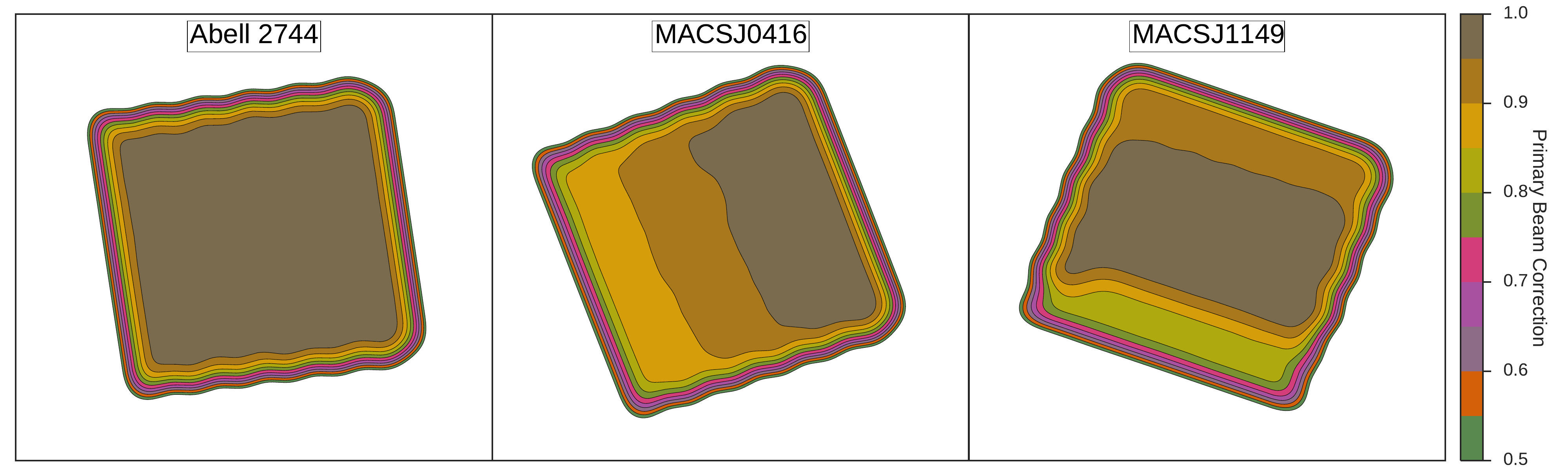}
\caption{Primary Beam (PB) correction images for each cluster, adopting natural weighting. Thus final PB-corrected images would correspond to Figures~\ref{fig:map_A2744}, \ref{fig:map_MACS0416} and \ref{fig:map_MACS1149} divided by the corresponding PB corrections shown here. Likewise, the final sensitivity corresponds to the measured sensitivity divided by the PB image. For A2744 the sensitivity is uniform across the mosaic, while for  MACSJ0416 there is a clear sensitivity gradient from top right to bottom left, and in MACSJ1149 the upper and lower regions of the mosaic have sensitivities substantially lower than in the uniform middle portion (See \S\ref{sec:PB} for details). 
\label{fig:pb_map}}
\end{figure*}

Various factors can affect the mosaic sensitivity across different pointings. These include the intrinsic primary beam response shape, as well as any factor that can decrease the sensitivity in a given pointing (e.g., higher $T_{\rm sys}$, less data). The PB correction images for each cluster (Figure~\ref{fig:pb_map}) should contain all relevant information regarding fluctuations from the nominal sensitivity. 

We find that the PB correction for A2744 is roughly constant and high across the mosaic ($\approx$1), demonstrating that the final sensitivity is uniform over the entire mosaic. For MACSJ0416 and MACSJ1149, however, the PB correction images are not smooth, indicating sensitivity variations by as much as $\sim$15--20\% across the mosaics in both cases. 

For MACSJ0416, the sensitivity loss appears to result from the flagging of several baselines owing to shadowing (i.e., when the primary beam of one antenna is partly covered by another) during the most sensitive, lowest PWV execution. It should be noted that all of the executions observed for MACSJ0416 start from the top right pointing and end with the bottom left pointing, such that they map the mosaic from top to bottom and west to east. During the most sensitive execution, the first pointing was initiated at an elevation of $\approx$41$^{\circ}$ while the last concluded at $\approx$27$^{\circ}$. The latter results in significant shadowing when using a compact array such as C36-2. This particular elevation difference across the mosaic will lead to sensitivity fluctuations for two reasons. First, the shadowing flagging is higher in the bottom left pointing, resulting in less data ($14\%$ of the data in the last pointing were flagged by shadowing). At the same time, lower elevation observations result in higher phase scatter and higher $T_{\rm sys}$ values for the same pointings, further decreasing the sensitivity. The combination of these factors explains the observed drop in sensitivity toward the bottom left side of the mosaic for MACSJ0416. 

For MACSJ1149, there is a sharp drop in sensitivity in the bottom left pointings, as well as a mild gradient in the upper right pointings. The explanation for the sharp drop lies in the fact that, during one of the most sensitive executions, the last 30 pointings of the mosaic were not observed, resulting in an incomplete mosaic execution. Thus the bottom left portion of the mosaic only contains five of the six executions, yielding lower sensitivity. The weak sensitivity gradient in the upper right  pointings can be explained similarly to MACSJ0416, whereby one of the most sensitive executions began at a low elevation and was more strongly affected by shadowing and higher $T_{\rm sys}$ values.

In conclusion, A2744 has uniform sensitivity across the mosaic, while MACSJ0416 and MACSJ1149 do not. This varying sensitivity must be accounted for when studying the statistical properties of the detected sources.

\section{Results}

\subsection{Source extraction}

\begin{figure}[!htbp]
\centering
\includegraphics[width=\hsize]{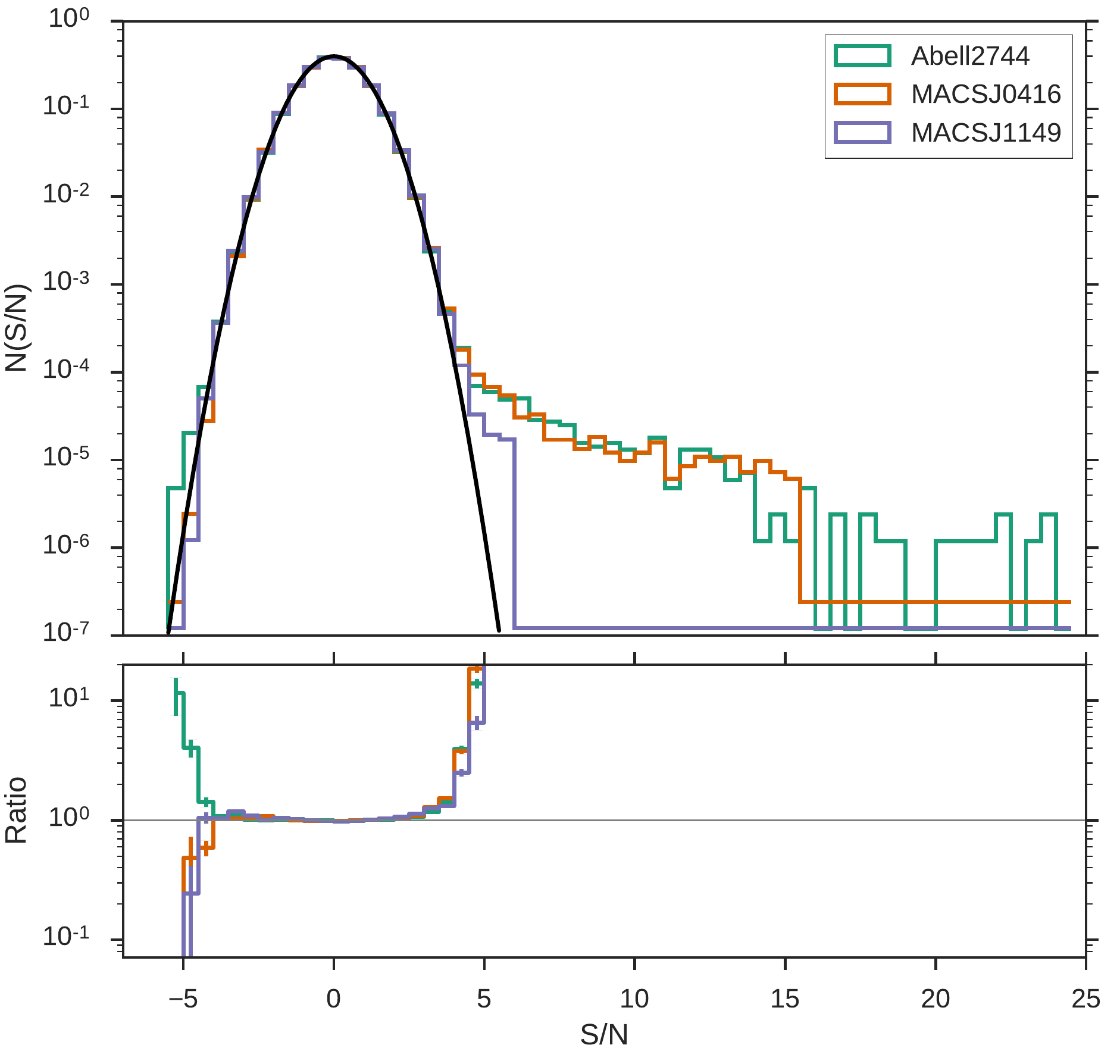}
\caption{Normalized histogram of the $S/N$ per pixel for the three cluster images adopting natural weighting. A2744, MACSJ0416, and MACSJ1149 are shown in green, orange and blue, respectively. The black curve shows the $S/N$ distribution expected for a Gaussian noise distribution, which is well matched to the image $S/N$ histograms, aside from possible deviations present near $S/N$$\sim$$-5.0$ and the clear positive tail denoting emission from real sources. In the bottom panel we show the ratio between the data counts and the Gaussian noise distribution for each bin. The error bars correspond to Poisson errors using the number of elements per bin. The deviation from unity in the ratio at $S/N$$\sim$$-5.0$ in the three cluster images can be explained by low-number statistics. Such behavior is also observed in the $S/N$ distribution tails of simulated images created with only Gaussian noise convolved with the same synthesized beams. 
\label{fig:sn_clusters}}
\end{figure}

\begin{table*}
\caption[]{High-significance ($\geq5\sigma$) continuum detections.
{\it Col. 1}: Source ID.
{\it Cols. 2-3}: Centroid J2000 position of ID in hh:mm:ss.ss+dd:mm:ss.ss.
{\it Cols. 4}: Positional error in arcseconds as given by the uv-plane fit.
{\it Col. 5}: Detection signal-to-noise.
{\it Col. 6}: Peak intensity and 1$\sigma$ error in mJy\,beam$^{-1}$.
\label{tab:gold}}
\centering
\begin{tabular}{ccccrc}
\hline     
{ID}&  {$\alpha_{\rm J2000}$} & {$\delta_{\rm J2000}$} & $\Delta\alpha$, $\Delta\delta$ & {$S/N$} & {$S_{\rm 1.1 mm, peak}$}\\
 &   [hh:mm:ss.ss] & [$\pm$dd:mm:ss:ss] & [$\arcsec$] &  & [mJy beam$^{-1}$]\\
\hline
\noalign{\smallskip}
A2744-ID01     & 00:14:19.80 & -30:23:07.66 & 0.011, 0.008 & 25.9 &$ 1.433 \pm 0.056 $\\
A2744-ID02     & 00:14:18.25 & -30:24:47.47 & 0.055, 0.015 & 14.4 &$ 1.292 \pm 0.091 $\\
A2744-ID03     & 00:14:20.40 & -30:22:54.42 & 0.038, 0.024 & 13.9 &$ 0.798 \pm 0.058 $\\
A2744-ID04     & 00:14:17.58 & -30:23:00.56 & 0.020, 0.015 & 13.8 &$ 0.932 \pm 0.068 $\\
A2744-ID05     & 00:14:19.12 & -30:22:42.20 & 0.056, 0.040 &  7.7 &$ 0.655 \pm 0.086 $\\
A2744-ID06     & 00:14:17.28 & -30:22:58.60 & 0.076, 0.080 &  6.5 &$ 0.574 \pm 0.089 $\\
A2744-ID07     & 00:14:22.10 & -30:22:49.67 & 0.035, 0.030 &  6.2 &$ 0.455 \pm 0.074 $\\
MACSJ0416-ID01 & 04:16:10.79 & -24:04:47.53 & 0.037, 0.037 & 15.4 &$ 1.010 \pm 0.066 $\\
MACSJ0416-ID02 & 04:16:06.96 & -24:03:59.96 & 0.130, 0.107 &  6.8 &$ 0.406 \pm 0.062 $\\
MACSJ0416-ID03 & 04:16:08.81 & -24:05:22.58 & 0.116, 0.064 &  5.8 &$ 0.389 \pm 0.067 $\\
MACSJ0416-ID04 & 04:16:11.67 & -24:04:19.44 & 0.148, 0.156 &  5.1 &$ 0.333 \pm 0.066 $\\
MACSJ1149-ID01 & 11:49:36.09 & +22:24:24.60 & 0.116, 0.102 &  5.9 &$ 0.442 \pm 0.074 $\\
\hline
\end{tabular}
\end{table*}

The search for sources was performed in the PB-uncorrected continuum cleaned images, which allows the detection of sources as a function of signal-to-noise independent of the final (non-uniform) sensitivity reached in different regions of the mosaics.
The first step was to find the rms value that best represented each image. We limited the region used for source detection to that where the PB sensitivity is higher than 0.5 of the peak sensitivity (PB$\geq0.5$). After selecting all the pixels with ${\rm PB}\geq0.5$, we performed a sigma clip at $5\sigma$ to remove clear bright 1.1\,mm sources and used the remaining pixels to obtain an initial estimate of the rms. With this initial rms, we searched for all pixels with signal-to-noise ($S/N$) $\geq$ 5 and used the clustering algorithm \texttt{DBSCAN} in \texttt{Python:scikit-learn} \citep{Pedregosa2011} to group the pixels that corresponded to unique sources. Each initial source was fit with a six-parameter 2d Gaussian function and these corresponding models were removed from the continuum maps to obtain an initial source-free image. The final rms was measured in each source-free image and is presented in Table~\ref{tab:imaging_results}.

\subsection{High-significance continuum detections}
\label{Sec:high_significance_cont_detections}
\begin{figure*}[!htbp]
\includegraphics[width=\hsize/3]{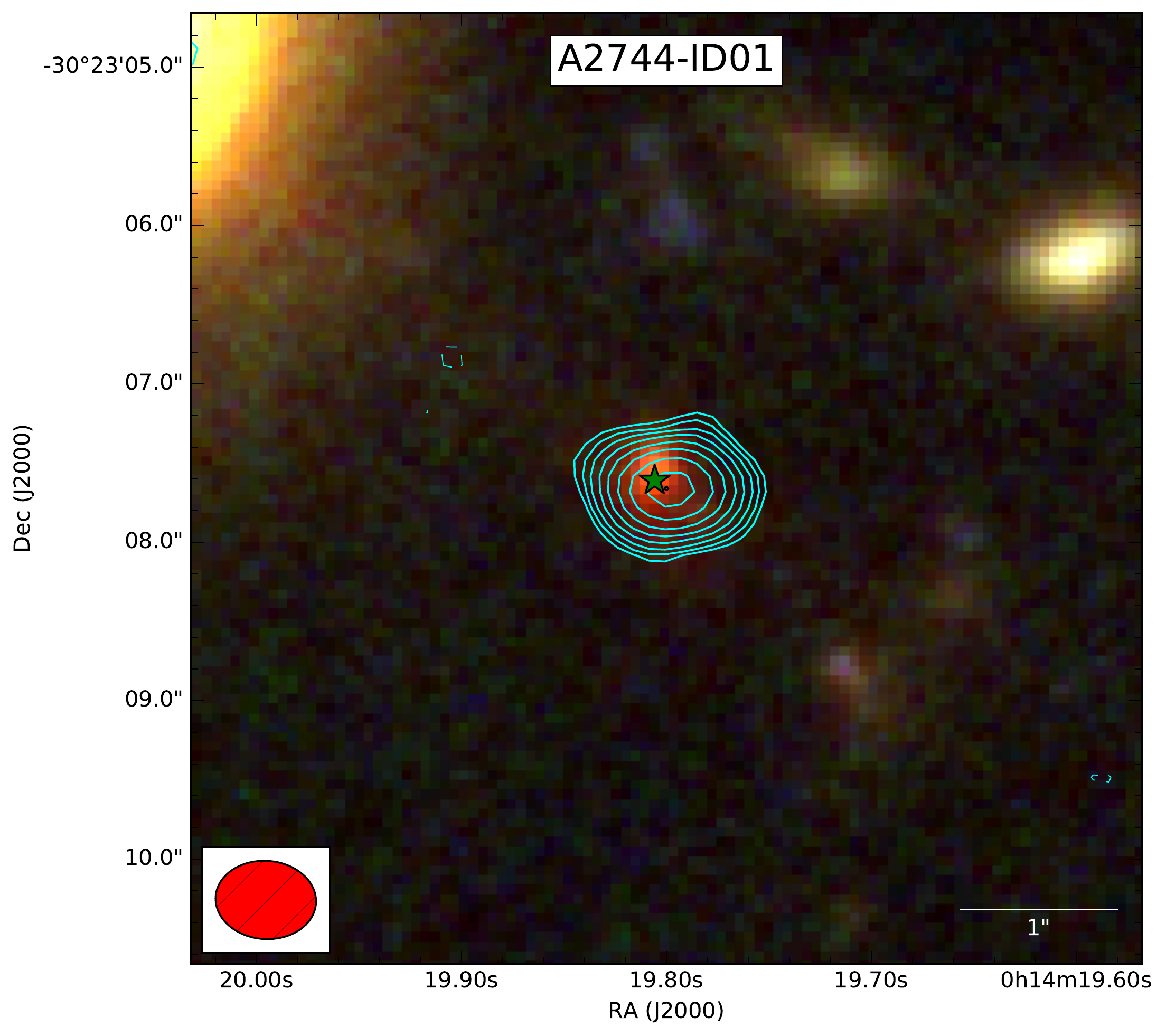}
\includegraphics[width=\hsize/3]{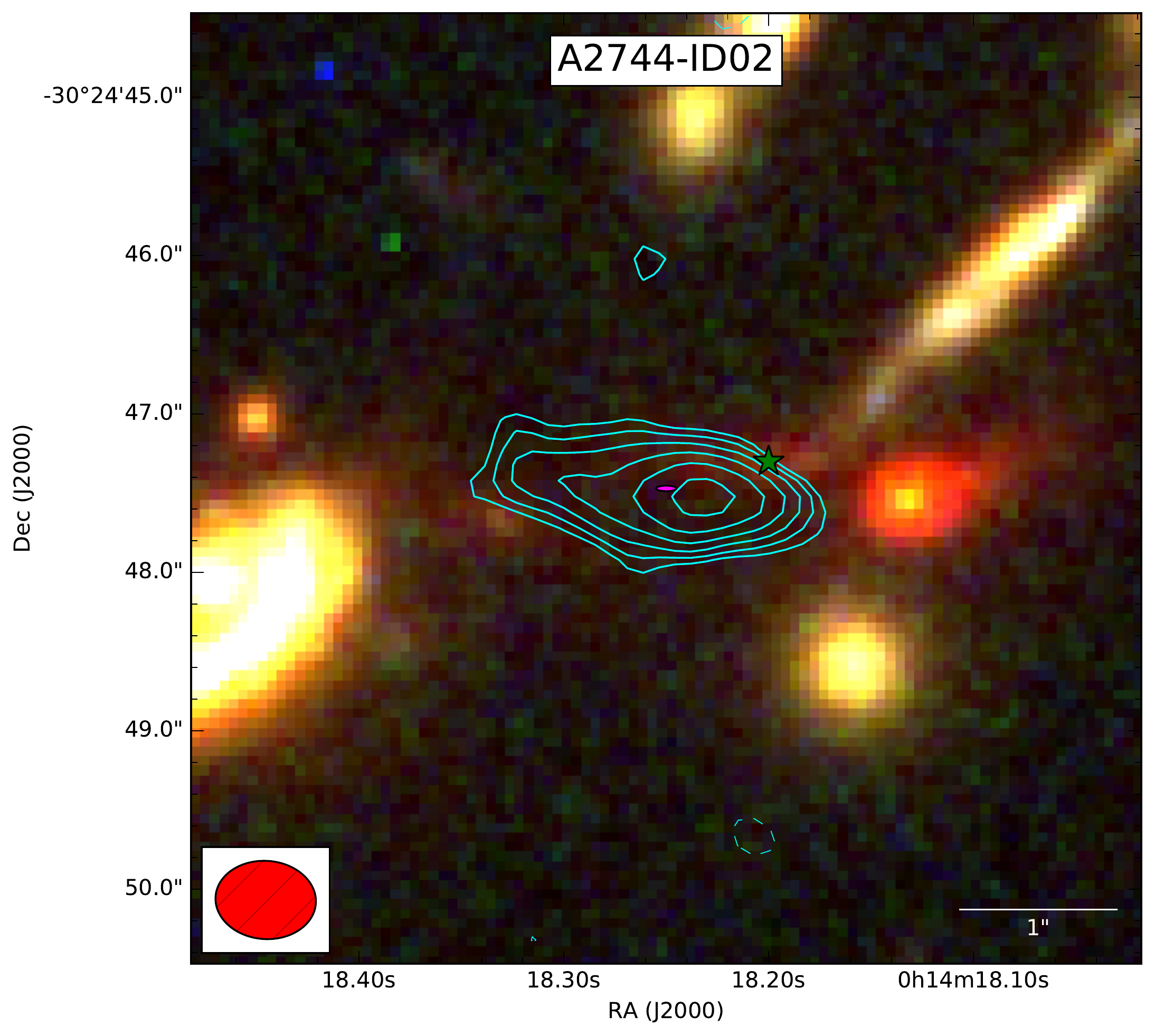}
\includegraphics[width=\hsize/3]{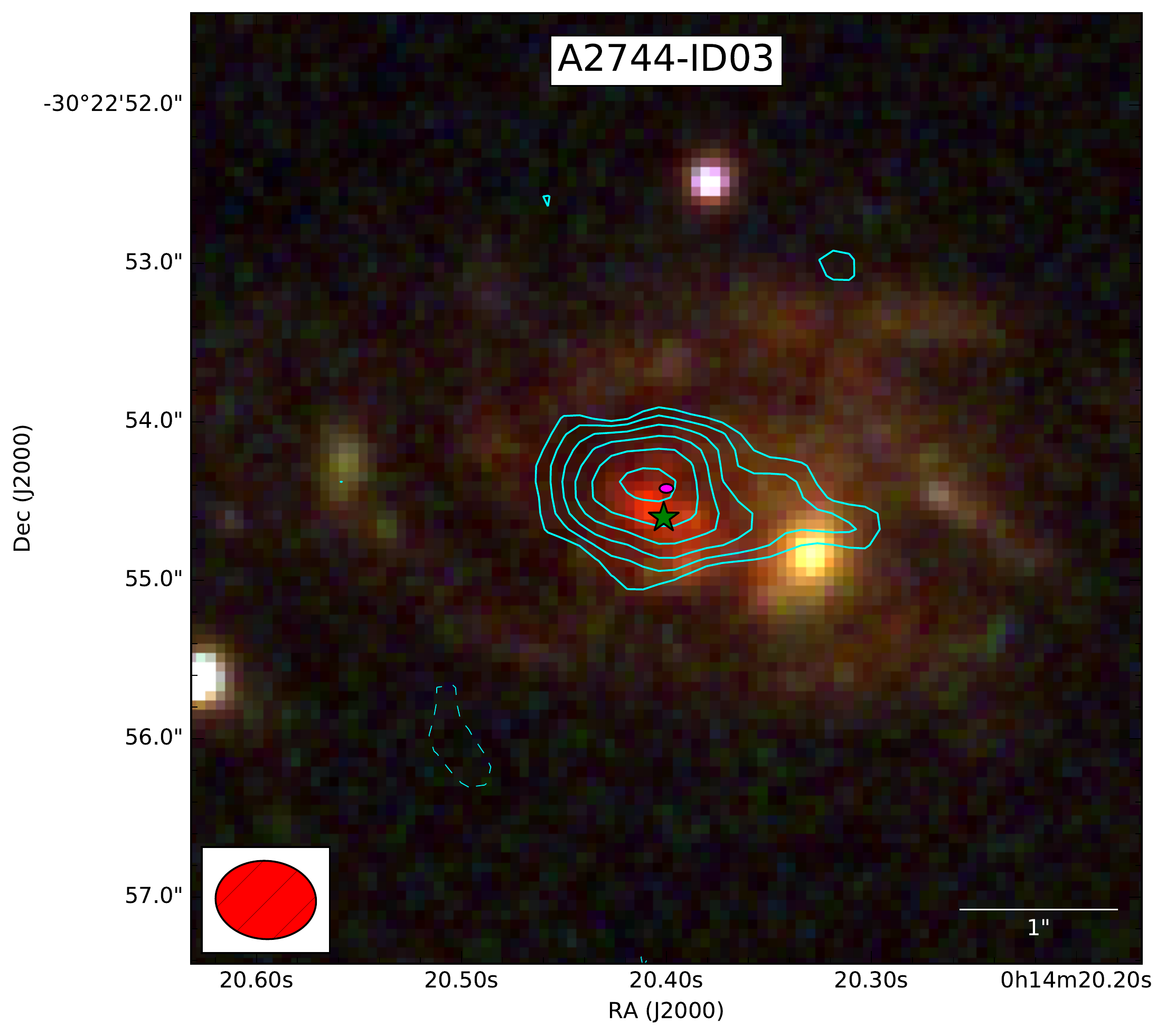}
\includegraphics[width=\hsize/3]{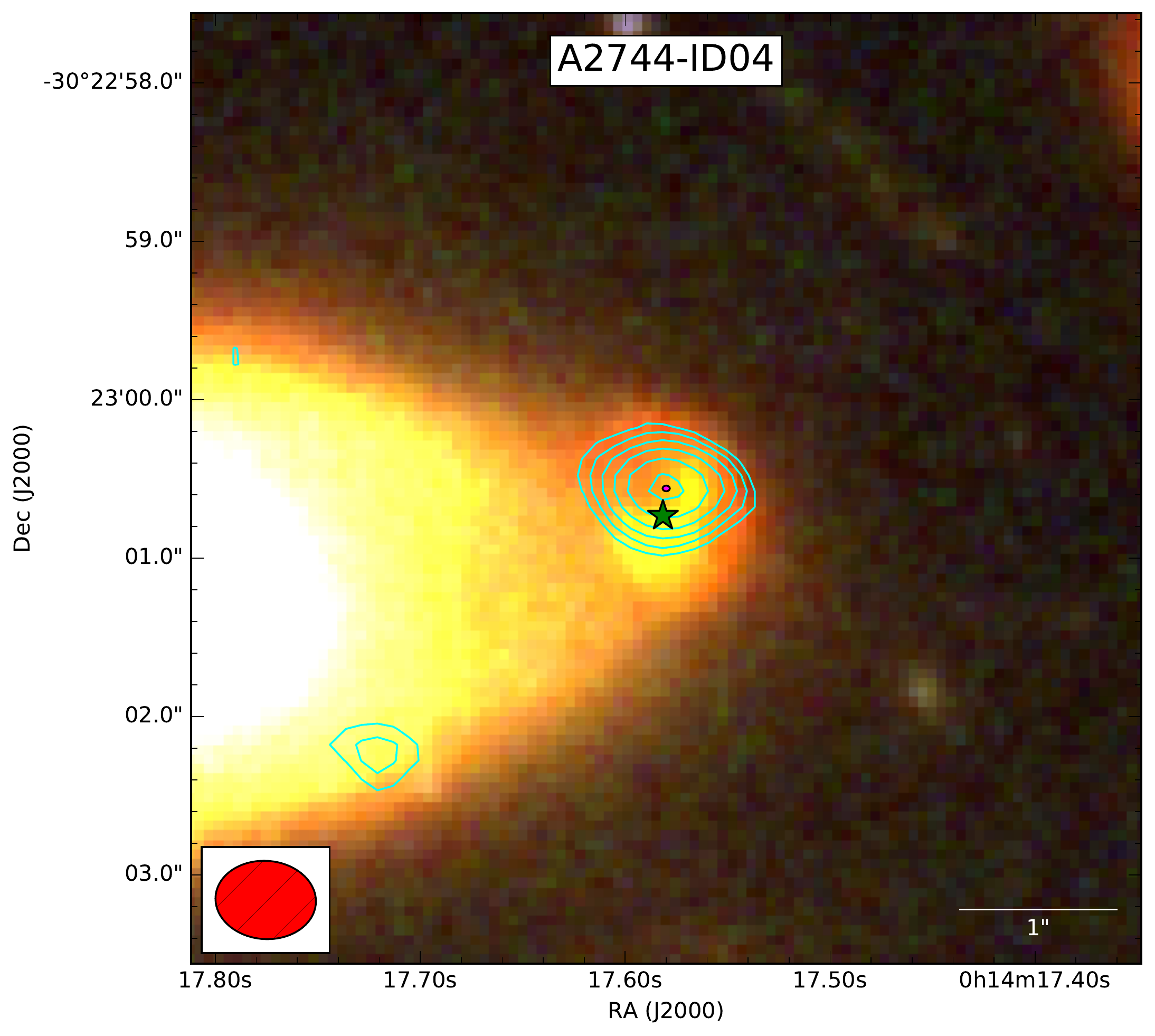}
\includegraphics[width=\hsize/3]{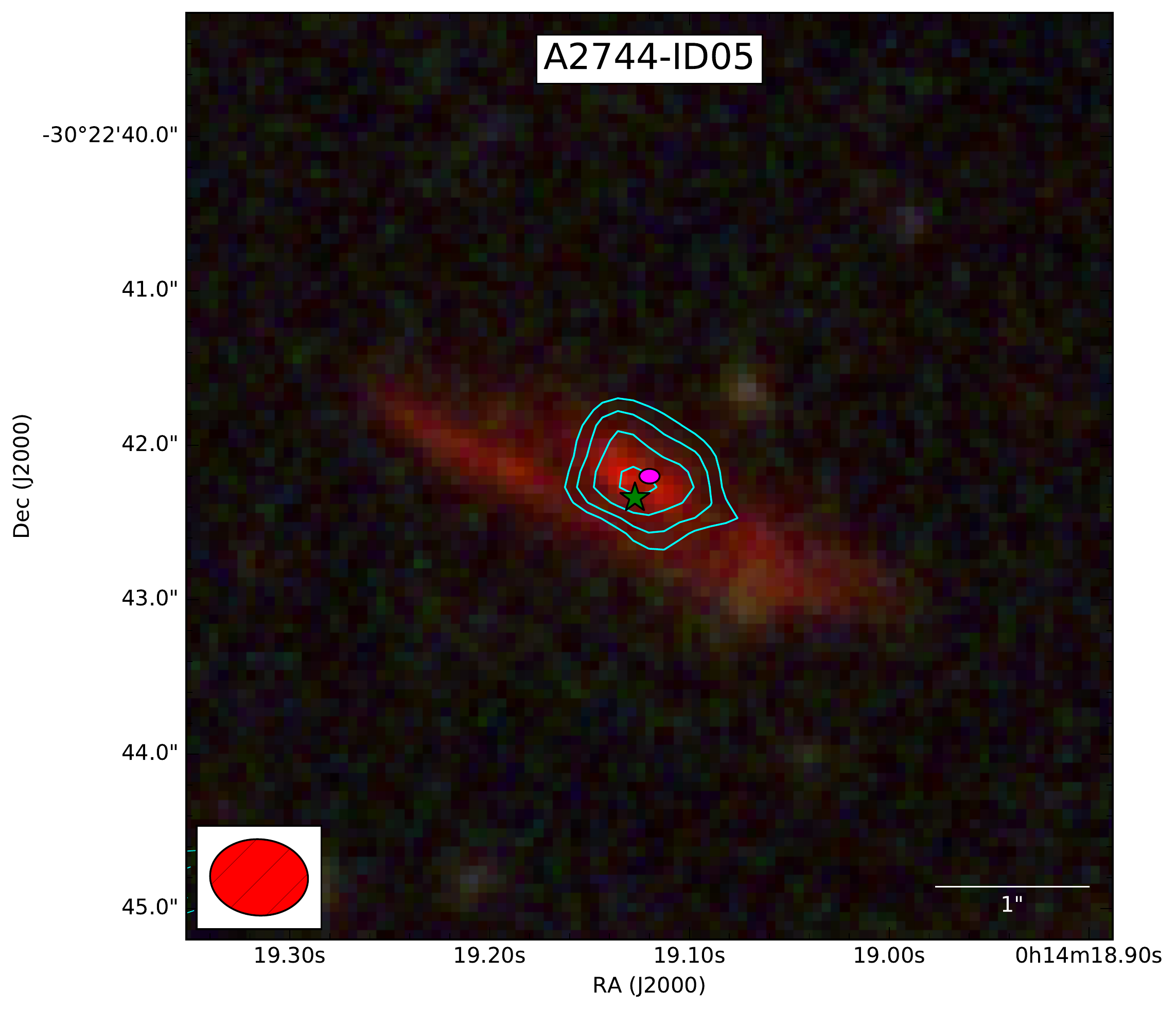}
\includegraphics[width=\hsize/3]{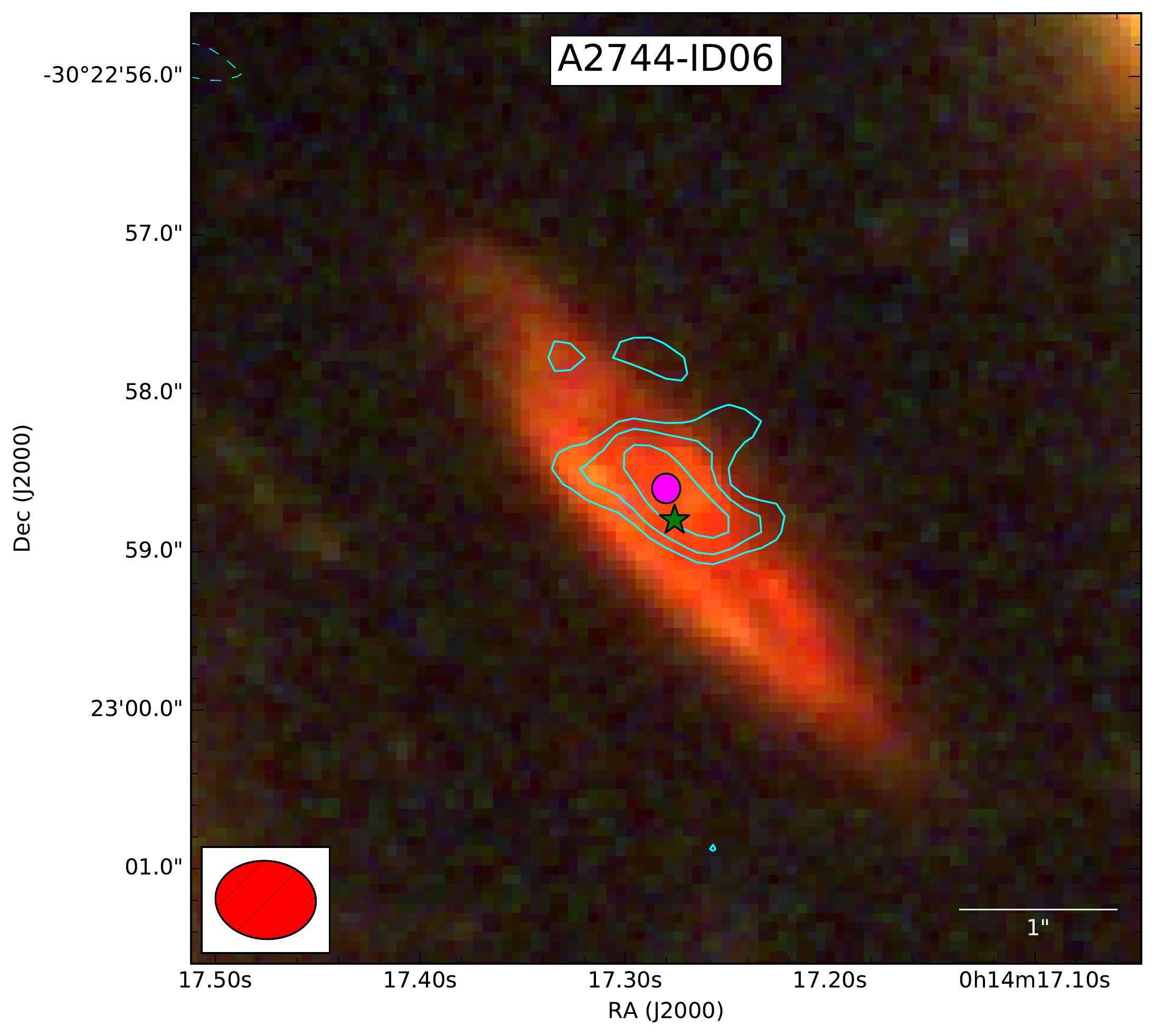}
\includegraphics[width=\hsize/3]{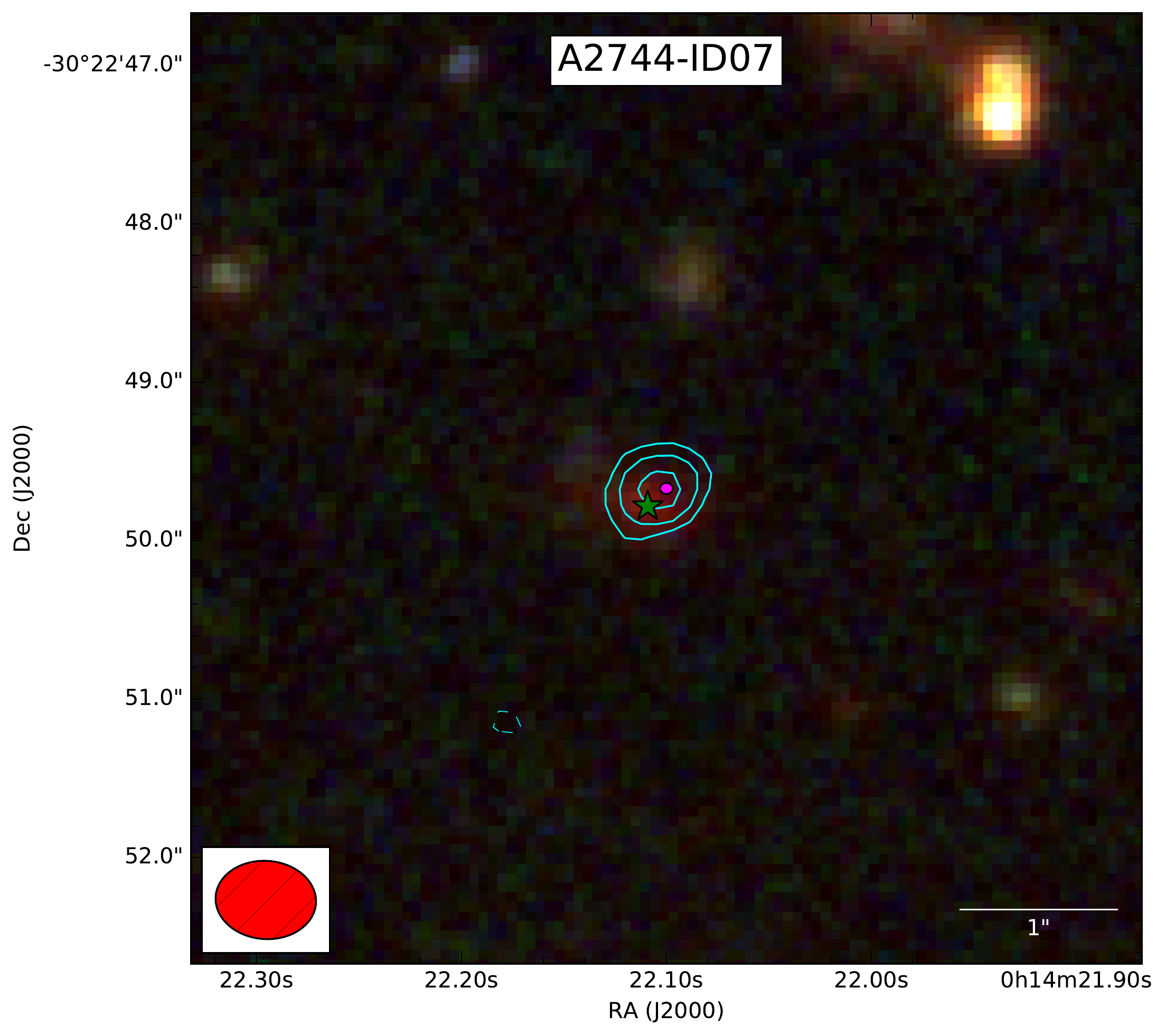}
\includegraphics[width=\hsize/3]{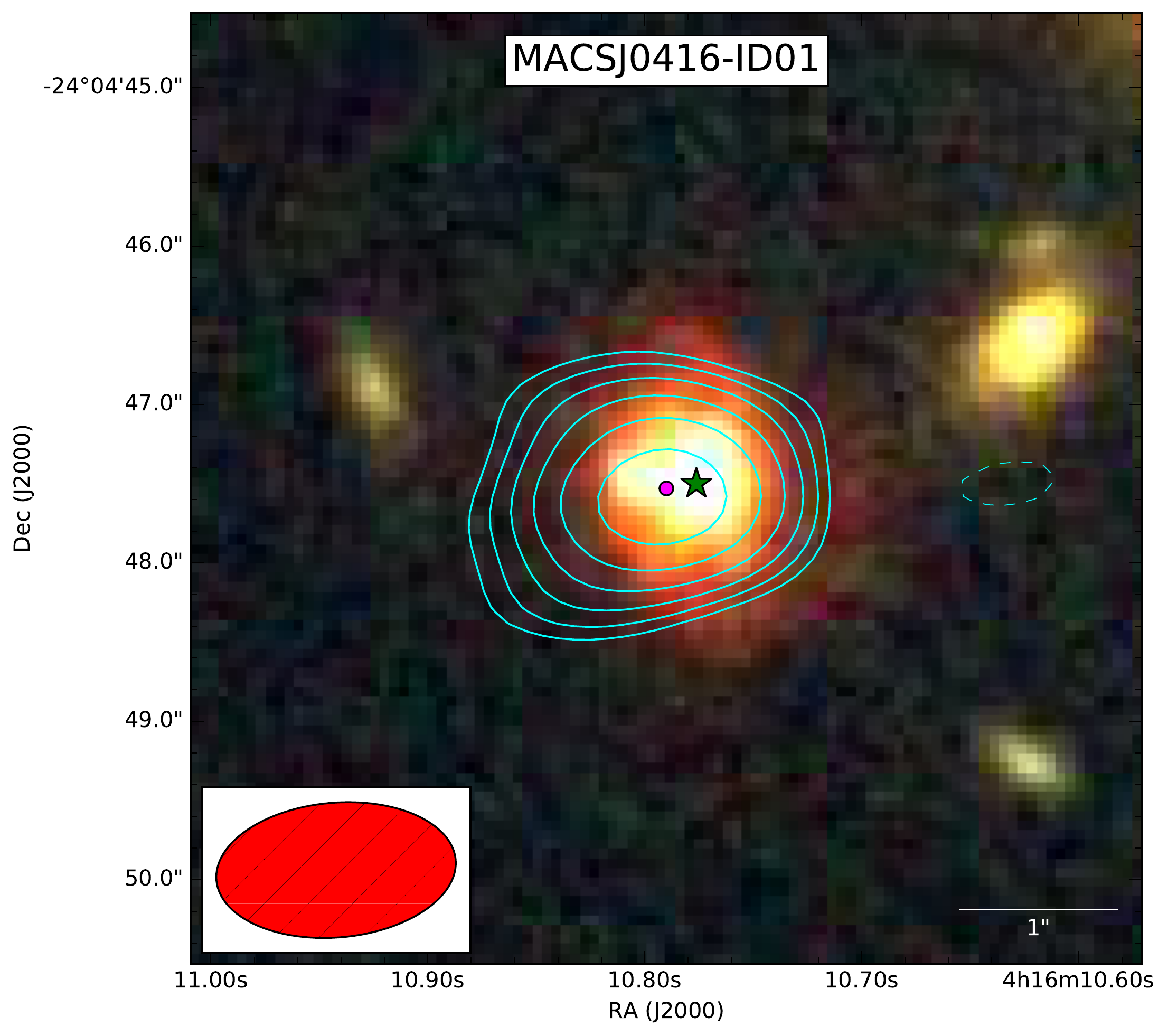}
\includegraphics[width=\hsize/3]{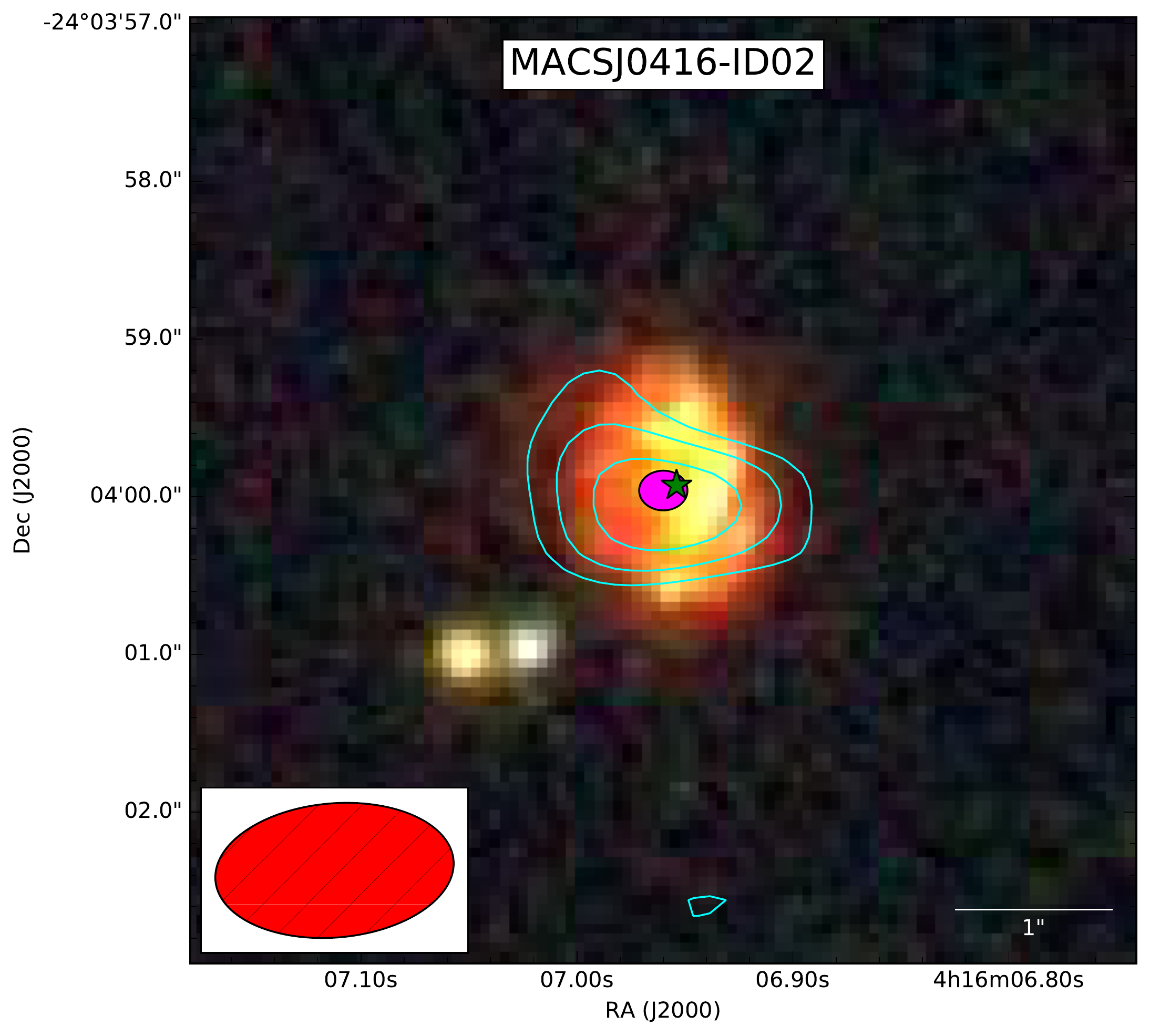}
\includegraphics[width=\hsize/3]{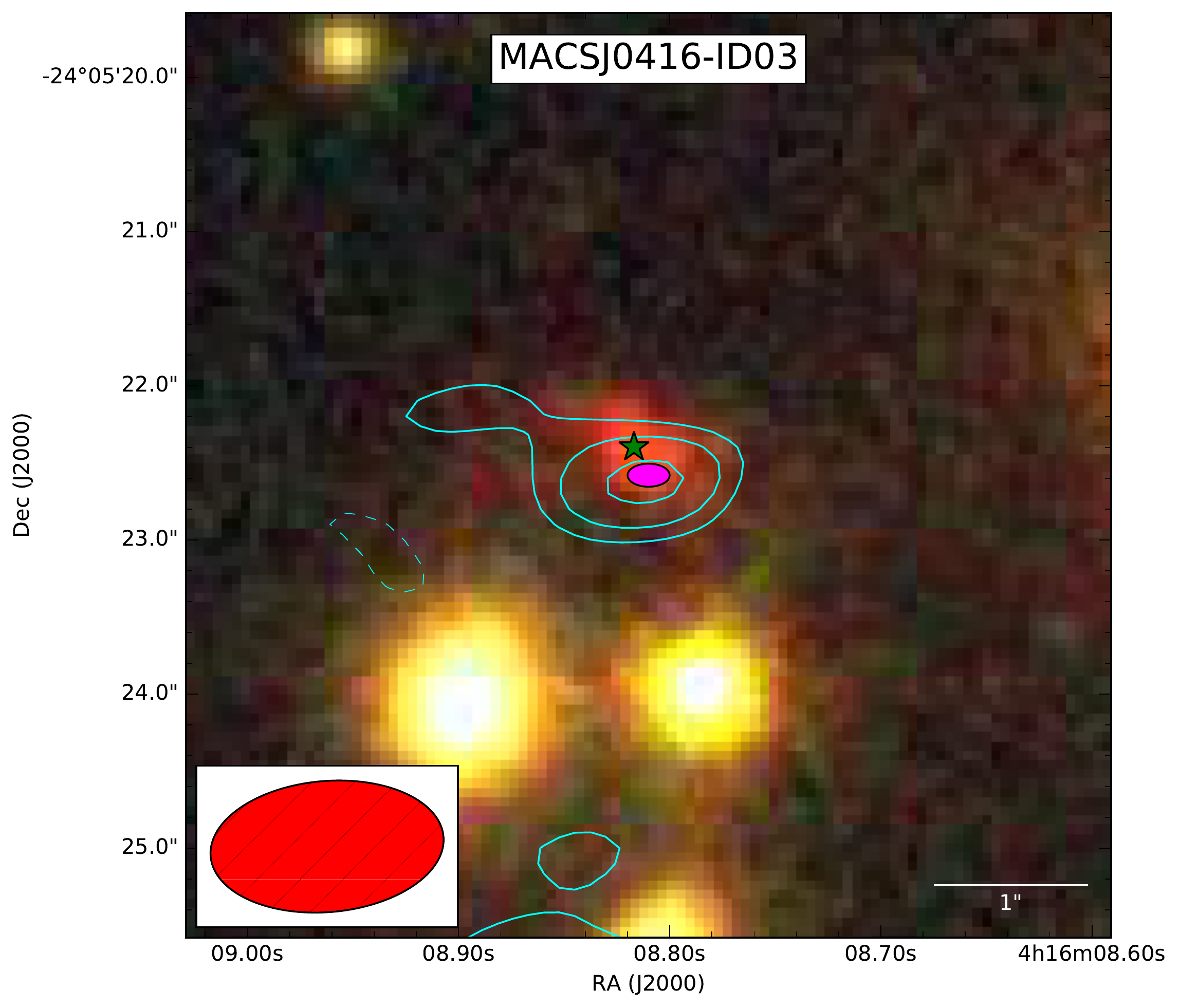}
\includegraphics[width=\hsize/3]{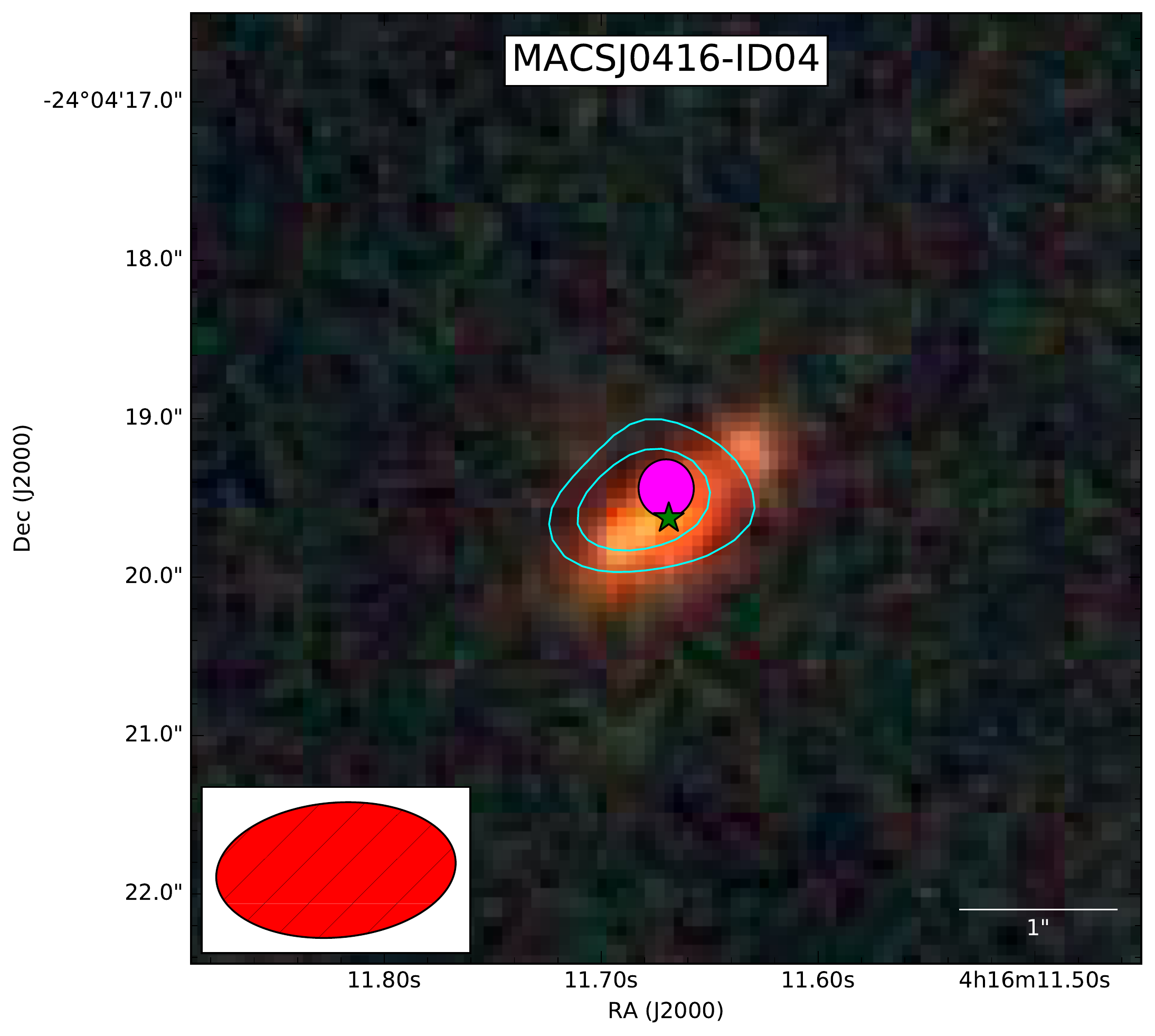}
\includegraphics[width=\hsize/3]{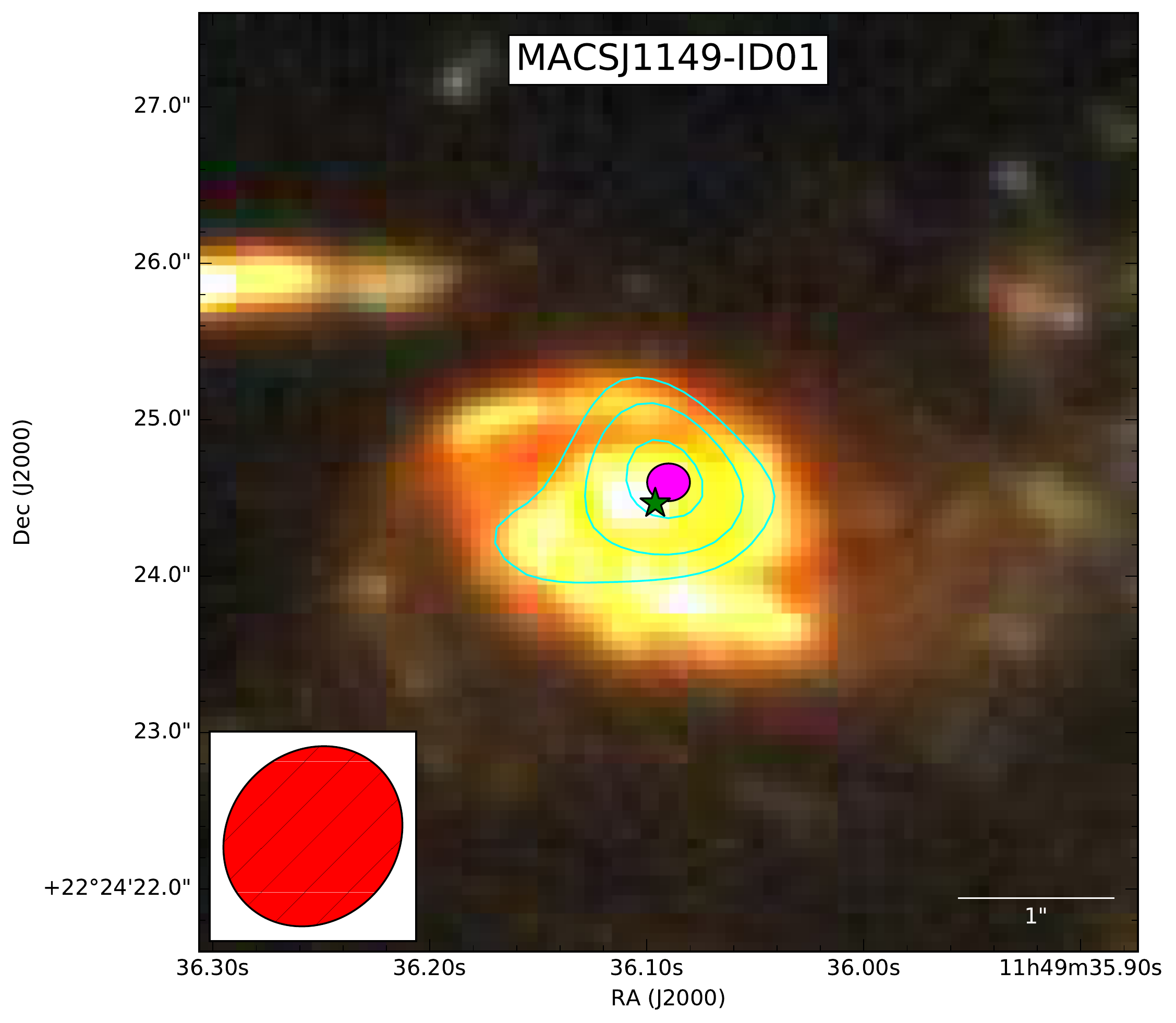}
\caption{6\arcsec$\times$6\arcsec color image cutouts of the NIR counterparts to the ALMA detections, with blue corresponding to bands $F435W$$+$$F060W$, green to bands $F810W$$+$$F105W$ and red to bands $F125W$$+$$F140W$$+$$F160W$. The cyan contours denote the ALMA 1.1\,mm emission, whereby the levels are displayed on a logarithmic scale at $\pm$ 3, 4, 5.3, 7.1, 9.5, 12.6, 16.8, 22.5, 30 $\times\sigma$, with $\sigma$ being the primary beam corrected noise level at the position of the source. The ALMA synthesized beam corresponding to each observation is shown in the lower left corner of each cutout. A green star represents the position of the associated NIR counterpart, while the magenta ellipse shows the $1\sigma$ range of the center of the ALMA emission based on the $uv$-plane fit. In all cases, the positional error associated with the NIR position is small compared to that of the ALMA emission. 
\label{fig:counterparts}}
\end{figure*}

To obtain a final list of detections, we selected all pixels with $S/N$ $>$ 4 and performed the same modeling as described above, however at this stage we also fit a 2d-Gaussian function with only three free parameters, fixing the shape to that of the synthesized beam and allowing changes only in flux density and position. This was done to check to what extent the detected sources may be resolved. We adopted a traditional $S/N$ cut of $\geq$ 5 to select "secure" detections, which we present in Table~\ref{tab:gold}. The positional error is estimated by fitting the sources in the uv-plane (See $\S$4.2.1). Based on the deep {\it HST} $F160W$ imaging, a counterpart search was performed by selecting the sources in the {\it HST} catalogs with separations to the ALMA source coordinates smaller than the corresponding ALMA synthesized beams. All but one of these sources have clear near-IR (NIR) counterparts and the majority are exceptionally red (see Figure \ref{fig:counterparts} for color images); the association with such rare sources strongly confirms the reliability of the mm detections.

The only source where a counterpart match is challenging is A2744-ID02, as the ALMA detection sits in a region where emission is quite faint even in the $F160W$ filter and undetected in the bluer bands, exhibiting a highly dust extincted spectral energy distribution. As seen in Figure \ref{fig:counterparts}, the counterpart of A2744-ID02 appears to be extended in the East-West axis, with 2--3 faint NIR objects ("clumps") lying on either side of the extended 1.1\,mm emission and fainter "bridge" emission extending both in between these clumps as well as toward a bright compact peak of NIR emission $\sim1\farcs5$ to the west of the ALMA source. All of the emission is red in NIR colors, similar to the other ALMA-FFs counterparts. The resolved ALMA emission appears to be elongated roughly coincident with a suppression in the $F160W$ emission, suggesting that the ALMA source may arise from a dusty region that divides the faint $F160W$ clumps, which may represent less-obscured regions from a single extended object with strong and variable extinction. Interestingly, at longer wavelengths ($K_{s}$ and IRAC 3.6--8.0\,$\mu$m), the flux at the ALMA position increases relative to the clumps, such that by 8\,$\mu$m the peak emission is in fact centered almost exactly on the ALMA position. This scenario is  supported by the fact that the clumps have similar photometric redshifts of $z_{\rm ph}\approx2.5$. We select the nearest detected clump ($\sim0\farcs7$ offset) as the counterpart to the ALMA detection. A full characterization of the NIR counterpart galaxies will be discussed in N. Laporte et al. 2016, in preparation, where we find that all of the NIR counterpart galaxies have photometric redshifts of $z$$\geq$1 (none are members of the lensing galaxy clusters.

Figure\,\ref{fig:sn_clusters} shows the histogram of $S/N$ values measured in all pixels with ${\rm PB}$ $\geq$ 0.5 for each cluster. The noise distribution of our images is well-matched to a Gaussian distribution in the low $S/N$ regime, while an excess of positive signal is seen clearly beyond $S/N\gtrsim 4$--5, indicative of real signal from extragalactic sources. Weak hints of deviations from the Gaussian distribution in the measured pixel $S/N$ are observed in the image of A2744 around $-5\sigma$, suggesting there might be some small, additional systematic errors that have not been accounted for; such deviations appear to increase when we include regions with $PB < 0.5$, although in general these are few compared to the large positive excess seen in this field. Further investigations into the cause of these small deviations were unfruitful, and thus we only caution that there may be unaccounted for systematic errors at the 5--10\% level.

\section{Discussion}\label{sec:discuss}

\begin{table*}
\caption[]{Flux density measurements for high-significance continuum detections. 
{\it Col. 1}: Source ID.
{\it Col. 2}: Integrated flux density and 1$\sigma$ statistical error assuming a PS model, in mJy.
{\it Col. 3}: Integrated flux density and 1$\sigma$ statistical error assuming a 6-parameter extended Gaussian model (EXT), in mJy.
{\it Col. 4}: Reduced $\chi^{2}/DOF$ for PS model.
{\it Col. 5}: Reduced $\chi^{2}/DOF$ for EXT model.
{\it Col. 6}: Integrated flux density and 1$\sigma$ statistical error from $uv$ fitting, in mJy.
\label{tab:gold_flux}}
\centering
\begin{tabular}{cccccc}
\hline     
{ID}&{F$_{\rm Int,\,PS}$ [mJy]} & {F$_{\rm Int,\,EXT}$ [mJy]} & {$\chi^2_{\rm red,\,PS}$} & {$\chi^2_{\rm red,\,EXT}$} & {F$_{\rm uv-fit}$ [mJy]} \\
\hline
\noalign{\smallskip}
A2744-ID01     &$ 1.495 \pm 0.081 $&$ 1.545 \pm 0.081 $& 1.7 & 1.6   &$1.570\pm0.073$\\
A2744-ID02     &$ 1.656 \pm 0.131 $&$ 3.262 \pm 0.213 $& 10.3 & 2.2  &$2.816\pm0.229$\\
A2744-ID03     &$ 1.074 \pm 0.084 $&$ 1.979 \pm 0.137 $& 7.9 & 2.0   &$1.589\pm0.125$\\
A2744-ID04     &$ 1.012 \pm 0.099 $&$ 1.173 \pm 0.111 $& 2.0 & 1.7   &$1.009\pm0.074$\\
A2744-ID05     &$ 0.859 \pm 0.125 $&$ 1.422 \pm 0.188 $& 3.0 & 0.7   &$1.113\pm0.135$\\
A2744-ID06     &$ 0.822 \pm 0.129 $&$ 2.274 \pm 0.274 $& 12.1 & 11.9 &$1.283\pm0.241$\\
A2744-ID07     &$ 0.500 \pm 0.108 $&$ 0.716 \pm 0.149 $& 1.5 & 1.4   &$0.539\pm0.082$\\
MACSJ0416-ID01 &$ 1.144 \pm 0.089 $&$ 1.356 \pm 0.111 $& 3.0 & 0.6   &$1.319\pm0.103$\\
MACSJ0416-ID02 &$ 0.455 \pm 0.081 $&$ 0.626 \pm 0.111 $& 2.8 & 2.0   &$0.574\pm0.132$\\
MACSJ0416-ID03 &$ 0.375 \pm 0.091 $&$ 0.477 \pm 0.114 $& 4.7 & 4.3   &$0.411\pm0.072$\\
MACSJ0416-ID04 &$ 0.367 \pm 0.089 $&$ 0.456 \pm 0.111 $& 3.0 & 2.9   &$0.478\pm0.166$\\
MACSJ1149-ID01 &$ 0.500 \pm 0.104 $&$ 0.635 \pm 0.125 $& 1.1 & 0.5   &$0.579\pm0.134$\\
\hline
\end{tabular}
\end{table*}

\begin{figure*}[!htbp]
\includegraphics[width=\hsize/2]{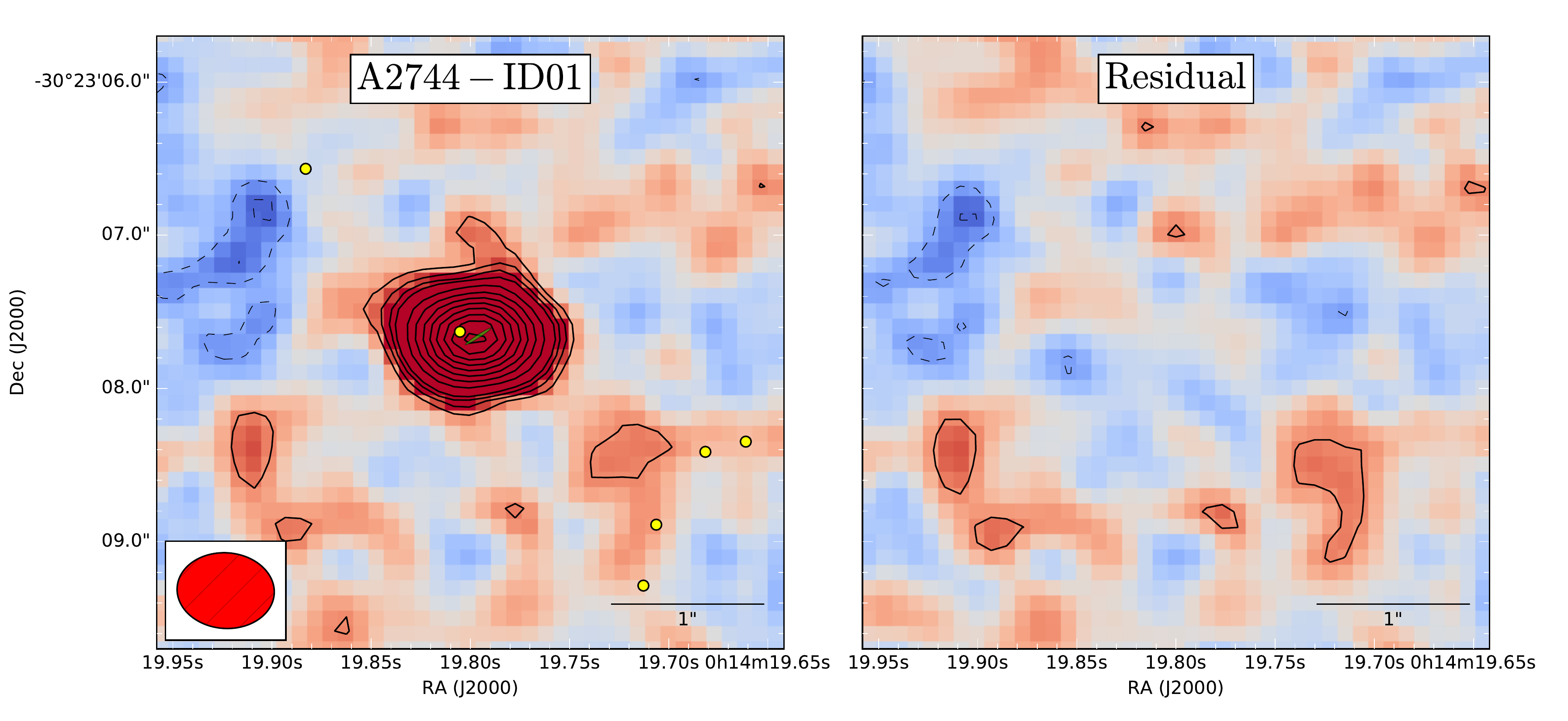}
\includegraphics[width=\hsize/2]{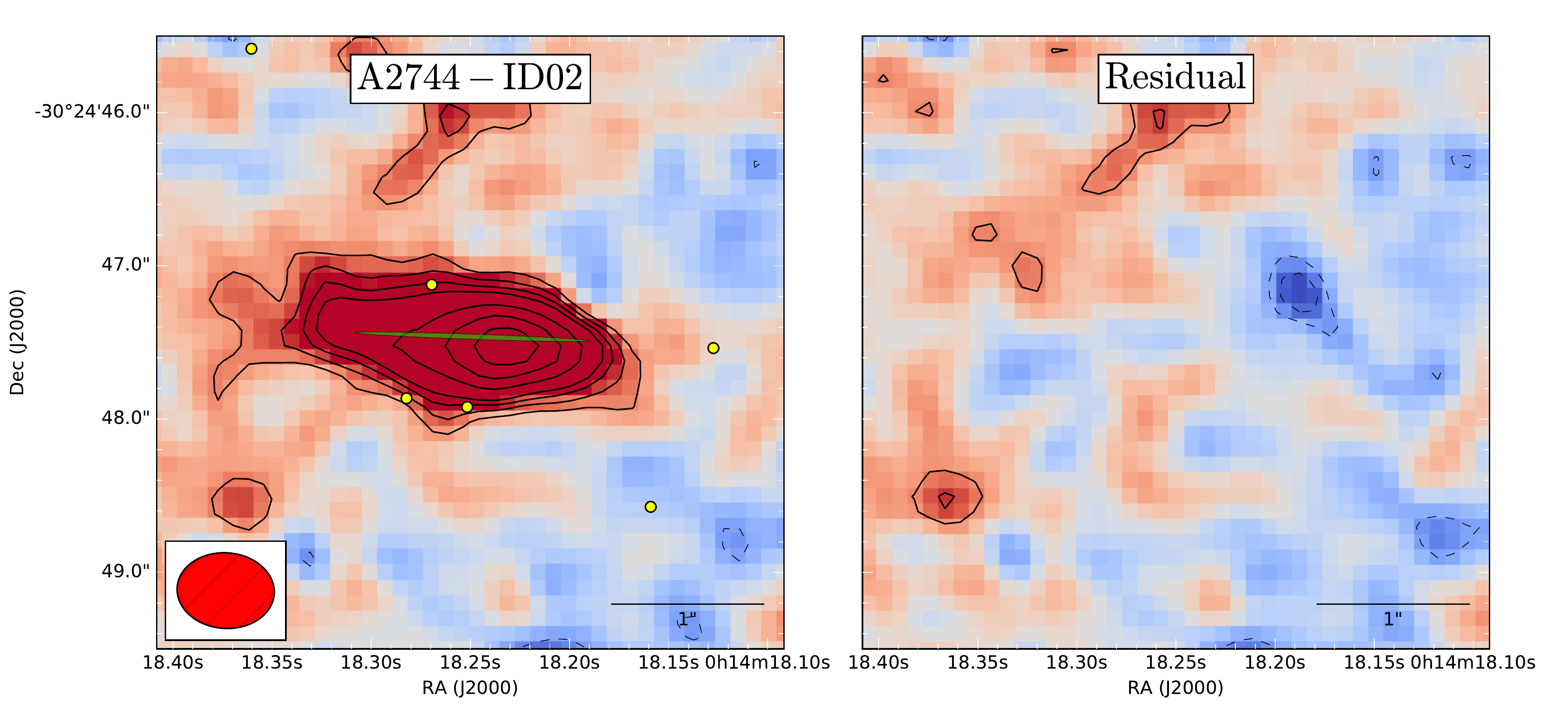}
\includegraphics[width=\hsize/2]{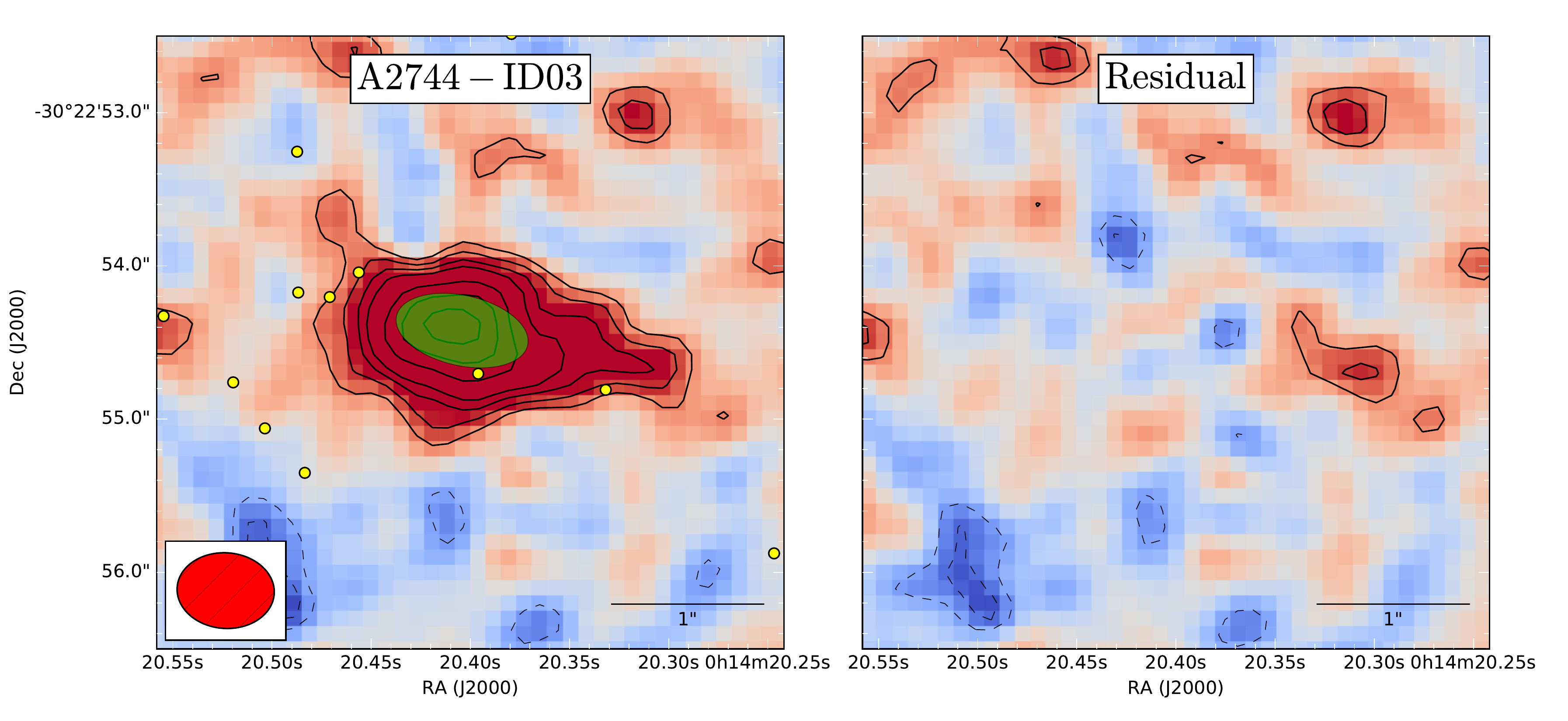}
\includegraphics[width=\hsize/2]{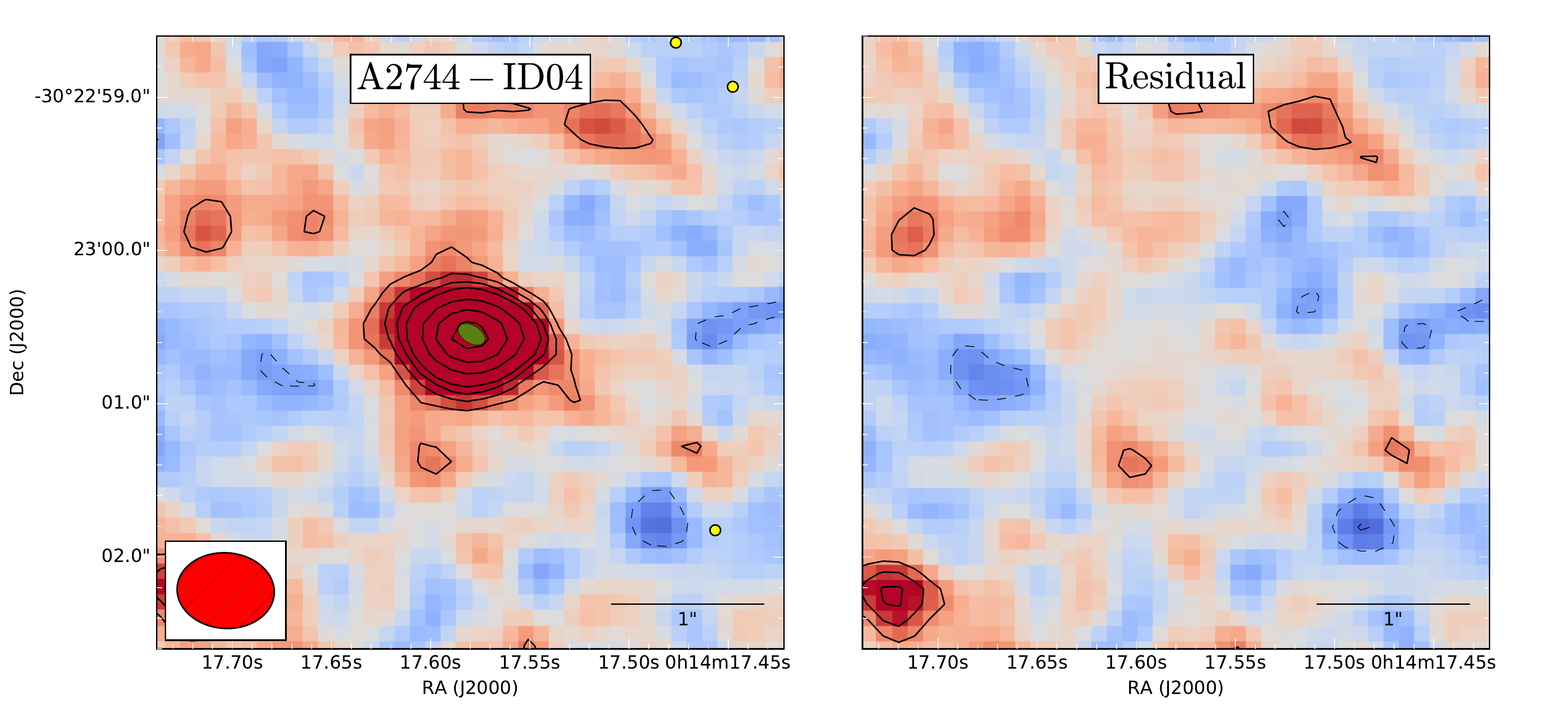}
\includegraphics[width=\hsize/2]{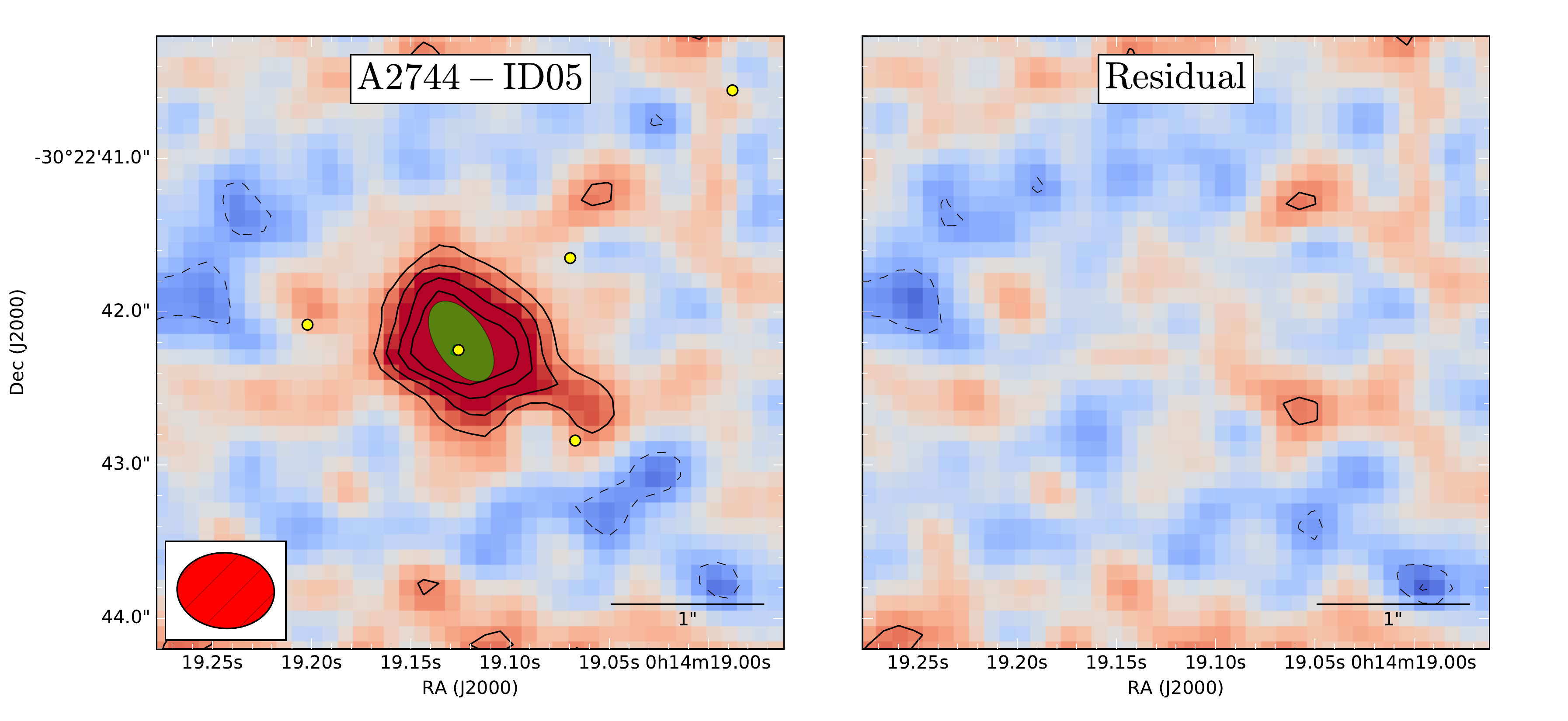}
\includegraphics[width=\hsize/2]{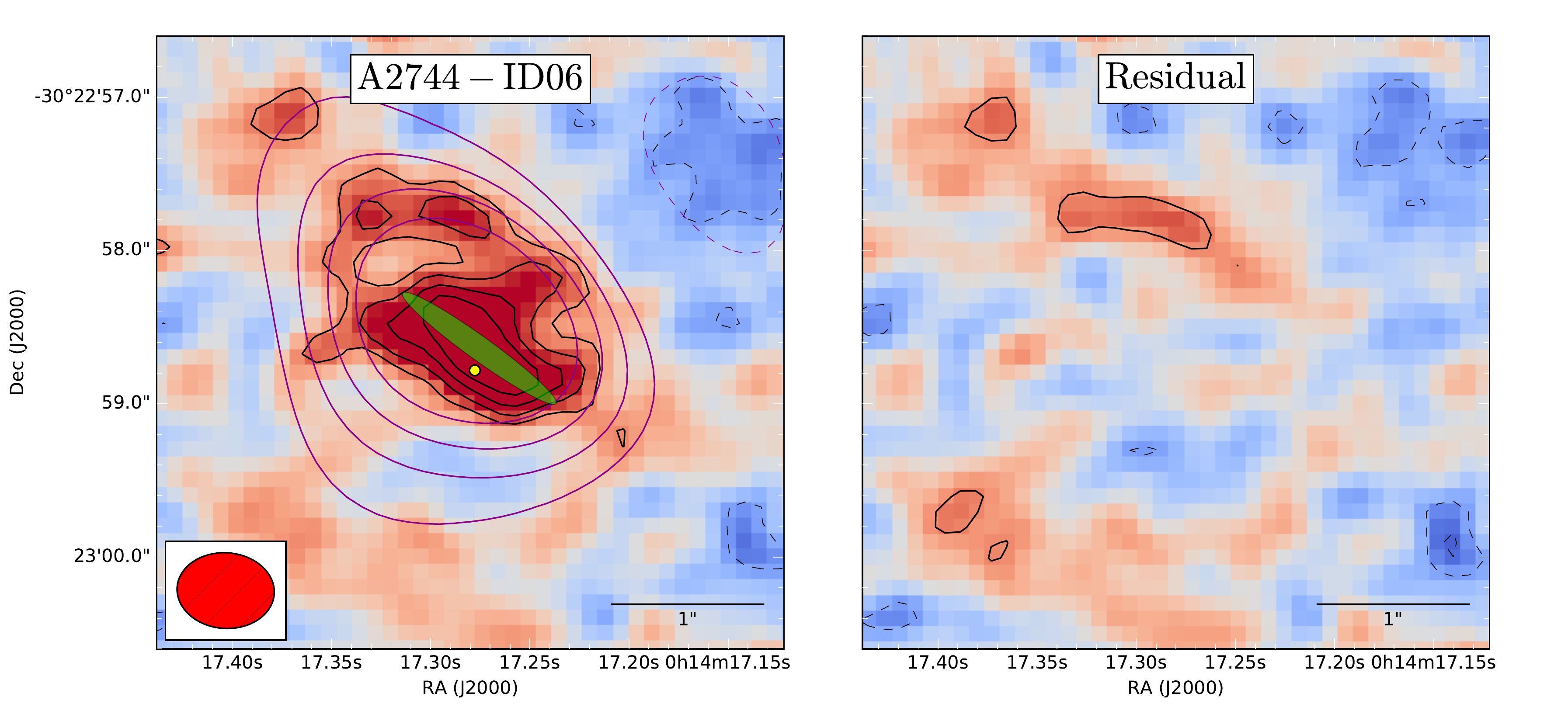}
\includegraphics[width=\hsize/2]{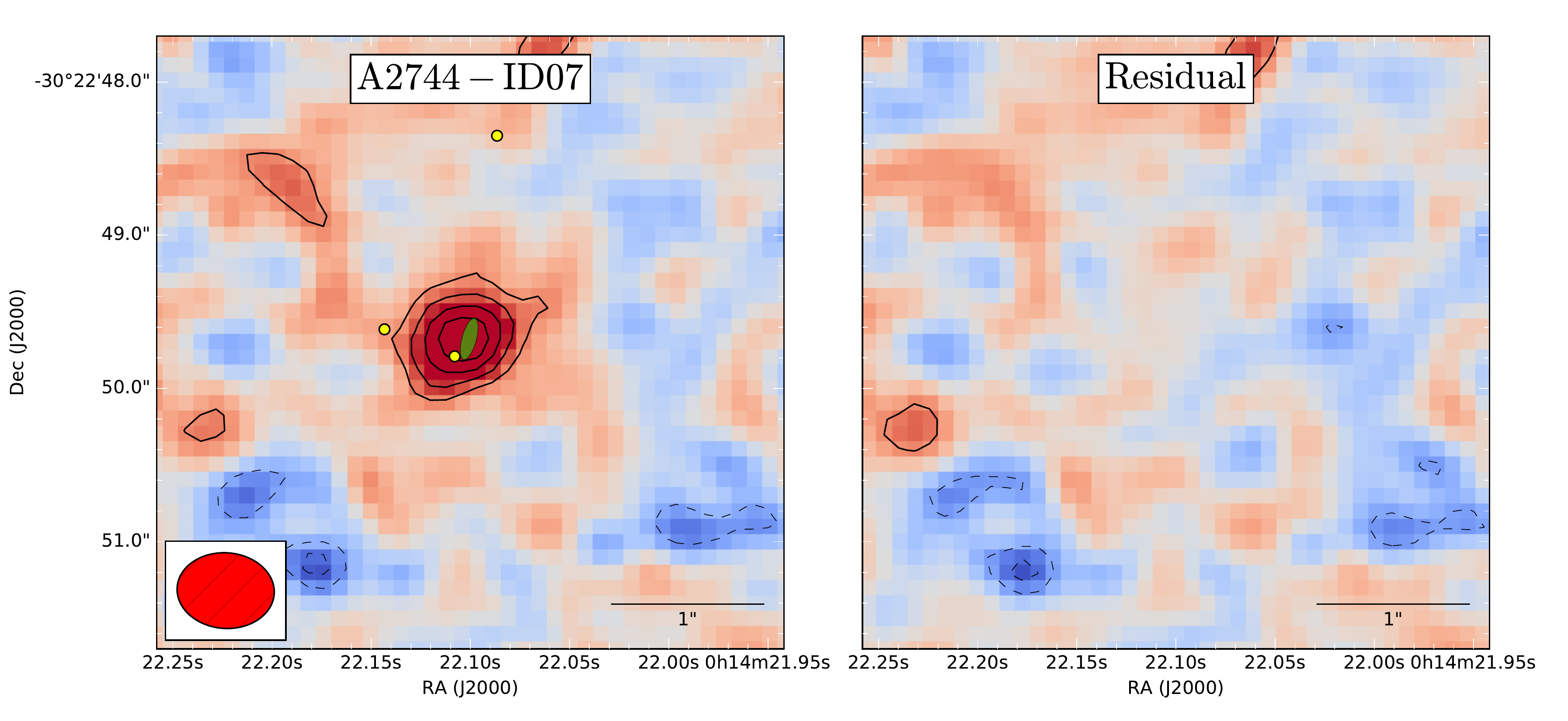}
\includegraphics[width=\hsize/2]{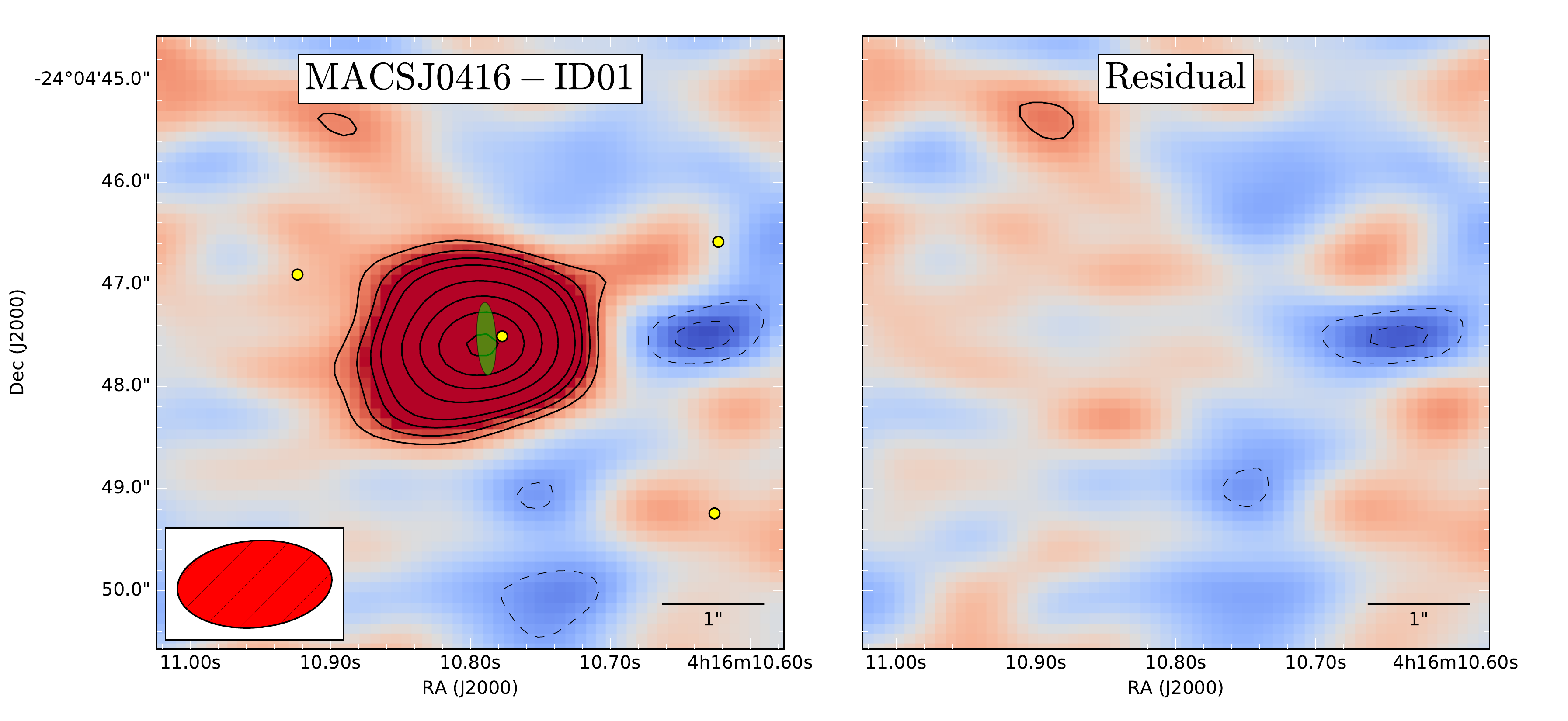}
\includegraphics[width=\hsize/2]{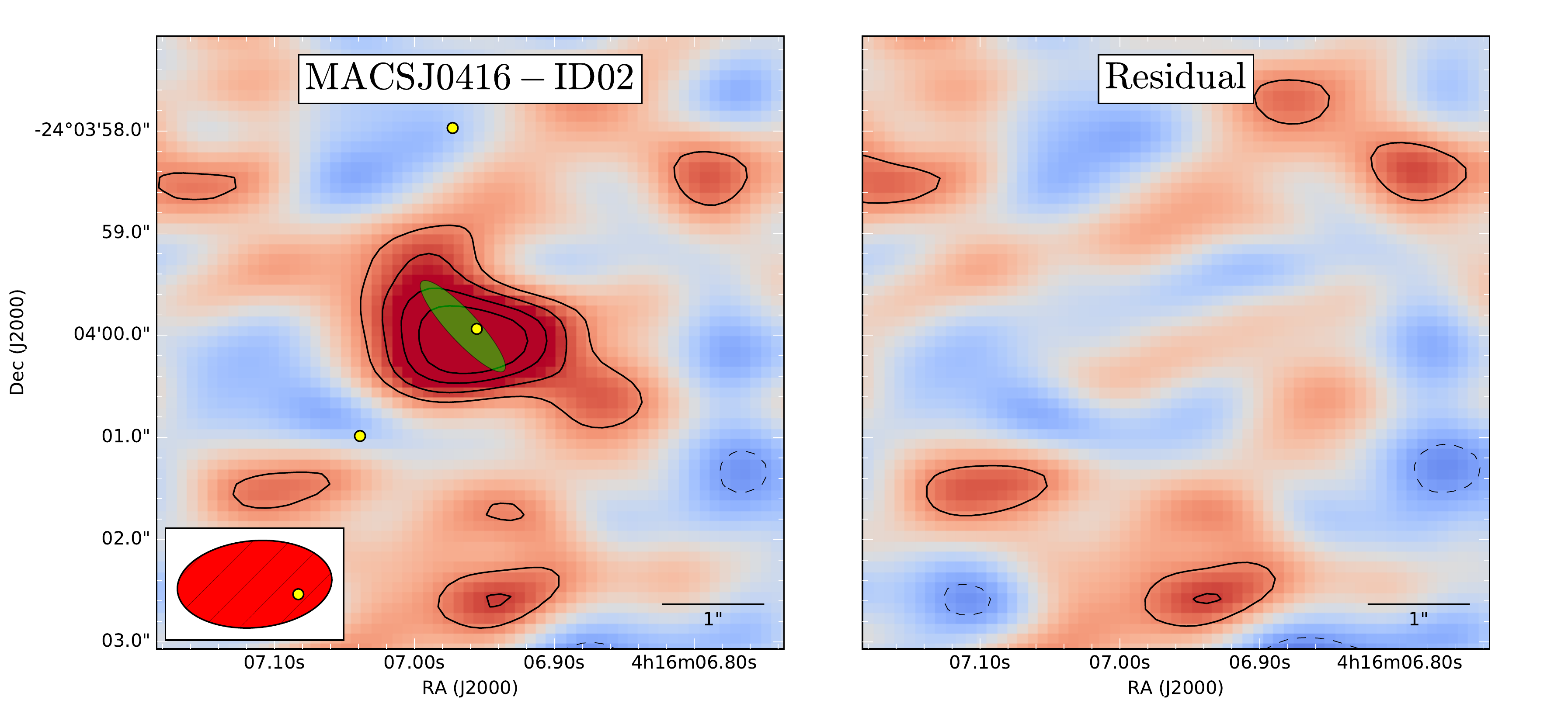}
\includegraphics[width=\hsize/2]{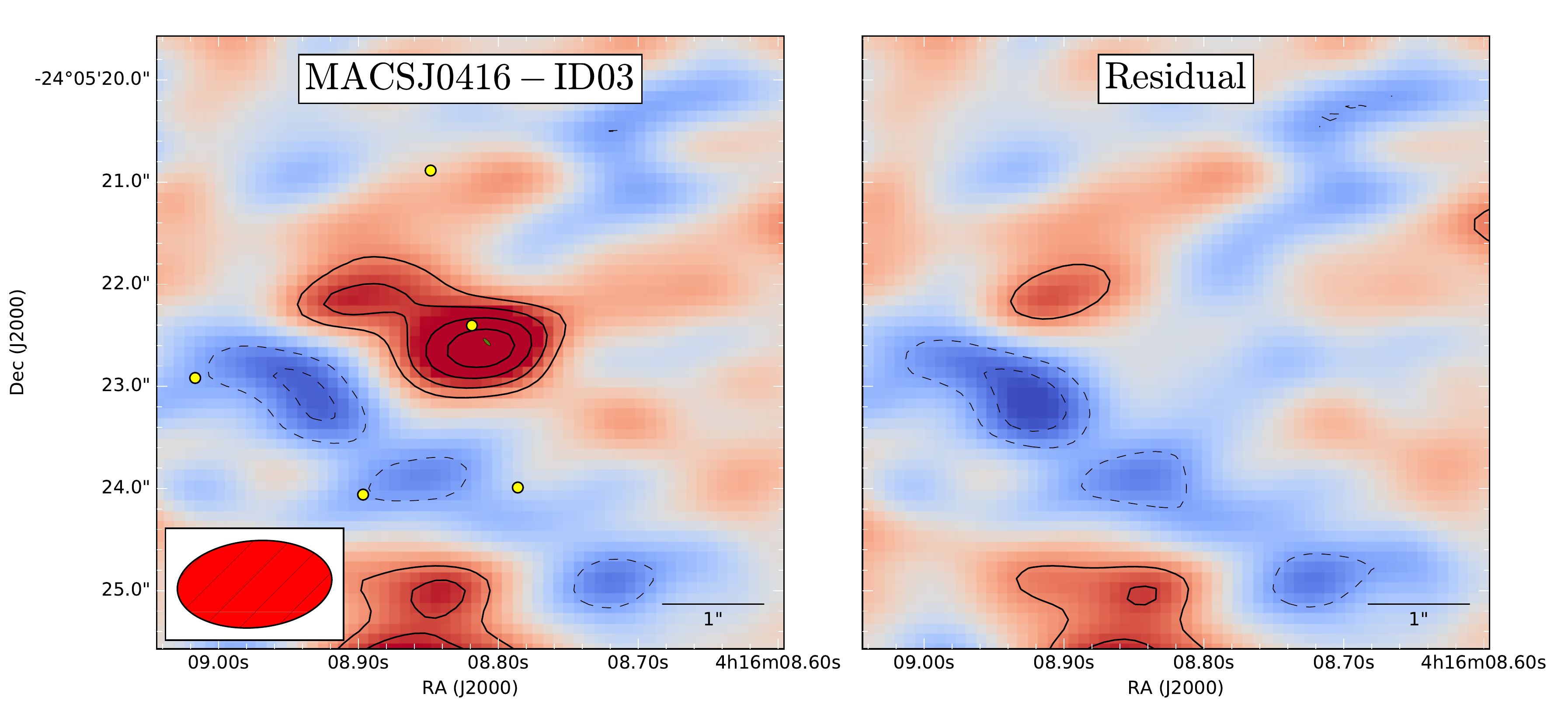}
\caption{Cutout images of the secure continuum detections in the FFs clusters. \textit{left:}  1.1\,mm continuum emission image adopting natural weighting, with colors as in Figure~\ref{fig:map_A2744}. 
Solid black curves show the positive $S/N$ contours with natural weighting, starting from $\pm2\sigma$ up to $\pm5\sigma$ in steps of $1\sigma$, and then above $\pm5\sigma$ in steps of $2.5\sigma$. Dashed curves show the negative $S/N$ contours.
In the bottom left corner we show the corresponding synthesized beam. For A2744-ID06 we additionally overlay solid purple curves showing the $S/N$ contours from the corresponding Taper weighted image.
The best-fit 2-d elliptical Gaussian model is shown as a green region whose size denotes where the emission is half of the maximum. The yellow points represent the position of optical/NIR detected galaxies. \textit{right:} Residual of the 1.1\,mm continuum emission after the best-fit model is subtracted from the $uv$ visibilities. The color scale is identical to the left-hand side.
\label{fig:cont_fits1}}
\end{figure*}

\begin{figure*}[!htbp]
\includegraphics[width=\hsize/2]{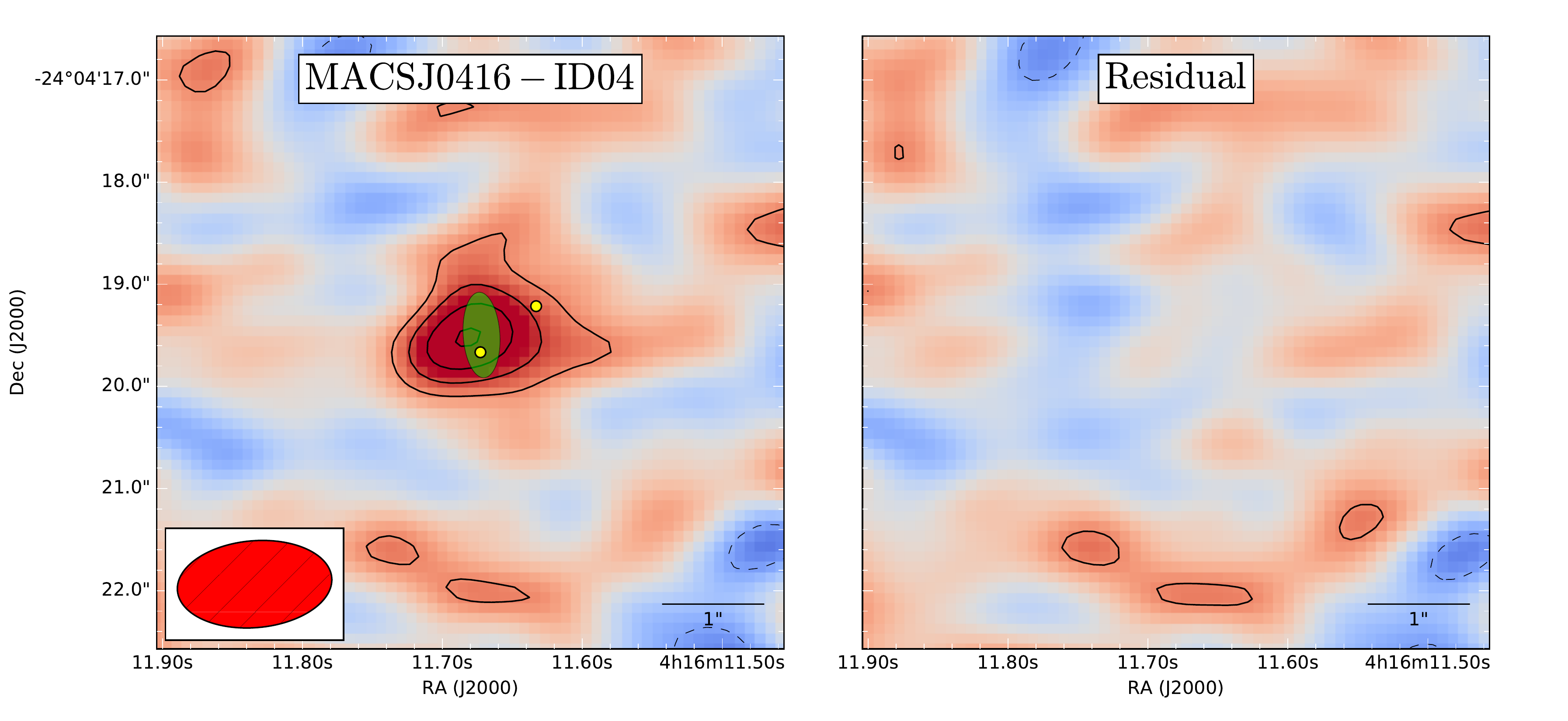}
\includegraphics[width=\hsize/2]{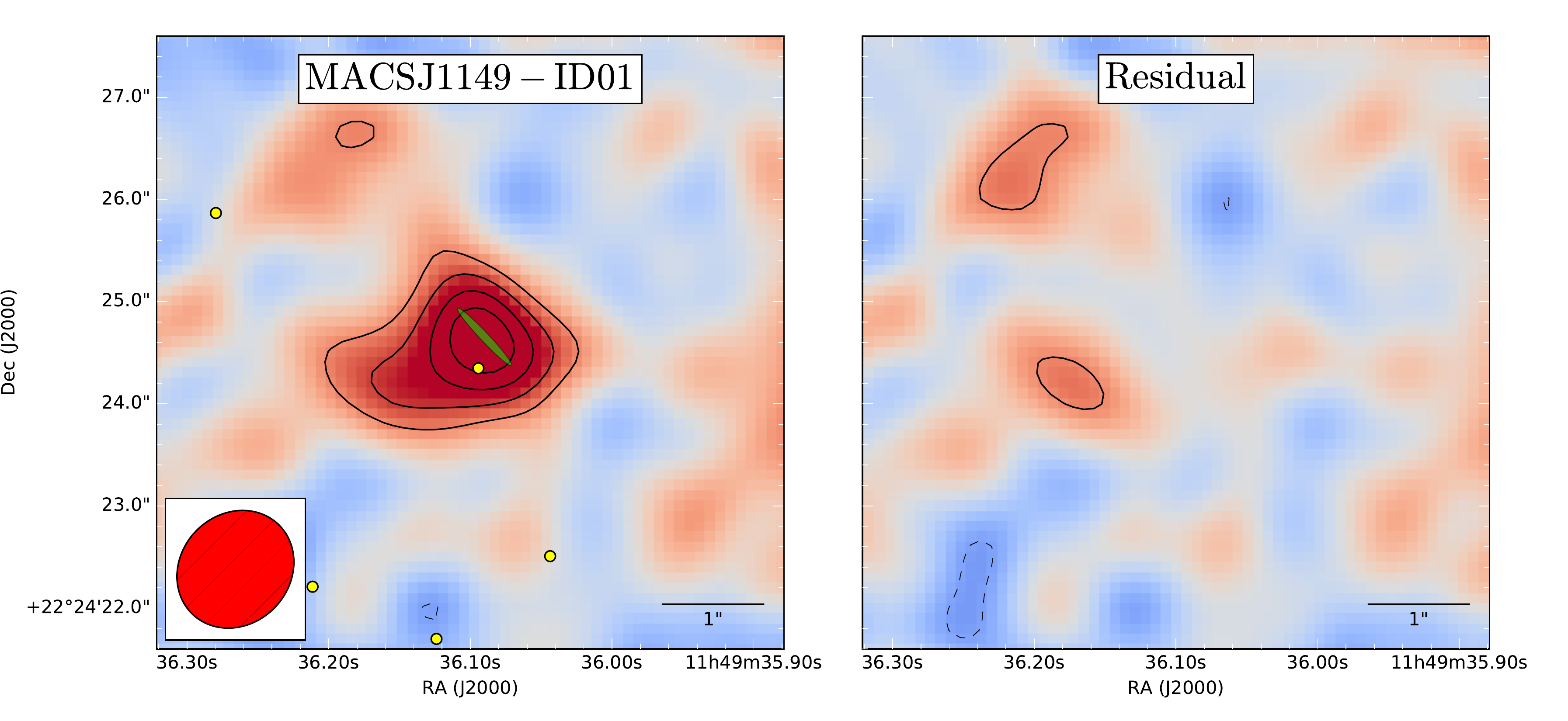}
\caption{Continuation of Figure~\ref{fig:cont_fits1}.
\label{fig:cont_fits2}}
\end{figure*}

\subsection{Purity}\label{sec:purity}

In order to establish the purity levels of our detections, we need to estimate how many false detections we expect per $S/N$ for a given observation. We use the task \texttt{Simobserve} in {\sc casa} to simulate observations with similar properties as the ones obtained in our campaign. For the simulations the thermal noise is added using an atmospheric profile for the ALMA site \citep{Pardo2001} and assuming the observations are performed when the field transits (this assumption is not critical for the purposes here). We adopt the nominal second (MACSJ0416 and MACSJ1149) and third (A2744) most compact ALMA antenna configurations (C36-2 and C36-3) during cycle 2, which are the closest matches to the ones used in the real observations. The other parameters such as frequency coverage, PWV and integration time are set to achieve an rms similar to the ones obtained for each cluster field. After construction of the simulations, we used the noisy measurement set files to create images in the same manner as used to produce the real observations. These images were then analyzed with the same algorithm used to detect sources (\S\ref{sec:imaging}). 

We performed 30 simulations per antenna configuration and recorded the number of detections with $S/N\geq5$ recovered by our code. The number of detections was then renormalized as a function of $S/N$ for each antenna configuration. For all antenna configurations, we obtain $N\ll1$ simulated-noise detections for $S/N\geq5$, meaning that our $S/N\geq5$ sample should all be true detections assuming that the noise properties of the simulations are similar to those in the observations.

The deviation from a perfect Gaussian distribution ratio at $S/N$$\sim$$-5.0$ in the data distribution observed in Figure \ref{fig:sn_clusters} and discussed in $\S$\ref{Sec:high_significance_cont_detections} can be explained by low-number statistics, as the same behavior is observed in the simulated observations histograms when Gaussian noise is assumed. It is important to note that because of the stochastic behavior in the extreme tails of the distribution, extra care needs to be taken when using the negative pixels as a noise reference, as some fraction of noise distributions will never be completely symmetrical, even when Gaussian noise is assumed. This effect can easily lead to misinterpretations when using the negative pixels as a reference to estimate the purity of continuum detections. For instance, in MACSJ0416 and MACSJ1149, we find no negative sources with $S/N$$\leq$$-5.0$, while in A2744 we find two sources with $S/N$$\leq$$-5.0$ compared to seven sources with $S/N$$\leq$$+5.0$. The implied purity for A2744 is thus $\sim$0.7 based on negative count symmetry. However, based on the NIR counterparts found for all the ALMA sources with $S/N\geq5$, we estimate a "true" detection purity closer to $\approx$1 for the three clusters, consistent with the simulations. Although the difference between the purity estimated from negative counts and the "true" purity is not statistically significant (only $1.1\sigma$), it exemplifies the extra care needed when using the negative counts for purity estimates.

\subsection{Flux estimates}\label{sec:flux}
\subsubsection{Image fit}

We obtained three flux density measurements for each source. The first two methods fit for the flux density in the image plane assuming a point source (PS) and an extended source (6 parameters Gaussian function, hereafter EXT), respectively, while the third method fits for the flux density in the $uv$-plane. Fitting the sources in the image plane is substantially more straightforward, but inherits certain dependencies based on the weighting method used to construct the images. 

To measure the flux densities from the images, we fit an elliptical Gaussian to each source. For a PS, we fixed the size and angle of the elliptical Gaussian to that of the synthesized beam, and only allow for changes in flux and position. For an EXT, we allow all the six parameters describing the 2-d Gaussian to vary. For both fits, a reduced $\chi^2$ is estimated using all available image pixels that satisfy $S/N>2$ (to limit the fit to pixels associated with the source). The degrees of freedom are estimated as $DOF = N-P$, where $N$ is the number of pixels with $S/N>2$ and $P$ is the number of free parameters in the fit (three for a PS and six for an EXT).
The measured flux densities and reduced $\chi^2$ values are presented in Table~\ref{tab:gold_flux}. These values were measured in the natural weighting images.

The $1\sigma$ uncertainties in the flux densities measurements are of the order of 0.1 mJy. We should expect a number density of $\approx10^{4.8}\,\,{\rm deg}^{-2}$ sources with flux densities $>0.084$ mJy \citep{Fujimoto2016}. Even for the largest beam obtained in MACSJ1149 (Table~\ref{tab:imaging_results}) we obtain $\sim140$ beams per source at such flux density levels, therefore expecting a negligible flux boosting from confused sources in our results.

\subsubsection{$uv$-plane fit}

To measure the flux in the $uv$-plane we used \texttt{uvmcmcfit},\footnote{\url{https://github.com/sbussmann/uvmcmcfit}} which is a \texttt{Python} implementation to fit models to interferometric data in the $uv$ plane. The code allows one to extract the maximum amount of information from the observations, particularly if the sources appear marginally resolved.

\texttt{uvmcmcfit} is designed to allow for de-lensing of sources in the case of galaxy-galaxy or galaxy-group magnification. We chose to first model the observed sources as if no magnification were present. This approach should return the shape and flux of the sources in the lensed image plane. Reconstructions to the source plane of each detected galaxy are beyond the scope of the current work and will be presented in J. Gonzalez-Lopez et al. (2016, in prep.).

For each source, we adopt the simplest model, assuming that the FIR emission can be fit with an elliptical Gaussian described by the following parameters: the total intrinsic flux density, the position of the source (RA,DEC), the effective radius defined as $r_{s}=\sqrt{a_{s}\times b_{s}}$ ($b$ and $a$ are the major and minor axis respectively), the axial ratio ($q_{s}=b_{s}/a_{s}$), and the position angle in degrees east of north ($\phi_{s}$). The effective radius is a good measurement of the total size of the sources.

The goodness of fit for a given set of model parameters is determined from the maximum likelihood estimate $L$ given by

\begin{equation}
L = \sum_{u,v}\left(\frac{|V_{\rm ALMA}-V_{\rm model}|^2}{\sigma^2}+\log{2\pi\sigma^2}\right),
\end{equation}

\noindent where $\sigma$ is the $1\sigma$ uncertainty level for each $uv$ complex visibility (hereafter visibility; Fourier transform of the intensity distribution on the sky) given by the associated weights. In practice, each source is typically observed by only a handful of pointings from the full mosaic, and each of these pointings will have an associated weight that we must apply to its visibilities to combine with those from neighboring pointings. Because well-determined weights for the visibilities of each pointing are needed, we performed the following procedure for each source before fitting. 

We estimate an initial source position based on the continuum images and select all pointings which lie within 19\arcsec (PB$\geq$0.1) of the initial position. We then shift the phase center of all the selected pointings to the position of the source. We correct the amplitudes of the visibilities by the initial PB correction of each pointing with respect to the position of the source as ${\rm data}_{i} = {\rm data}_{i}/{\rm PB}_{i}$, where $i$ is the index of each pointing. The weights of the visibilities are then corrected by the same PB correction as for the amplitudes as ${\rm weight}_{i} = {\rm weight}_{i}\times{\rm PB}^{2}_{i}$. We concatenate all the pointings into a new dataset, which should now contain a correct set of relative weights for the visibilities, since the appropriate factor was applied to the weights given by the calibration. 
Finally, we scale the weights such that:
\begin{equation}
\sum {\rm weight}\times{\rm real}^{2} + {\rm weight}\times{\rm Im}^2 = N_{\rm visibilities}.
\end{equation}

\texttt{uvmcmcfit} uses \texttt{EMCEE} \citep{Foreman-mackey2013} to sample the posterior probability density function (PDF) of the model parameters. Between 10,000--240,000 iterations were required, depending on the speed of convergence. We used a 'burn-in' phase of 5,000 iterations to identify the 'best-fit' model parameters.
The flux density given by the posterior PDF is presented in the last column of Table~\ref{tab:gold_flux}. The errors presented correspond to the 1 $\sigma$ range of the posterior PDF. The best-fit models are presented in Figures~\ref{fig:cont_fits1} and \ref{fig:cont_fits2}.

\begin{figure}[!htbp]
\centering
\includegraphics[width=\hsize]{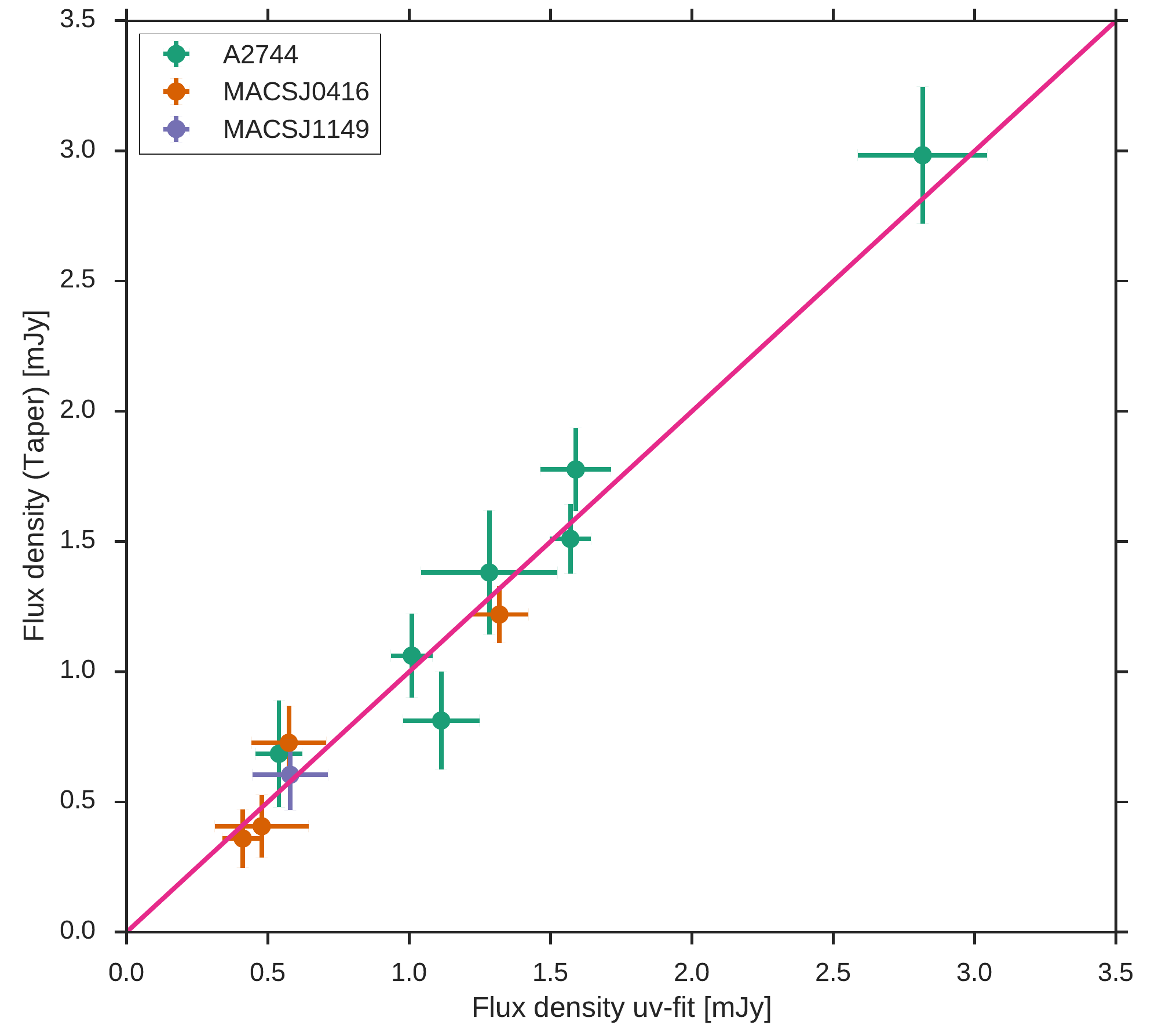}
\caption{Comparison of the flux density measured using the $uv$-plane fit method with that measured in Taper images. We see that the $uv$-plane fit recovers most of the emission detected in the lower resolutions images, within the corresponding errors. 
\label{fig:fluxes_comparison}}
\end{figure}

To test that we are recovering most of the observed flux density with the $uv$-plane fit, we created a "Taper" image for each observation. This Taper image also adopts natural weighting but includes a uvtaper with an outertaper equal to 1\farcs5, which yields a beam size $\gtrsim$1\farcs5 but substantially worse rms sensitivity. We apply the same method used to measure the flux density in the natural weighted images on the Taper images. In Fig. \ref{fig:fluxes_comparison} we present the flux density measured using the $uv$-plane fit in the natural image and the F$_{\rm Int,\,EXT}$ measured in the Taper images. We see that the two estimates agree within the errors, demonstrating that we apparently recover a substantial amount (if not all) of the extended flux with the $uv$-plane fit method. We recover the total flux density even in the complex case of A2744-ID06 (see Fig. \ref{fig:cont_fits1}), where the Taper image emission is substantially more extended than the natural weighted emission. 
Given these findings, we adopt the F$_{uv-fit}$ flux estimates for all sources.

\subsection{Source sizes}\label{sec:size}

\begin{table*}
\caption[]{Source size measurements for high-significance continuum detections.
{\it Col. 1}: Source ID. 
{\it Col. 2}: Effective radius defined as $r_{\rm s}=\sqrt{a_{\rm s}\times b_{\rm s}}$, in arcseconds. 
{\it Col. 3}: The axial ratio, $q_{\rm s}=b_{\rm s}/a_{\rm s}$.
{\it Col. 4}: The position angle in degrees east of north, $\phi_{\rm s}$.
{\it Col. 5}: Magnification value ($\mu$) estimated using the available lensing models and assuming $z=2\pm1$ for the sources without spectroscopic redshift. For MACSJ0146-ID01 and MACSJ0146-ID02, we adopt $z_{\rm spec}=2.086$ and $z_{\rm spec}=1.953$ from GLASS, respectively.
{\it Col. 6}: Demagnified effective radius, $r_{\rm s, demag}$ estimated as $r_{\rm s}/\sqrt{\mu}$, in arcseconds. Upper limits correspond to $2\sigma$.
{\it Col. 7}: Demagnified flux density, $F_{\rm uv-fit, demag}$, estimated as $F_{\rm uv-fit}/\mu$, in mJy.
{\it Col. 8}: Extended emission classification, with 0 for point-like, 1 for marginally extended, and 2 for significantly extended.
\label{tab:gold_size}}
\centering
\begin{tabular}{cccccccc}
\hline     
{ID} & {$r_{\rm s}$ [$\arcsec$]} & {$q_{\rm s}$} & {$\phi_{\rm s}$ [$^{\circ}$]} & {$\mu$} & $r_{\rm s, demag}$ [$\arcsec$] & {$F_{\rm uv-fit, demag}$ [mJy]} & Extension. \\
\hline
\noalign{\smallskip}
A2744-ID01 & $ 0.05 \pm 0.01 $ &  $ 0.47 \pm 0.21 $ &  $ 110 \pm 26 $ & $2.8^{+1.3}_{-0.6}$ & $0.03^{+0.01}_{-0.01}$ & $0.557^{+0.163}_{-0.176}$ & 1 \\[5pt]
A2744-ID02 & $ 0.23 \pm 0.04 $ &  $ 0.17 \pm 0.05 $ &  $ 85 \pm 2 $ & $1.7^{+0.6}_{-0.4}$ & $0.17^{+0.04}_{-0.03}$ & $1.630^{+0.534}_{-0.373}$ & 2 \\[5pt]
A2744-ID03 & $ 0.26 \pm 0.03 $ &  $ 0.58 \pm 0.13 $ &  $ 81 \pm 11 $ & $1.9^{+0.3}_{-0.3}$ & $0.19^{+0.03}_{-0.03}$ & $0.849^{+0.161}_{-0.133}$ & 2 \\[5pt]
A2744-ID04& $ 0.06 \pm 0.02 $ &  $ 0.62 \pm 0.24 $ &  $ 84 \pm 51 $ & $2.7^{+1.2}_{-1.0}$ & $0.04^{+0.02}_{-0.01}$ & $0.372^{+0.197}_{-0.112}$ & 1 \\[5pt]
A2744-ID05 & $ 0.19 \pm 0.05 $ &  $ 0.66 \pm 0.23 $ &  $ 60 \pm 42 $ & $1.6^{+0.3}_{-0.3}$ & $0.16^{+0.04}_{-0.04}$ & $0.685^{+0.197}_{-0.122}$ & 2 \\[5pt]
A2744-ID06 & $ 0.26 \pm 0.08 $ &  $ 0.3 \pm 0.17 $ &  $ 50 \pm 10 $ & $2.2^{+0.7}_{-0.8}$ & $0.18^{+0.06}_{-0.06}$ & $0.577^{+0.285}_{-0.164}$ & 2 \\[5pt]
A2744-ID07 & $ 0.07 \pm 0.04 $ &  $ 0.56 \pm 0.24 $ &  $ 85 \pm 57 $ & $1.6^{+0.2}_{-0.2}$ & $<0.12$ & $0.345^{+0.069}_{-0.059}$ & 0 \\[5pt]
MACSJ0416-ID01 & $ 0.23 \pm 0.06 $ &  $ 0.61 \pm 0.24 $ &  $ 99 \pm 71 $ & $1.8^{+0.1}_{-0.5}$ & $0.18^{+0.05}_{-0.04}$ & $0.773^{+0.239}_{-0.107}$ & 2 \\[5pt]
MACSJ0416-ID02 & $ 0.32 \pm 0.15 $ &  $ 0.58 \pm 0.23 $ &  $ 63 \pm 40 $ & $2.2^{+0.3}_{-0.4}$ & $0.22^{+0.10}_{-0.10}$ & $0.259^{+0.086}_{-0.062}$ & 2 \\[5pt]
MACSJ0416-ID03 & $ 0.10 \pm 0.07 $ &  $ 0.62 \pm 0.22 $ &  $ 97 \pm 51 $ & $1.5^{+0.4}_{-0.4}$ & $<0.20$ & $0.267^{+0.093}_{-0.063}$ & 0 \\[5pt]
MACSJ0416-ID04 & $ 0.37 \pm 0.21 $ &  $ 0.56 \pm 0.28 $ &  $ 81 \pm 62 $ & $1.8^{+0.4}_{-0.5}$ & $0.28^{+0.17}_{-0.16}$ & $0.269^{+0.122}_{-0.096}$ & 1 \\[5pt]
MACSJ1149-ID01 & $ 0.28 \pm 0.13 $ &  $ 0.61 \pm 0.23 $ &  $ 92 \pm 44 $ & $4.2^{+5.4}_{-2.1}$ & $0.13^{+0.10}_{-0.07}$ & $0.137^{+0.148}_{-0.083}$ & 2 \\[5pt]
\hline
\end{tabular}
\end{table*}

\begin{figure*}[!htbp]
\includegraphics[width=\hsize/2]{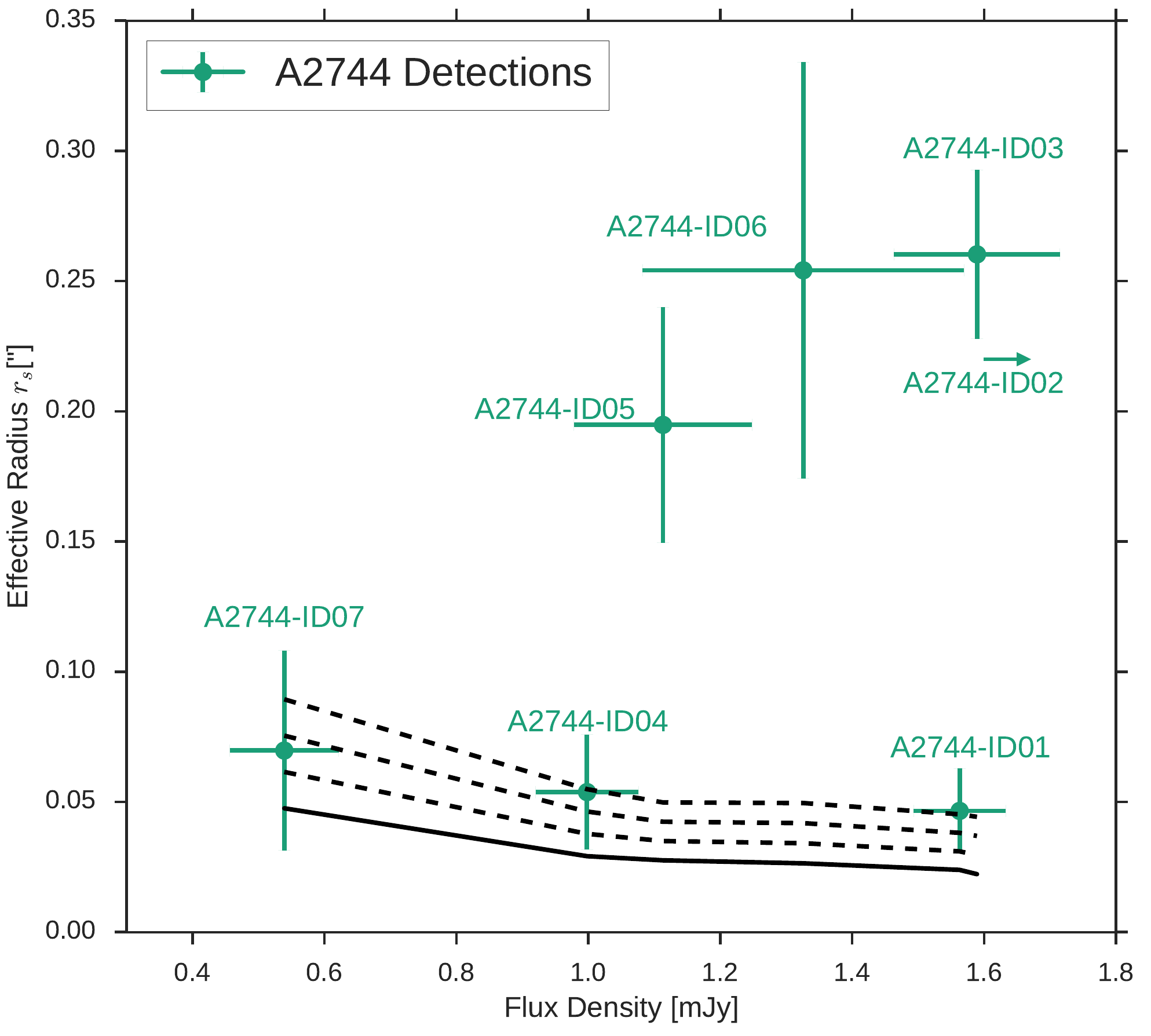}
\includegraphics[width=\hsize/2]{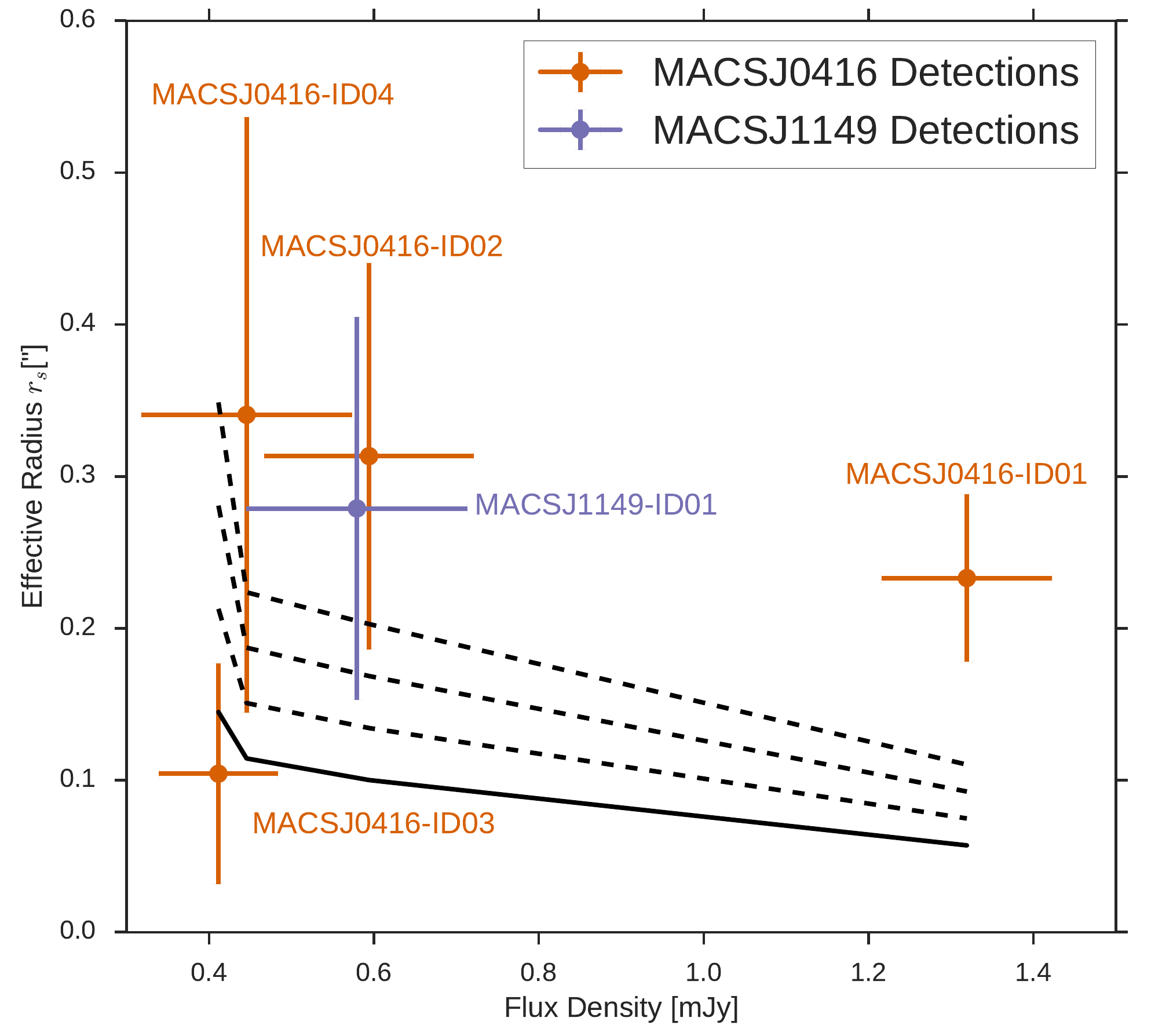}
\caption{Measured effective radii $r_{\rm s}$ as a function of flux density for the ALMA sources. We plot sources from A2744 (\textit{left}) and MACS0416$+$MACS1149  (\textit{right}) separately, since the latter have factors of $\approx$2 larger beam sizes. The solid black line in each plot corresponds to the average value of $r_{\rm s}$ measured for simulated point sources ingested into the data. The dashed black lines likewise correspond to limits of 1, 2 and 3 times the standard deviation ($\sigma$) above the mean for simulated point sources, respectively, as a function of flux density. For visualization purposes, we only plot simulated lines for MACSJ0416 in the right panel, noting that the limits for MACSJ1149 are roughly identical. Additionally, A2744-ID02 is not shown in the left plot, since its flux density is much higher ($\sim$2.8\,mJy) and it is clearly extended.
Based on the criteria in $\S$\ref{sec:size}, two sources are considered point-like, three sources are marginally extended, and seven are significantly extended.
\label{fig:size_point_source}}
\end{figure*}

\begin{figure}[!htbp]
\centering
\includegraphics[width=\hsize]{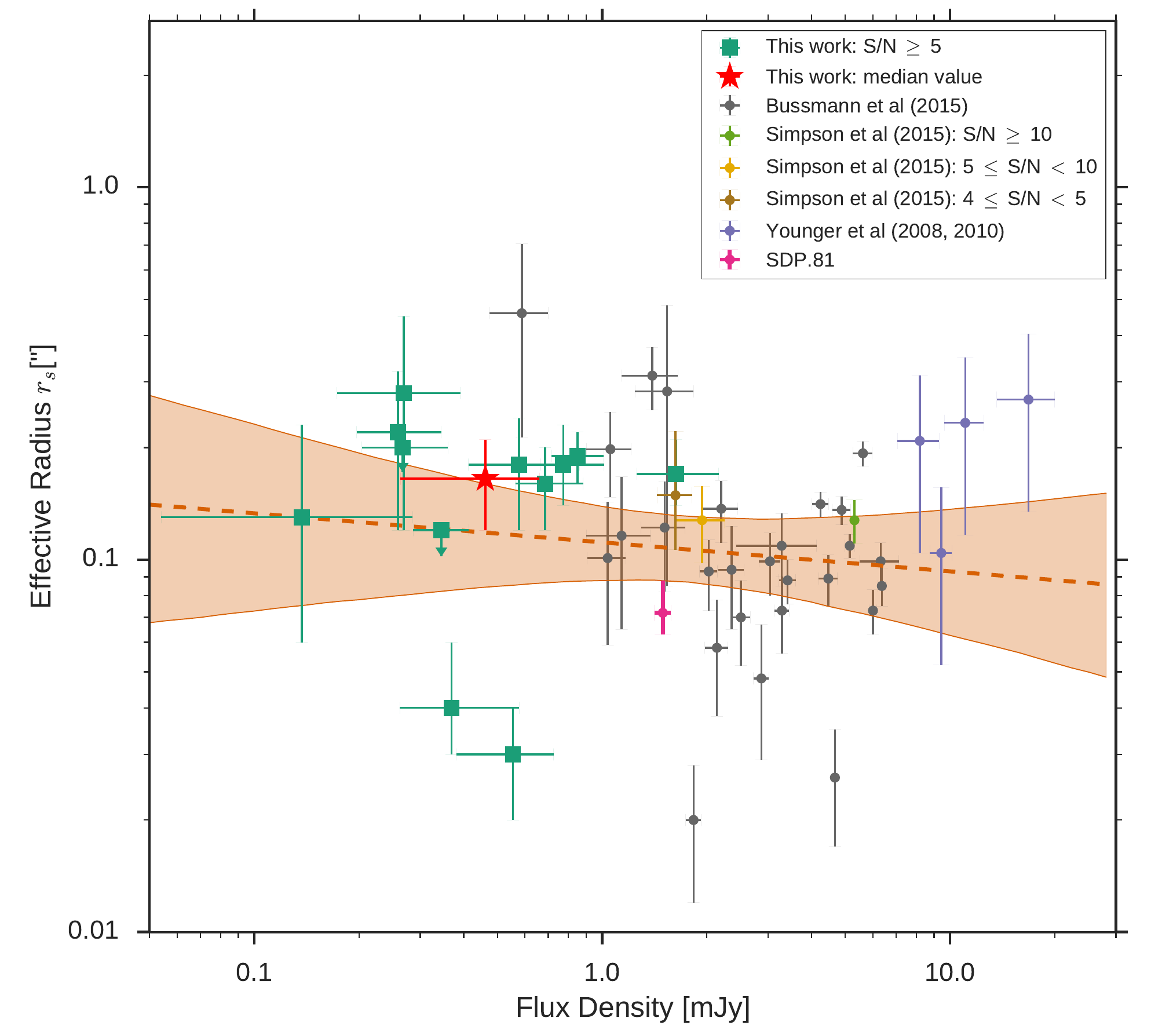}
\caption{Estimated demagnified sizes, $r_{\rm s, demag}$, for the ALMA-detected sources in the three FFs clusters studied here, as measured by fits in the $uv$-plane, plotted alongside measured sizes for several literature samples obtained using interferometric data \citep{Younger2008, Younger2010, Valtchanov2011, Rybak2015, Simpson2015, Bussmann2015}. The estimated demagnified 1.1\,mm flux densities are shown; for the literature sample, when necessary we have converted to this wavelength assuming a slope of $\beta=1.8$.
We see that the ALMA-FFs sources studied here exhibit a large dispersion in extent compared to brighter flux density sources; notably $\sim$50\% of the ALMA-FFs sources appear comparable to the largest literature sample sources. The dashed line shows the best fit regression to all samples, with $\log{r_{\rm s, demag}} = m\times\log{F_{\rm 1.1\,mm}} + b$ with $m=-0.08\pm0.10$ and $b=-0.95\pm0.05$. The orange shaded region shows the $2\sigma$ range of possible fits to the data. The measured sizes show no obvious trend with flux density.
\label{fig:fluxes_sizes}}
\end{figure}

To estimate the sizes of the observed sources (in the natural weighting image plane), we used the 'best-fit' models that were fit in the previous subsection. The source size parameters were estimated from the posterior PDF, in the same manner as for the flux density. The estimated parameters are listed in Table~\ref{tab:gold_size}. 

The observed effective radii, $r_{\rm s}$, range from $0.05\pm0.01\arcsec$ up to $0.37\pm0.21\arcsec$. The reliability of these measurements is a function of their measured signal-to-noise, as it is harder to measure the emission extent at low $S/N$. In order to verify which sources are truly extended, we ingested fake point sources with a range of flux densities directly into the observed visibility data for each galaxy cluster. The ingested flux densities were chosen to recover similar $S/N$ and flux density values as for the detected sources. Once the sources were ingested, the sizes where measured in the same manner as done for the real detections.  

Figure \ref{fig:size_point_source} shows the measured effective radii for each detected source as well for the full range of ingested point sources as a function of flux density. As noted previously, the achievable size limit for a point source is less accurate at lower $S/N$, making it more challenging to measure the emission extent. To account for this, we define a source size as significantly extended when its measured effective radius lies $\approx$2$\sigma$ above the mean size measured for point sources with the same flux density (denoted by the second dashed line in Fig. \ref{fig:size_point_source}). Sources with measured extents above this 2$\sigma$ threshold but error bars that drop below it are considered marginally extended, while all remaining sources (i.e., any within $\approx$2$\sigma$ of the mean) are considered point-like. only two sources (A2744-ID07 and MACSJ0416-ID03) are fully consistent with being point sources, while seven  are clearly significantly extended. The 2$\sigma$ line for MACSJ1149 is a bit lower than the one plotted for MACSJ0416 so the source is classified as extended despite of its lower error bar going below the 2$\sigma$ line. We consider a further three (A2744-ID01, A2744-ID04 and MACSJ0416-ID04) to be significantly extended, although their error bars put them close to the threshold for the point source size distribution. The extent classifications are listed in the last column of Table~\ref{tab:gold_size}.

To test the reliability of the measured sizes for sources with low signal-to-noise which were found to be extended, we simulated 10 extended sources with the same size properties as MACSJ0416-ID04, the detection with the lowest $S/N$ in all three clusters. The assumed properties are $r_{\rm s}=0\farcs37$, $q_{\rm s}=0.56$ and $\phi_{\rm s}=81^{\circ}$. After fitting the sizes of the ingested sources, the average parameter measurements were $r_{\rm s}=0\farcs36\pm0\farcs04$, $q_{\rm s}=0.55\pm0.08$ and $\phi_{\rm s}=83.5^{\circ}\pm13.4^{\circ}$. These all lie within $1\sigma$ of the input parameters for the ingested sources, demonstrating that reliable size measurements can be made for extended sources with $S/N$ values similar to even our weakest detections. Unfortunately, a full characterization of the reliability of size measurements for low signal-to-noise sources using \texttt{uvmcmcfit} over all parameter space is well beyond the scope of this paper, and we refer interested readers to \citet{Bussmann2013,Bussmann2015} for such details.

In order to analyze the delensed size distribution of the extended sources, we need to estimate the magnification values of each galaxy behind the galaxy cluster, which relies both on the source redshift and the cluster lensing model. MACSJ0416-ID01 and MACSJ0416-ID02 are the only sources with measured redshifts, $z=2.086$ and $z=1.953$ respectively, obtained from the Grism Lens-Amplified Survey from Space \citep[GLASS, ][]{Treu2015}. For the rest of the sources, we will assume that they lie at $z$$=$2 and use the range $z$$=$1--3 as an estimate of the error associated with this assumption. This average redshift and dispersion are consistent with the redshift distribution of published  1.1\,mm detected galaxies found by ALMA to date \cite{Simpson2014, Dunlop2016}. Since none of the sources are close to the critical curves of the clusters, as seen in Fig. \ref{fig:map_A2744}, \ref{fig:map_MACS0416} and \ref{fig:map_MACS1149}, the redshift uncertainty will not be critical to the magnification estimate.

As stated above, each of these cluster fields has a set of lensing models that were created by several independent teams using different assumptions and techniques 
\citep{Bradac2005,Bradac2009,Diego2015,Jauzac2015a,Jauzac2015b,Jauzac2016,Johnson2014,Kawamata2016,Liesenborgs2006,Merten2009,Merten2011,Sebesta2015,Williams2014,Zitrin2009,Zitrin2013}.

We estimate the magnification values as in \cite{Coe2015}, by taking the 16th, 50th and 84th percentiles of the magnification values given by the full range from all the models in the FFs archives. We propagate the $z$$=$1--3 redshift uncertainty for the sources which lack spectroscopic redshifts into the magnification errors.
For A2744 and MACS0416 we only use the v3 or newer models, as the v3 models correspond to those made using the deep Frontier Fields images for modeling. For MACSJ1149, no v3 models are available, so we use the latest available model from each team.

In Table~\ref{tab:gold_size} we present the derived magnification values for the sources. Most of the magnification values lie around $\mu\sim2$, with the highest reaching a value of $\mu=4.2$. For MACSJ0416-ID01 and MACSJ0416-ID02, the galaxies with spectroscopic redshifts, the magnification errors are produced solely by the systematic errors among the lens models of different teams. A comparison of the magnification errors between sources should demonstrate that our adopted redshift uncertainty is not a dominating factor in the magnification error budget.   

With magnification estimates in hand, we proceed to determine the intrinsic parameters of the detected sources. We correct the effective radius as $r_{s}/\sqrt{\mu}$ and the flux density as $F_{\rm uv-fit}/\mu$. The obtained values are presented in Table~\ref{tab:gold_size}. The errors in the lensing-corrected values are estimated based on the $1\sigma$ range of the resulting distributions of $r_{s}/\sqrt{\mu}$ and $F_{\rm uv-fit}/\mu$ using the parent distributions. 

With this information, we now compare our demagnified angular sizes and flux densities with those of bright sources recently constrained by ALMA and previous interferometric studies in Figure~\ref{fig:fluxes_sizes}. 
The comparison samples correspond to: four bright sources observed with the Submillimeter Array \citep[SMA][]{Ho2004} presented by \cite{Younger2008,Younger2010}; the lensed galaxy SDP.81 observed by \cite{ALMA2015} and with an intrinsic size estimated by \mbox{\cite{Valtchanov2011,Rybak2015}}; objects from the UKIDSS ultra deep survey (UDS) part of the SCUBA-2 cosmology legacy survey and the un-lensed objects from the {\it Herschel} Multi-tiered Extragalactic Survey (HerMES), both observed with ALMA \citep{Simpson2015,Bussmann2015}. 

We do not include strongly lensed galaxies identified in wide field surveys, since they can affected by the size bias. These type of surveys select galaxies by observed flux density, which is the product of the intrinsic flux density and magnification. This typically favors sources with high magnification values, which in turn biases sources to be preferentially smaller. This effect is produced by the small region of high magnification in the source plane, as small sources near the source plane caustics will have a higher flux-weighted magnification than more extended sources \citep{Serjeant2012,Hezaveh2012,Wardlow2013,Spilker2016}. The ALMA-FFs sample is much less affected by the size bias, since none of our detections lie close to the critical curves and therefore do not have high magnification values.

It is immediately clear that the samples studied here probe observed flux densities up to an order of magnitude fainter than those measured in the comparison samples. Roughly 2/3 of the ALMA-FFs sources have effective radii of $r_{\rm s, demag}$$\gtrsim$0\farcs16, compared to the average of $<r_{\rm s}>$$\sim$0\farcs1 measured in brighter samples. This high fraction of more extended sources could imply that the sub-mJy population is intrinsically more extended than the brighter compact sources. However, once we factor in the two point-like sources ($r_{\rm s}$$\lesssim$0\farcs05) and the three marginally extended sources, there appears to be considerable dispersion among the fainter ALMA-FFs population. To investigate this issue further, we fit a linear regression between flux density and effective radius as $\log{r_{\rm s}} = m\times\log{F_{\nu}} + b$ to the ALMA-FFs and HerMES samples, where the same code was used to determine $r_{\rm s}$. For simplicity, we convert the upper limits for the point-like sources in our sample into measured sizes. The best-fit regression values are $m=-0.08\pm0.10$ and $b=-0.95\pm0.05$, demonstrating that there is no obvious size evolution. In Figure~\ref{fig:fluxes_sizes} we show the fit relation (dashed line) together with its $2\sigma$ range (orange region).

\subsection{Positional Offsets}

\begin{table*}
\caption[]{High-significance ($\geq5\sigma$) continuum detections NIR counterpart positions and offsets.
{\it Col. 1}: Source ID.
{\it Cols. 2-3}: Centroid J2000 position of ID in hh:mm:ss.ss+dd:mm:ss.ss for the NIR counterpart.
{\it Cols. 4}: Positional error in arcseconds.
{\it Col. 5}: Offset measured in the image plane in arcseconds.
{\it Col. 6}: Offset measured in the source plane in arcseconds.
\label{tab:counterpart_offset}}
\centering
\begin{tabular}{cccccc}
\hline     
{ID}&  {$\alpha_{\rm J2000}$} & {$\delta_{\rm J2000}$} & $\Delta\alpha$, $\Delta\delta$ &{Image plane offset} & {Source plane offset} \\
 &   [hh:mm:ss.ss] & [$\pm$dd:mm:ss:ss] & [$\arcsec$] & [$\arcsec$] &  [$\arcsec$]\\
\hline
\noalign{\smallskip}
A2744-ID01  &  00:14:19.8058 & -30:23:07.6094 & 0.002 , 0.002 & $ 0.09 \pm 0.01 $ & $ 0.04 \pm 0.01 $\\
A2744-ID02  &  00:14:18.2000 & -30:24:47.3000 & 0.002 , 0.001 & $ 0.68 \pm 0.05 $ & $ 0.48 \pm 0.04 $\\
A2744-ID03  &  00:14:20.4013 & -30:22:54.6038 & 0.007 , 0.006 & $ 0.18 \pm 0.02 $ & $ 0.16 \pm 0.02 $\\
A2744-ID04  &  00:14:17.5816 & -30:23:00.7324 & 0.003 , 0.004 & $ 0.18 \pm 0.02 $ & $ 0.11 \pm 0.02 $\\
A2744-ID05  &  00:14:19.1273 & -30:22:42.3394 & 0.007 , 0.003 & $ 0.17 \pm 0.04 $ & $ 0.15 \pm 0.04 $\\
A2744-ID06  &  00:14:17.2759 & -30:22:58.8000 & 0.003 , 0.003 & $ 0.22 \pm 0.07 $ & $ 0.12 \pm 0.05 $\\
A2744-ID07  &  00:14:22.1091 & -30:22:49.7821 & 0.004 , 0.002 & $ 0.16 \pm 0.03 $ & $ 0.11 \pm 0.02 $\\
MACSJ0416-ID01  &  04:16:10.7762 & -24:04:47.5002 & 0.001 , 0.001 & $ 0.20 \pm 0.03 $ & $ 0.18 \pm 0.03 $\\
MACSJ0416-ID02  &  04:16:06.9539 & -24:03:59.9277 & 0.002 , 0.003 & $ 0.14 \pm 0.08 $ & $ 0.12 \pm 0.06 $\\
MACSJ0416-ID03  &  04:16:08.8169 & -24:05:22.3970 & 0.003 , 0.003 & $ 0.22 \pm 0.06 $ & $ 0.17 \pm 0.06 $\\
MACSJ0416-ID04  &  04:16:11.6688 & -24:04:19.6271 & 0.002 , 0.002 & $ 0.23 \pm 0.12 $ & $ 0.18 \pm 0.08 $\\
MACSJ1149-ID01  &  11:49:36.0961 & +22:24:24.4659 & 0.002 , 0.001 & $ 0.20 \pm 0.09 $ & $ 0.25 \pm 0.14 $\\

\hline
\end{tabular}
\end{table*}

\begin{figure}[!htbp]
\centering
\includegraphics[width=\hsize]{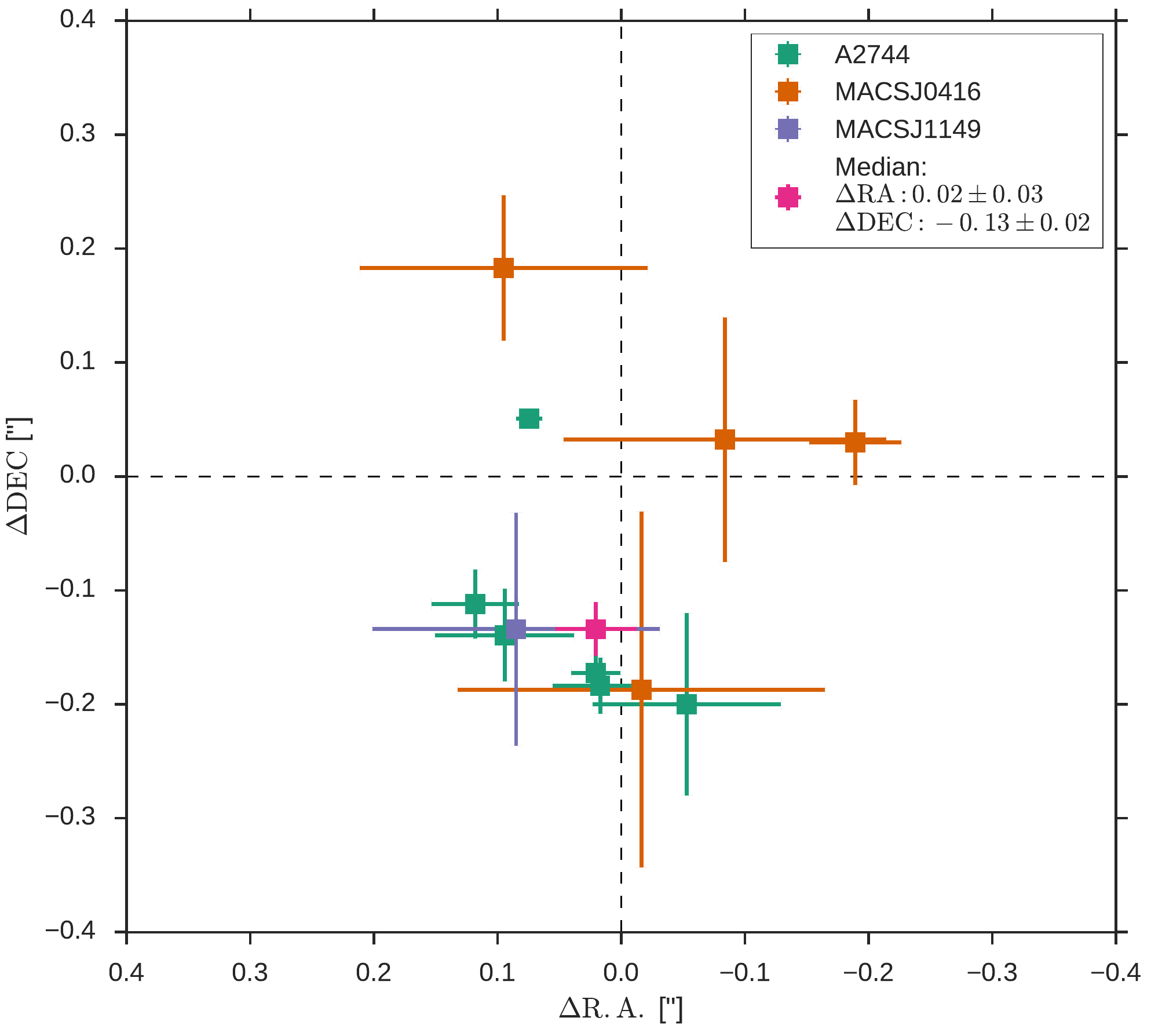}
\caption{Positional offsets between the ALMA detections and the nearest optical/NIR detections. We find good agreement between the positions of the sources in the FIR and in the optical rest-frame.  The high-significance continuum detections demonstrate that the astrometric consistency between the {\it HST} and ALMA reference frames appears robust to $\approx$0\farcs1 and that the NIR and FIR/mm emission are nearly co-spatial for most of the cases. 
\label{fig:offsets}}
\end{figure}

Because the ALMA-detected sources may be optically faint and/or red, we use the deep {\it HST} WFC3 $F160W$ image in the FFs to search for NIR counterparts. In general, we adopt the closest galaxy as the true counterpart (typical separations smaller than $\sim0.2\arcsec$). Once the counterparts are selected, we fit a simple 2-d elliptical Gaussian to the NIR emission to more accurately measure the positions of the counterparts centroids in the same manner as the positions of the ALMA sources.

In Figure~\ref{fig:offsets}, we present the measured positional offsets between the 1.1\,mm and NIR emission of the detected galaxies. A2744-ID02 was left out of this analysis, as it appears to arise from a very obscured part of an extended counterpart galaxy, as described in $\S$\ref{Sec:high_significance_cont_detections}. This offset appears real and highlights an important case when the FIR/mm and optical/NIR are not co-spatial, similar to physical offsets seen in the local starburst galaxy NGC4038/9 \citep{Wang2004,Klaas2010}. The intrinsic offset between the FIR/mm and optical/NIR emission can be enlarged by lensing when observed in the image plane, as in our case. Since one of the goals of this analysis is to check the astrometric agreement between {\it HST} and ALMA, we therefore exclude this galaxy from our offset measurement.
The measured median offsets between the NIR and mm sources are $\Delta{\rm RA}$$=$0\farcs02$\pm$0\farcs03 and $\Delta{\rm DEC}$$=$$-$0\farcs13$\pm$0\farcs02. with a combined offset scatter of $\approx0\farcs1$. The errors in the offsets include the uncertainties of the NIR and ALMA emissions centers presented in Table \ref{tab:counterpart_offset} and plotted in Figure \ref{fig:counterparts}. These offsets are consistent with the astrometric offsets ($\approx0\farcs1$) measured for the Subaru catalogs to which the {\it HST} A2744 data was tied. \footnote{\url{https://archive.stsci.edu/pub/hlsp/frontier/abell2744/catalogs/subaru/astrometry/}}.

By comparing the observed positions of the phase calibrators used in the ALMA observations to those in the literature we find offsets of $\Delta{\rm RA}$=$<0\farcs01$ and $\Delta{\rm DEC}$=$<0\farcs01$ for A2744 and MACSJ1149 and $\Delta{\rm RA}$=$0\farcs012$ and $\Delta{\rm DEC}$=$<0\farcs016$ for MACSJ0416. Although the measured offset in MACSJ0416 is somewhat larger than those measured in the other two clusters, all are minimal compared to the observed offsets in the detected sources, indicating that the measured offsets are not due to astrometric problems in the ALMA observations.

The small scatter measured for the secure sample indicates that the NIR and FIR/mm emission are nearly co-spatial, except for the the special case of A2744-ID02 described above. These results also demonstrate that the astrometric consistency between the {\it HST} and ALMA reference frames appears robust to $\approx$0\farcs1. The measured offsets could be due to small astrometric offsets or true spatial offsets between the sources optical/NIR and FIR/mm emission. Under the assumption that we have a perfect astrometric agreement between ALMA and {\it HST}, the average offset for all the sources plotted in Fig.~\ref{fig:offsets} is $0\farcs17\pm0\farcs02$. To rule out the effects of lensing, we tracked the positions of the centroids in ALMA and {\it HST} to the source plane using the lensing models of Zitrin-NFWv3 and Zitrin-LTMv1 and measured intrinsic offsets. We selected Zitrin-NFW models since they were observed to better follow the median distribution of magnifications given by the combinations of all lens models \citet{Priewe2016}. This measurement includes the effect of lensing shear. Individual source plane offsets are given in Table~\ref{tab:counterpart_offset}, while the average demagnified offset is still $0\farcs14\pm0\farcs02$, which is $\approx$1.2\,kpc at $z=2$. The source plane offset errors include the positional uncertainties in both the NIR and ALMA emission but we have not propagated the systematic lens models uncertainties. For A2744-ID02, the demagnified offset to the brightest peak, assuming $z=2$, is $0\farcs48\pm0\farcs04$ including only the uncertainties in the position of both NIR and ALMA emissions. This offsets corresponds to $4.1\pm0.3$ kpc at $z=2$ similar than the spatial offset of $\sim4$ kpc measured in GN20 at $z=4.05$, one of the best studied SMGs at high redshift \citep{Hodge2015}. It is clear that these ALMA observations can open a window to study and resolve the obscured star-formation activity in galaxies at high redshift.

While A2744-ID02 is the most obvious case, an interesting aspect of the counterpart offsets is that for nine out of 12 ALMA-FFs sources (i.e., A2744-ID01, A2744-ID02, A2744-ID03, A2744-ID04, A2744-ID06, A2744-ID07, M0416-ID02, M0416-ID03, and M0416-ID04), the ALMA centroid position falls on a darker portion of the counterpart galaxy compared to the NIR peak (see Figure~\ref{fig:cont_fits1}). This effect has been noted other ALMA surveys \citep[e.g.,][]{Wiklind2014, Dunlop2016}, and attributed to the fact that the FIR/mm emission region likely suffers from strong dust extinction. This physical effect may be an important term in the remaining measured offsets, although the current error bars make this difficult to constrain for individual sources.

\section{Summary}
We have presented an analysis of the ALMA observations for the Frontier Fields galaxy clusters A2744, MACSJ0416, and MACSJ1149. 
The deepest 1.1\,mm ALMA natural weighted images made for the three clusters achieve rms sensitivities
of 55, 59 and 71 $\mu$Jy\,beam$^{-1}$ with natural weighting, respectively. The beam sizes range from $\approx$0\farcs5--1\farcs5 for the FFs clusters images, with A2744 achieving higher resolution due to observations partially made in a more extended configuration.

The mosaic sensitivity of the observations vary from cluster to cluster. A2744 shows a fairly uniform sensitivity across the mosaic. MACJS0416 and MACSJ1149 show lower sensitivities in some areas of the mosaics, produced mainly by an incomplete execution and shadowing during observations at low elevations. These sensitivity differences across the mosaic were accounted for when deriving source properties and purity studies of the detected sources. 

A total of 12 sources are detected with S/N$\geq$5 in the natural-weighted maps created for the three clusters. The range in observed flux densities goes from $0.411$ to $2.816$ mJy. Using the code \texttt{uvmcmcfit} we fit 2D Gaussian models in the uv-plane to the detected sources to estimate total flux densities and angular sizes of the galaxies at 1.1\,mm. The range of observed effective radii in the fit Gaussian models goes from $\lesssim$0\farcs05 to $0\farcs37\pm0\farcs21$. We estimated magnification values using the available lensing models, assuming that the sources lie at $z=2$ if no spectroscopic redshift is available. The resulting magnification values range between $\mu\sim$$=$1.5--4.2, and were used to correct the sizes and flux densities. We find that the demagnified (intrinsic) sizes of the FFs sample are consistent with brighter sources previously measured \citep{Simpson2015,Bussmann2015}. However, there is considerable dispersion in the FFs sample and roughly 2/3 of the FFs sources have demagnified sizes that are a factor of $\gtrsim$1.6 larger than the average of the brighter sources, implying that a substantial portion of the sub-mJy submm sources may be mildly more extended than their brighter counterparts. Larger samples are required to confirm this.

We find that all but one of the ALMA detections has a clear {\it HST} $F160W$ counterpart, with an observed scatter of $\sim0.1\arcsec$. The small scatter indicates that in general the mm and optical emission are effectively co-spatial and that the astrometry between the ALMA and {\it HST} observations is in reasonably good agreement. We note that there may be a possible offset of $\sim$$-$0\farcs13 in Declination (likely in the A2744 field). If we assume perfect astrometric agreement between ALMA and {\it HST}, the observed offset would represent a real spatial offset between the peak of the optical and FIR emission of $\sim1$\,kpc on average. For A2744-ID02, the offset between peak mm and NIR emission is substantially larger, $\sim$0\farcs5 or $\sim4$\,kpc in the source plane. The latter source provides an intriguing case to study spatially resolved highly obscured star-formation activity at high redshift.  

The continuum images used in this paper, as well as the sensitivity maps and visibilities are available for download at this web address (\url{http://www.astro.puc.cl/~jgonzal/ALMA_FF.html}).

\begin{acknowledgements}
This paper makes use of the following ALMA data: ADS/JAO.ALMA\#2013.1.00999.S.
ALMA is a partnership of ESO (representing its member states), NSF (USA) and
NINS (Japan), together with NRC (Canada) and NSC and ASIAA (Taiwan), in
cooperation with the Republic of Chile. The Joint ALMA Observatory is operated
by ESO, AUI/NRAO and NAOJ.\\
We acknowledge support from
CONICYT-Chile grants Basal-CATA PFB-06/2007 (JGL, FEB, RC, LI, NN), 
FONDECYT Regular 1141218 (JGL, FEB, RC), 
FONDECYT grant 3150238 (CRC), 
"EMBIGGEN" Anillo ACT1101 (FEB, NN), and
the Ministry of Economy, Development, and Tourism's Millennium Science
Initiative through grant IC120009, awarded to The Millennium Institute
of Astrophysics, MAS (FEB,CRC).\\
AZ is supported by NASA through Hubble Fellowship grant \#HST-HF2-51334.001-A awarded by STScI, which is operated by the Association of Universities for Research in Astronomy, Inc. under NASA contract NAS~5-26555.
AMMA acknowledges support from FONDECYT grant 3160776.
RD gratefully acknowledges the support provided by the BASAL Center for Astrophysics and Associated Technologies (CATA), and by FONDECYT grant N. 1130528.
This  work  was  partially supported  by  the Transregional  Collaborative  Research  Centre  TRR  33  (M.C.).
SK is supported by FONDECYT grant N. 3130488
We acknowledge support from European Research Council Advanced Grant FP7/669253 (NL)
This work utilizes gravitational lensing models produced by PIs Bradač, Natarajan \& Kneib (CATS), Merten \& Zitrin, Sharon, and Williams, and the GLAFIC and Diego groups. This lens modeling was partially funded by the HST Frontier Fields program conducted by STScI. STScI is operated by the Association of Universities for Research in Astronomy, Inc. under NASA contract NAS 5-26555. The lens models were obtained from the Mikulski Archive for Space Telescopes (MAST).
\end{acknowledgements}



\begin{thebibliography}{}

\bibitem[Alexander et al.(2005)]{Alexander2005} Alexander, D. M., Bauer, F. E., Chapman, S. C., Smail, I., Blain, A. W., Brandt, W. N., \& Ivison, R. J. 2005, \apj, 632, 736

\bibitem[ALMA Partnership et al.(2015)]{ALMA2015} ALMA 
Partnership, Vlahakis, C., Hunter, T.~R., et al.\ 2015, \apjl, 808, L4 

\bibitem[Atek et al.(2014)]{Atek2014} Atek, H., Richard, J., Kneib, J.-P., et al.\ 2014, \apj, 786, 60 

\bibitem[Barger et al.(2000)]{Barger2000} Barger, A. J., Cowie, L. L. \& Richards, E. A.  2000, \aj, 119, 2092

\bibitem[Blain et al.(1996)]{Blain1996} Blain, A. W. 1996, MNRAS, 283, 1340

\bibitem[Blain et al.(2002)]{Blain2002} Blain, A. W., Smail, I., Ivison, R. J., Kneib, J., \& Frayer, D. T. 2002, Phys. Rep., 369, 111

\bibitem[Brada{\v c} et al.(2005)]{Bradac2005} Brada{\v c}, M., Schneider, P., Lombardi, M., \& Erben, T.\ 2005, \aap, 437, 39 

\bibitem[Brada{\v c} et al.(2009)]{Bradac2009} Brada{\v c}, M., Treu, T., Applegate, D., et al.\ 2009, \apj, 706, 1201 

\bibitem[Briggs et al.(1999)]{Briggs1999} Briggs, D.~S., Schwab, 
F.~R., 
\& Sramek, R.~A.\ 1999, Synthesis Imaging in Radio Astronomy II, 180, 127 

\bibitem[Bussmann et al.(2013)]{Bussmann2013} Bussmann, R. S., P{\'{e}}rez-Fournon, I., Amber, S., et al. 2013, \apj, 779, 25 (B13)

\bibitem[Bussmann et al.(2015)]{Bussmann2015} Bussmann, R.~S., Riechers, D., Fialkov, A., et al.\ 2015, \apj, 812, 43 


\bibitem[Casey et al.(2014)]{Casey2014} Casey, C. M., Narayanan, D. \& Cooray, A. 2014, Physics Reports, 541, 45-161

\bibitem[Chapman et al.(2005)]{Chapman2005} Chapman, S. C., Blain, A. W., Smail, I., \& Ivison, R. J. 2005, \apj, 622, 772

\bibitem[Coe et al.(2015)]{Coe2015} Coe, D., Bradley, L., \& Zitrin, A.\ 2015, \apj, 800, 84 

\bibitem[Cowie et al.(2002)]{Cowie2002} Cowie, L. L., Barger, A. J. \& Kneib, J.-P.  2002, \aj, 123, 2197


\bibitem[Diego et al.(2015)]{Diego2015} Diego, J.~M., Broadhurst, 
T., Molnar, S.~M., Lam, D., \& Lim, J.\ 2015, \mnras, 447, 3130 

\bibitem[Dunlop et al.(2016)]{Dunlop2016} Dunlop, J.~S., McLure, R.~J., Biggs, A.~D., et al.\ 2016, arXiv:1606.00227 


\bibitem[Foreman-Mackey et al.(2013)]{Foreman-mackey2013} Foreman-Mackey, D., Hogg, D.~W., Lang, D., \& Goodman, J.\ 2013, \pasp, 125, 306 

\bibitem[Fujimoto et al.(2016)]{Fujimoto2016} Fujimoto, S., Ouchi, 
M., Ono, Y., et al.\ 2016, \apjs, 222, 1 

\bibitem[Hezaveh et al.(2012)]{Hezaveh2012} Hezaveh, Y.~D., Marrone, D.~P., \& Holder, G.~P.\ 2012, \apj, 761, 20 

\bibitem[Ho et al.(2004)]{Ho2004} Ho, P.~T.~P., Moran, J.~M., 
\& Lo, K.~Y.\ 2004, \apjl, 616, L1 

\bibitem[Hodge et al.(2015)]{Hodge2015} Hodge, J.~A., Riechers, D., Decarli, R., et al.\ 2015, \apjl, 798, L18 

\bibitem[Infante et al.(2015)]{Infante2015} Infante, L., Zheng, W., Laporte, N., et al.\ 2015, \apj, 815, 18 

\bibitem[Ikarashi et al.(2014)]{Ikarashi2014} Ikarashi, S. et al. 2011, \mnras, 415, 3081 (I14)


\bibitem[Jauzac et al.(2015a)]{Jauzac2015a} Jauzac, M., Jullo, E., 
Eckert, D., et al.\ 2015, \mnras, 446, 4132 

\bibitem[Jauzac et al.(2015b)]{Jauzac2015b} Jauzac, M., Richard, J., 
Jullo, E., et al.\ 2015, \mnras, 452, 1437 

\bibitem[Jauzac et al.(2016)]{Jauzac2016} Jauzac, M., Richard, J., Limousin, M., et al.\ 2016, \mnras, 457, 2029 


\bibitem[Johnson et al.(2014)]{Johnson2014} Johnson, T.~L., Sharon, K., Bayliss, M.~B., et al.\ 2014, \apj, 797, 48 


\bibitem[Kawamata et al.(2016)]{Kawamata2016} Kawamata, R., Oguri, M., Ishigaki, M., Shimasaku, K., \& Ouchi, M.\ 2016, \apj, 819, 114 

\bibitem[Klaas et al.(2010)]{Klaas2010} Klaas, U., Nielbock, M., Haas, M., Krause, O., \& Schreiber, J.\ 2010, \aap, 518, L44 

\bibitem[Laporte et al.(2015)]{Laporte2015} Laporte, N., Streblyanska, A., Kim, S., et al.\ 2015, \aap, 575, A92 

\bibitem[Liesenborgs et al.(2006)]{Liesenborgs2006} Liesenborgs, J., De Rijcke, S., \& Dejonghe, H.\ 2006, \mnras, 367, 1209 

\bibitem[Lotz et al.(2016)]{Lotz2016} Lotz, J.~M., Koekemoer, A., Coe, D., Grogin, N., Capak, P., Mack, J., Anderson, J., Avila, R., Barker, E. A., Borncamp, D., Brammer, G., Durbin, M., Gunning, H., Hilbert, B., Jenkner, H., Khandrika, H., Levay, Z., Lucas, R. A., MacKenty, J., Ogaz, S., Porterfield, B., Reid, N., Robberto, M., Royle, P., Smith, L. J., Storrie-Lombardi, L. J., Sunnquist, B., Surace, J., Taylor, D. C., Williams, R., Bullock, J., Dickinson, M., Finkelstein, S., Natarajan, P., Richard, J., Robertson, B., Tumlinson, J., Zitrin, A., Flanagan, K., Sembach, K., Soifer, B. T., \& Mountain, M. 2016, \apj, submitted (arXiv:1605.06567)


\bibitem[Magnelli et al.(2011)]{Magnelli2011} Magnelli, B., Elbaz, D., Chary, R. R., Dickinson, M., Le Borgne, D., Frayer, D. T. \& Willmer C. N. A. 2011, \aap, 528, A35

\bibitem[McLeod et al.(2015)]{Mcleod2015} McLeod, D.~J., McLure, R.~J., Dunlop, J.~S., et al.\ 2015, \mnras, 450, 3032 

\bibitem[McMullin et al.(2007)]{Mcmullin2007} McMullin, J.~P., Waters, B., Schiebel, D., Young, W., \& Golap, K.\ 2007, Astronomical Data Analysis Software and Systems XVI, 376, 127 

\bibitem[Merten et al.(2009)]{Merten2009} Merten, J., Cacciato, M., Meneghetti, M., Mignone, C., \& Bartelmann, M.\ 2009, \aap, 500, 681 

\bibitem[Merten et al.(2011)]{Merten2011} Merten, J., Coe, D., Dupke, R., et al.\ 2011, \mnras, 417, 333 



\bibitem[Miettinen et 
al.(2015)]{Miettinen2015} Miettinen, O., Smol{\v c}i{\'c}, V., Novak, M., et al.\ 2015, \aap, 577, A29  (M15)


\bibitem[Negrello et al.(2007)]{Negrello2007} Negrello, M., Perrotta, F., Gonz{\'{a}}lez-Nuevo, J., Silva, L., de Zotti, G., Granato, G. L., Baccigalupi, C., \& Danese, L. 2007, \mnras, 377, 1557 

\bibitem[Negrello et al.(2010)]{Negrello2010} Negrello, M., Hopwood, R., De Zotti, G., et al. 2010, Science, 330, 800



\bibitem[Noeske et al.(2007)]{Noeske2007} Noeske, K. G., Weiner, B. J., Faber, S. M., et al. 2007, \apj, 660, L43

\bibitem[Pardo et al.(2001)]{Pardo2001} Pardo, J.~R., Cernicharo, 
J., 
\& Serabyn, E.\ 2001, IEEE Transactions on Antennas and Propagation, 49, 1683 

\bibitem[Pedregosa et al.(2011)]{Pedregosa2011} Pedregosa et al., JMLR 12, pp. 2825-2830, 2011.

\bibitem[Priewe et al.(2016)]{Priewe2016} Priewe, J., Williams, L.~L.~R., Liesenborgs, J., Coe, D., \& Rodney, S.~A.\ 2016, arXiv:1605.07621 

\bibitem[Rawle et al.(2016)]{Rawle2016} Rawle, T.~D., Altieri, B., Egami, E., et al.\ 2016, \mnras, 459, 1626 

\bibitem[Richard et al.(2014)]{Richard2014} Richard, J., Jauzac, M., Limousin, M., Jullo, E., Cl{\'e}ment, B., Ebeling, H., Kneib, J.-P., Atek, H., Natarajan, P., Egami, E., Livermore, R. \& Bower, R. 2014, \mnras, 444, 268 

\bibitem[Rybak et al.(2015)]{Rybak2015} Rybak, M., McKean, J.~P., 
Vegetti, S., Andreani, P., \& White, S.~D.~M.\ 2015, \mnras, 451, L40 

\bibitem[Sebesta et al.(2015)]{Sebesta2015} Sebesta, K., Williams, 
L.~L.~R., Mohammed, I., Saha, P., 
\& Liesenborgs, J.\ 2015, arXiv:1507.08960 

\bibitem[Serjeant(2012)]{Serjeant2012} Serjeant, S.\ 2012, \mnras, 424, 2429 

\bibitem[Simpson et al.(2014)]{Simpson2014} Simpson, J.~M., Swinbank, A.~M., Smail, I., et al.\ 2014, \apj, 788, 125 

\bibitem[Simpson et al.(2015)]{Simpson2015} Simpson, J.~M., Smail, I., Swinbank, A.~M., et al.\ 2015, \apj, 799, 81 (S15)

\bibitem[Smail et al.(1997)]{Smail1997} Smail, I., Ivison, R. J., \& Blain, A. W. 1997, \apj, 490, L5

\bibitem[Smail et al.(2004)]{Smail2004} Smail, I., Chapman, S. C., Blain, A. W. \& Ivison, R. J. 2004, \apj, 616, 71 

\bibitem[Spilker et al.(2016)]{Spilker2016} Spilker, J., Marrone, D., Aravena, M., et al.\ 2016, arXiv:1604.05723 


\bibitem[Swinbank et al.(2010)]{Swinbank2010} Swinbank, A. M., Smail, I., Longmore, S. et al. 2010, Nature, 464, 733

\bibitem[Treu et al.(2015)]{Treu2015} Treu, T., Schmidt, K.~B., Brammer, G.~B., et al.\ 2015, \apj, 812, 114 

\bibitem[Valtchanov et al.(2011)]{Valtchanov2011} Valtchanov, I., 
Virdee, J., Ivison, R.~J., et al.\ 2011, \mnras, 415, 3473 

\bibitem[Viero et al.(2013)]{Viero2013} Viero, M. P., Moncelsi, L., Quadri, R. F. et al. 2013, \apj, 
779, 32

\bibitem[Wang et al.(2004)]{Wang2004} Wang, Z., Fazio, G.~G., Ashby, M.~L.~N., et al.\ 2004, \apjs, 154, 193 

\bibitem[Wardlow et al.(2013)]{Wardlow2013} Wardlow, J. L., Cooray, A., De Bernardis, F., et al. 2013, \apj, 762, 59


\bibitem[Wiklind et al.(2014)]{Wiklind2014} Wiklind, T., Conselice, C.~J., Dahlen, T., et al.\ 2014, \apj, 785, 111 

\bibitem[Williams et al.(2014)]{Williams2014} Williams, R.~J., Wagg, J., Maiolino, R., et al.\ 2014, \mnras, 439, 2096 

\bibitem[Younger et al.(2008)]{Younger2008} Younger, J.~D., Fazio, 
G.~G., Wilner, D.~J., et al.\ 2008, \apj, 688, 59 


\bibitem[Younger et al.(2010)]{Younger2010} Younger, J.~D., Fazio, 
G.~G., Ashby, M.~L.~N., et al.\ 2010, \mnras, 407, 1268 


\bibitem[Zheng et al.(2014)]{Zheng2014} Zheng, W., Shu, X., Moustakas, J., et al.\ 2014, \apj, 795, 93 

\bibitem[Zitrin et al.(2009)]{Zitrin2009} Zitrin, A., Broadhurst, 
T., Umetsu, K., et al.\ 2009, \mnras, 396, 1985 

\bibitem[Zitrin et al.(2013)]{Zitrin2013} Zitrin, A., Meneghetti, 
M., Umetsu, K., et al.\ 2013, \apjl, 762, L30 

\bibitem[Zitrin et al.(2014)]{Zitrin2014} Zitrin, A., Zheng, W., Broadhurst, T., et al.\ 2014, \apjl, 793, L12 



\end{thebibliography}
\end{document}